\documentclass[twocolumn,twocolappendix]{aastex701} 

\usepackage{amsmath,amstext}
\usepackage[T1]{fontenc}
\usepackage{txfonts} 
\usepackage[figure,figure*]{hypcap} 

\usepackage{hyperref}
\hypersetup{colorlinks=true, linkcolor=blue, filecolor=blue, urlcolor=blue}

\usepackage{graphicx}
\usepackage{subcaption}
\usepackage[export]{adjustbox}
\usepackage{float}

\newcommand*{\kms}{km\,s$^{-1}$}
\newcommand{\Msun}{\mbox{$M_{\odot}$}}

\newcommand{\rell}{r_{\rm half,ell}}

\newcommand{\ntot}{56}
\newcommand{\nmem}{49}
\newcommand{\nmix}{65}

\newcommand{\vsys}{-13.0^{+1.1}_{-1.1}}
\newcommand{\vdisp}{4.7^{+1.5}_{-1.3}}
\newcommand{\mass}{5.9^{+3.7}_{-3.4} \times 10^5}
\newcommand{\ml}{660 \pm 590}

\newcommand{\pmra}{0.299 \pm 0.052}
\newcommand{\pmdec}{-1.097 \pm 0.075}

\newcommand{\met}{-2.45^{+0.12}_{-0.13}}
\newcommand{\metdisp}{0.30^{+0.15}_{-0.11}}

\newcommand{\rapo}{43.2^{+0.5}_{-0.5}}
\newcommand{\rperi}{14.7^{+1.9}_{-1.6}}
\newcommand{\e}{0.49^{+0.04}_{-0.05}}

\shorttitle{Willman 1}
\shortauthors{Chiu et al.}

\begin{document}

\title{Willman 1 Revisited: The Kinematics, Chemistry, and Orbital Properties of a Potentially Disrupting Dwarf Galaxy}

\correspondingauthor{Camille~Chiu}
\email{camille.chiu@yale.edu}

\author[0000-0001-5915-4264]{Camille~Chiu}
\affiliation{Department of Astronomy, Yale University, New Haven, CT 06520, USA}
\email{camille.chiu@yale.edu}

\author[0000-0002-7007-9725]{Marla~Geha}
\affiliation{Department of Astronomy, Yale University, New Haven, CT 06520, USA}
\email{marla.geha@yale.edu}

\author[0000-0003-1697-7062]{William~Cerny}
\affiliation{Department of Astronomy, Yale University, New Haven, CT 06520, USA}
\email{william.cerny@yale.edu}

\author[0000-0002-3204-1742]{Nitya~Kallivayalil}
\affiliation{Department of Astronomy, University of Virginia, Charlottesville, VA 22904, USA}
\email{njk3r@virginia.edu}

\author[0000-0002-3188-2718]{Hannah~Richstein}
\affiliation{Department of Astronomy, University of Virginia, Charlottesville, VA 22904, USA}
\email{hr8jz@virginia.edu}

\author[0000-0001-9061-1697]{Christopher T. Garling}
\affiliation{Department of Astronomy, University of Virginia, Charlottesville, VA 22904, USA}
\email{txa5ge@virginia.edu}

\author[0000-0003-2892-9906]{Beth Willman}
\affiliation{LSST Discovery Alliance, 933 North Cherry Avenue, Tucson, AZ 85719, USA}
\email{bwillman@lsst-da.org}

\begin{abstract}

The ultra-faint Milky Way satellite Willman 1 (W1; $M_V=-2.6$; $r_{\rm half}\sim27$ pc) was the first stellar overdensity found via resolved stars in the Sloan Digital Sky Survey, yet its classification as either a dwarf galaxy or star cluster remains ambiguous. Using new Keck/DEIMOS spectroscopy, Hubble Space Telescope photometry, and orbital modeling, we re-examine the nature of W1. From our updated sample of \ntot\ member stars, we find that past analyses included four binaries and seven nonmembers, identified here using {\it Gaia} proper motions and updated velocities. We continue to find a velocity dispersion consistent with previous analyses, measuring $\sigma_v = \vdisp$ \kms\ from \nmem\ stars out to $3~r_{\rm half}$. If W1 is in equilibrium, this suggests a dynamical mass of $\mass \Msun$ and a mass-to-light ratio of ${\rm (M/L)}_V = \ml$. Based on Ca II triplet measurements, we estimate an iron abundance of ${\rm [Fe/H]}=\met$ and a metallicity dispersion of $\sigma_{\rm [Fe/H]} = \metdisp$ dex. We confirm that W1 does not exhibit mass segregation inside $\sim$1 $r_{\rm half}$. Our best-fit orbital model predicts that W1 is at apocenter, implying that W1 was closer to the Milky Way in the recent past, reaching a pericentric distance $\lesssim$25 kpc from the Galactic center $\sim$0.3 Gyr ago. Given its internal kinematics, metallicity spread, and lack of mass segregation, we conclude that W1 is a galaxy. However, given its orbit and structural properties, which suggest that W1 might be tidally disrupted, and the difficulty of identifying a pure member sample, we caution that the measured internal velocity dispersion may not accurately reflect the dynamical mass of this system.
\end{abstract}

\section{Introduction}

The advent of wide-field photometric surveys such as the Sloan Digital Sky Survey (SDSS) and the Dark Energy Survey (DES) has led to the rapid discovery of several dozen Milky Way satellite galaxies, including populations of ultra-faint dwarf galaxies and ultra-faint star clusters \citep[e.g.,][]{Willman2005, Belokurov2007, Drlica2015, Koposov2015, Torrealba2016, Cerny2024, Homma2024}. The faintest and most compact of these stellar systems, with stellar masses of only a few hundred solar masses and half-light radii of only tens of parsecs, blur the line between star clusters and dwarf galaxies \citep[e.g.,][]{Smith2024,Cerny2026}. The classification of an object as a dwarf galaxy or star cluster typically hinges on one or both of the following diagnostics \citep{willman2012}: (1) a kinematic distribution that cannot be explained solely by its baryonic component, implying the presence of large quantities of dark matter; (2) a spectroscopically observed spread in ${\rm [Fe/H]}$ metallicity, indicating multiple epochs of star formation. Beyond these two criteria, the presence of mass segregation has also been proposed as a discriminator \citep{baumgardt2022}.

One such system that straddles the boundary between dwarf galaxies and star clusters is the Milky Way satellite Willman 1 (W1; SDSSJ1049+5103). W1 was first identified as an overdensity of old, metal-poor stars in SDSS DR2 by \cite{Willman2005} and was the lowest-luminosity galaxy known at the time of discovery \citep[$M_V=-2.5$;][]{Willman2005}. Despite tentative evidence for mass segregation, \cite{willman2006} initially leaned toward a dwarf galaxy classification for W1 based on an observed metallicity spread between three red giant branch (RGB) stars. This classification was supported by Keck/DEIMOS spectroscopy by \cite{Martin2007}, who identified seven RGB members with a significant metallicity spread. However, follow-up spectroscopy from the Hobby--Eberly Telescope by \cite{Siegel2008} determined that at least two and up to five of these identified RGB stars were likely Milky Way foreground dwarf stars, significantly decreasing the observed metallicity spread. From a reanalysis with Keck/DEIMOS spectroscopy and a careful characterization of foreground contamination, \cite{willman2011} identified 45 member stars, 40 of which were classified as high confidence. This included an additional RGB member and two horizontal branch members whose metallicity spread was once again indicative of a dwarf galaxy. These spectroscopic results are supported by the photometric Hubble Space Telescope (HST)-based metallicity distribution function measured by \cite{fu2023}, who determined a significant metallicity spread of $\sigma_{\rm [Fe/H]} = 0.65^{+0.10}_{-0.09}$ dex.

Assuming equilibrium, \cite{willman2011} estimated the dynamical mass of W1 to be $\sim$$3.9 \times 10^5 M_{\odot}$ based on a velocity dispersion of $4.8 \pm 0.8$\,\kms, implying a mass-to-light ratio of $\sim$770. Due to this apparent high dark matter content and relatively close proximity at a distance of $\sim$39 kpc, W1 remains one of the most promising targets in the search for dark matter annihilation signals in gamma rays \citep[e.g.,][]{Aliu2009,Bringmann2009,Li2021,McDaniel2023} and X-rays \citep[e.g.,][]{Nieto2010,Loewenstein2010,Loewenstein2012,saeedi2020}.

Yet, previous studies of W1 found evidence for dynamical disequilibrium, including stellar multidirectional tails and a tentative excess of stars at large half-light radii \citep{willman2006,willman2011}. Specifically, \cite{willman2011} reported an irregular kinematic distribution where stars within the effective half-light radius have a systematically higher velocity than those beyond; this disturbed morphology might point to tidal interactions. If W1 is not in dynamical equilibrium and indeed is undergoing tidal disruption, then the dynamical mass inferred from its internal velocity dispersion may be misleading.

The main difficulty in accurately classifying W1 and determining its properties lies in the kinematic and spectroscopic similarities between the foreground stars in the Milky Way and the members of W1 \citep{Siegel2008,willman2011,saeedi2020}. With an estimated systemic radial velocity of $-12.8 \pm 1.0$ \kms, the velocities of W1 stars substantially overlap with the velocities of Milky Way stars \citep{willman2011}. If W1 is dynamically disturbed, spatial and kinematic cuts, which are often implemented to supplement photometric color-magnitude diagram (CMD) cuts, might instead obscure the odd structural and kinematic features of W1. On the other hand, due to the already small sample size of stars, just a few Milky Way contaminants in the member sample could artificially alter the velocity dispersion and, thus, the inferred dynamical mass of W1, as well as the metallicity spread that is used to characterize it as a dwarf galaxy in the first place. 

To address these uncertainties, this paper revisits W1 and presents a comprehensive study of its structural, kinematic, chemical, and orbital properties. First, we describe our uniform analysis of both new and archival Keck/DEIMOS spectroscopy for W1 and our incorporation of \emph{Gaia} astrometry (\S\ref{sec_data}). We then discuss the probabilistic membership selection performed to reduce the presence of Milky Way foreground contamination (\S\ref{sec_mem}). From this member sample, we measure the properties of W1 including its velocity dispersion, inferred dynamical mass, and metallicity distribution (\S\ref{sec:results}). With HST/Advanced Camera Survey (ACS) imaging of W1, we summarize previous results for W1's star formation history and look for evidence of mass segregation (\S\ref{sec:mass_segregation}). We next model the orbital history of W1 and simulate mock tidal streams (\S\ref{kinematics}). We conclude with a discussion of W1's classification as either a dwarf galaxy or star cluster and its tidal disruption status (\S\ref{dis}).

\section{Observations and Data Reduction}\label{sec_data}

Our kinematic and metallicity analysis is based on a homogeneous reduction of archival Keck/DEIMOS spectroscopic data (\S\ref{ssec:deimos_redux}, \S\ref{ssec:deimos_redux_feh}), half of which are published here for the first time.  We supplement this with ground-based photometry and {\it Gaia} astrometry (\S\ref{ssec:ground_gaia}).

\subsection{Keck/DEIMOS Spectroscopy:  Kinematics}\label{ssec:deimos_redux}

We present homogeneously reduced archival data of W1 from the DEIMOS spectrograph \citep{faber03a} on the Keck II 10 m telescope.   Twelve multislit masks were taken of W1 between 2006 and 2023 (see Table~\ref{tab:masks}). One of these masks was first presented in \citet[][PI: Chapman]{Martin2007} and four masks were published in \citet[][PI: Brodie]{willman2011}. The remaining seven masks are presented here for the first time (PIs: Geha, Rich).  All data were taken with the 1200G grating, providing a spectral resolving power of $R \sim 6000$ at 8500\,\AA. The spectral dispersion of this setup is roughly 0.33\,\AA\ per pixel with a resulting spectral resolution of 1.37\,\AA\ (FWHM).   A summary of observations is presented in Table~\ref{tab:masks}.

A detailed discussion of the data reduction is presented in \citet{geha_paper1} and briefly discussed here. Raw science frames and associated calibrations were obtain through the Keck Observatory Archive\footnote{https://koa.ipac.caltech.edu}.   Data were reduced to one-dimensional wavelength-calibrated spectra using the open-source python-based data reduction code {\tt PypeIt} \citep{pypeit2020}.  {\tt PypeIt} reduces the eight individual DEIMOS detectors as four mosaic images, where each red/blue pair of detectors is reduced together.  

Stellar radial velocities were measured using the {\tt dmost} package \citep{geha_paper1}\footnote{https://github.com/marlageha/dmost}.  {\tt dmost} first pre-processes spectra to identify and remove extragalactic objects using a modified version of the automated redshift software {\tt Marz} \citep{marz}.  For the remaining spectra, {\tt dmost} forward models the one-dimensional spectrum for each source from a given exposure with both a stellar template from the PHOENIX library \citep{Husser2013} and a synthetic telluric absorption spectrum from TelFit \citep{telfit2014}. The velocity is determined for each science exposure through a Markov Chain Monte Carlo (MCMC) procedure constraining both the radial velocity of the target star as well as a wavelength shift of the telluric spectrum needed to correct for slit miscentering \citep[see, e.g.,][]{sohn2007}.   We reject velocities where the posterior shape is far from Gaussian, largely spectra with signal-to-noise ratio (${\rm S/N}$) $<3$.  The final radial velocity for each star is derived through an inverse-variance weighted average of the velocity measurements from each individual exposure.    In cases where less than half of individual science exposures for a given star produce a reliable velocity measurement,  we attempt to measure a velocity based on the one-dimensional coadded spectrum following the same method above.    In both cases, a velocity error scaling and velocity floor is applied to the random errors, based on an assessment from thousands of repeat DEIMOS measurements as discussed in \citet{geha_paper1}.   The velocity error floor is 1.1 \kms\ for a velocity measured on a single mask.

There are 474 unique targets with extracted DEIMOS spectra, plus 28 sources extracted serendipitously in addition to the target source.   We have photometry and were able to measure radial velocities for 340 of these sources.   We find that 125 of these sources are extragalactic, including 14 broad-line quasar spectra.   The remaining 215 sources have velocities and spectra consistent with stars, although we expect only a fraction of these to be W1 members (see \S\ref{ssec:membership}). We note that 12 of these sources have velocity uncertainties > 15 \kms, above the threshold considered as a secure velocity measurement by \citet{geha_paper1}, and they are thus not considered for membership.  We therefore consider 203 stars with measured radial velocities for membership. The radial velocities for the 203 stars and redshifts for the 125 extragalactic sources are available in Table A3 and A4 of \citet{geha_paper1}, respectively.

\subsubsection{Flagging Velocity Variables}\label{ssec:deimos_binaries}

The orbital motions of unresolved binary stars systems when measured at a single epoch can inflate a system's inferred velocity dispersion. To identify such velocity variable stars, we evaluate whether velocities measured at different times are consistent with random fluctuations from a constant value.  For each star, we first calculate the weighted mean velocity, the equivalent of assuming a constant velocity model. We then calculate the $\chi^2$ value for this model, including random and systematic error components. Following \citet{maxted2001}, we then calculate the probability, $p$,  of obtaining the observed value of $\chi^2$ or higher from random fluctuations around a constant value.  $p$ is evaluated for the appropriate number of degrees of freedom (the number of observations minus one).  We define a star as a velocity variable if $\log_{10}p < -1$.  Of the 340 stars with measured velocities, 63 have measurements on multiple DEIMOS mask pointings separated by more than one year. We find 16 of these stars to be velocity variables, six of which are members. Their positions on the CMD are shown in Figure \ref{fig:membership}a.

\begin{deluxetable*}{r C C C C C C C C}
\tablewidth{\columnwidth}
\tabletypesize{\footnotesize}
\tablecaption{Summary of Willman 1 Keck/DEIMOS Observations.\label{tab:masks}}
\tablehead{\colhead{Mask} & \colhead{Date} & \colhead{N$_{\rm exp}$} & \colhead{$\Sigma t_{\rm exp}$} &
    \colhead{PIs} & \colhead{Slit width} & \colhead{$N_{\rm slits}$} & \colhead{$N_{\rm good}/N_{\rm slits}$} & \colhead{References} \\
       \colhead{} & \colhead{} & \colhead{} & \colhead{(sec)} & \colhead{} & \colhead{(arcsec)}&
       \colhead{} & \colhead{} & \colhead{}}
\startdata
\hline
 203WiSB &   20060527 &            3 &        3600 & Chapman &       0.7 &  63 &     0.62 & \text{\cite{Martin2007}} \\
 W1\_1    &   20061120 &            5 &        9000 & Brodie  &       1.0   & 113 &     0.48 & \text{\cite{willman2011}} \\
 W1\_2    &   20061121 &            5 &        9000 & Brodie  &       1.0   & 101 &     0.38 & \text{\cite{willman2011}}\\
 W1\_3    &   20061122 &            3 &        5400 & Brodie  &       1.0   &  92 &     0.41 & \text{\cite{willman2011}}\\
 W1\_4    &   20070320 &            3 &        5400 & Brodie  &       0.7 & 123 &     0.07 & \text{\cite{willman2011}}\\
 W1\_5    &   20170425 &            7 &       12000 & Geha    &       0.7 &  71 &     0.28 & {\rm New~Data; ~this~work}\\
 W1\_6    &   20170426 &            5 &        9000 & Geha    &       0.7 &  66 &     0.42 & {\rm New~Data; ~this~work}\\
 W1\_9    &   20210408 &            5 &        9000 & Geha    &       0.7 &  70 &     0.4 & {\rm New~Data; ~this~work}\\ 
Wil1\_1B  &   20230124 &            6 &        7200 & Rich    &       0.8 &  39 &    0.38     & {\rm New~Data; ~this~work}\\
wil1\_5   &   20230124 &            6 &        7200 & Rich    &       0.8 &  33 &      0.45    & {\rm New~Data; ~this~work}\\
Wil1\_2B  &   20230125 &            6 &        7200 & Rich    &       0.8 &  31 &    0.42    & {\rm New~Data; ~this~work}\\
wil1\_6   &   20230125 &            6 &        7200 & Rich    &       0.8 &  29 &      0.34  & {\rm New~Data; ~this~work}
\enddata
\tablecomments{List of DEIMOS masks present in this work: (1) DEIMOS mask name, (2) date observed, (3) number of exposures reduced in this work, (4) total integrated exposure time, (5) Principal Investigator (PI) name, (6) slit width, (7) number of slits in a mask, (8) fraction of targeted objects in a mask with a measured velocity, and (9) published reference.}
\end{deluxetable*}


\subsection{Keck/DEIMOS Spectroscopy:  Metallicities}\label{ssec:deimos_redux_feh}

We use {\tt dmost} to measure the equivalent width of several stellar absorption lines.   We first determine equivalent widths for NaI:\,$\lambda$8183, 8185\,\AA{} and MgI:\,$\lambda$8807\,\AA{}.  The strengths of the NaI and MgI lines are used as proxies for surface gravity to evaluate membership in W1.   We model these lines as a double or single Gaussian, respectively, and integrate the resulting fitted parameters to determine equivalent widths.  These quantities are used at the end of \S\ref{ssec:membership} to validate membership selection.

We measure the equivalent width of the Ca II triplet (CaT) lines ($\lambda$8498, 8542, 8662\,\AA{}) as described in \citet{geha_paper1}.   We simultaneously model the CaT lines with a Gaussian-plus-Lorentzian profile (for stars at ${\rm S/N}$ > 15 per spectral pixel) or a Gaussian profile (for stars at ${\rm S/N}$ < 15 per spectral pixel).  We integrate the resulting fits to determine total equivalent width, which we refer to as CaT EW.    From these CaT EW measurements, we derive ${\rm [Fe/H]}$ measurements for all candidate RGB stars ($M_V < 3$) assuming the empirical luminosity-dependent EW–${\rm [Fe/H]}$ calibration from \citet{Navabi2025} which is an update to the \citet{Carrera2013} calibration using the same functional form.  We adopt the form of the calibration based on the absolute $V$-band magnitude of each star, which we estimate by transforming our MegaCam $g$- and $r$-band photometry (see \S\ref{ssec:ground_gaia}).   As detailed in \citet{geha_paper1}, an error floor of 0.05\,\AA{} is added to the CaT measurements based on repeat measurements, while a 0.1\,dex error floor is included in the ${\rm [Fe/H]}$ values to account for inherent scatter in the Navabi et al.~calibration itself, noting this term dominates at ${\rm S/N}>25$.

\subsection{Ground-based Imaging and Matching to {\it Gaia}}\label{ssec:ground_gaia}

We require photometry for all kinematic measurements to assess membership in W1 (see Figure \ref{fig:sample}, top panel).   We use the Canada--France--Hawaii Telescopes (CFHT) MegaCam photometric catalog from \citet{munoz2018a} and adopt the fitted density profile and structural parameters from this analysis (see Table~\ref{table_properties}). These data are a full magnitude deeper than that available from the DESI Legacy Imaging Surveys DR9 \citep[][]{Dey2019}. The \citet{munoz2018a} $g$- and $r$-band magnitudes are corrected for foreground Galactic extinction using \citet{Schlegel1998}. Given the small magnitude uncertainties reported by \citet[][$< 0.01$ mag]{munoz2018a}, we add a systematic photometric uncertainty of 0.02 mag in quadrature.

A important improvement to previous studies of W1 is the available \emph{Gaia} DR3 catalog \citep{GaiaDR3}. While \emph{Gaia} will only cover W1 stars above the subgiant branch ($r \lesssim 20$), it significantly reduces contamination in this region due to the addition of proper-motion and parallax measurements.    We match our DEIMOS catalog to \emph{Gaia} DR3 with a $1.5''$ matching radius, resulting in 65 matches that are considered for membership.

\section{Membership Sample}\label{sec_mem}

Assessing whether a given star belongs to W1 or is part of the foreground Milky Way is particularly challenging because of the kinematic overlap between these two populations.  We begin by describing our membership assessment method (\S\ref{ssec:membership}).  We then present our final member sample and compare to previous member samples in the literature (\S\ref{ssec:final_members}).

\subsection{Membership Determination}\label{ssec:membership}

To evaluate the likelihood that a given star is a member of W1, we assess five criteria for membership: (1) the star's position on the CMD, $P_{\rm CMD}$; (2) the star's heliocentric radial velocity, $P_{\rm vel}$; (3) the star's proper motion, $P_{\rm pm}$; (4) the star's parallax, $P_{\rm parallax}$; and (5) the star's ${\rm [Fe/H]}$ iron abundance, $P_{\rm [Fe/H]}$. Criteria (3) and (4) are only considered for stars with parallax and proper motion measurements available from \emph{Gaia} DR3 \citep[][see also \S\ref{ssec:ground_gaia}]{GaiaDR3}.  

\begin{figure*}[t!]
\includegraphics[width = \textwidth, trim=1cm 0cm 0cm 0cm]{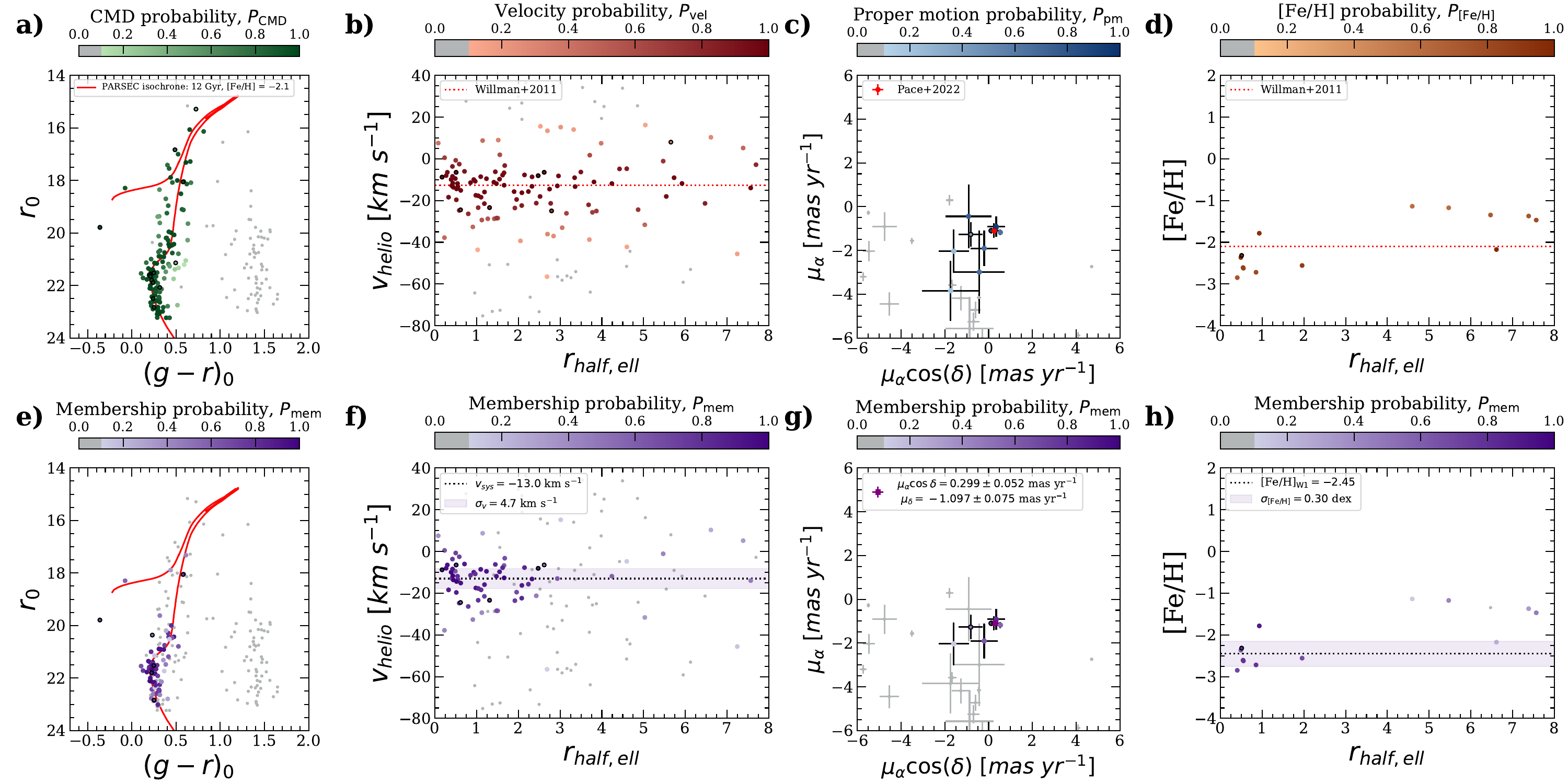}
\caption{From our five criteria for membership (panels (a), (b), (c), and (d)), the membership probability of each star was determined (panels (e), (f), (g), and (h)). In each plot ({\it left to right:} CMD, velocity, proper-motion, and metallicity profile), all stars with photometric and radial velocity measurements are color-coded according to the membership probability of the criterion in question, with clear nonmembers ($P_{\rm criterion} < 0.1$) colored gray. Stars with variable velocities are circled in black. The priors for each marker are delineated in the top panels \citep{willman2011,Pace2022}, and the properties derived from our member sample are shown in the bottom panels (see \S\ref{sec:results}, Table \ref{table_properties}).}
\label{fig:membership}
\end{figure*}

Due to tentative evidence for multidirectional tails and tidal stripping \citep[][]{Martin2007,willman2011}, no spatial cuts were performed on the data as to not introduce assumptions about W1's spatial spread of stars, although we limit our sample to 3 $\rell$ to determine physical properties (see \S\ref{sec:results}).  We additionally consider two surface gravity indicators (NaI and MgI), but find these do not improve membership discrimination for the case of W1 and discuss this below. The membership probability, $P_{\rm mem}$, of a star is defined as the product of all five criteria (Figure \ref{fig:membership}):

\begin{eqnarray}
    P_{\rm mem} \propto P_{\rm CMD} \times P_{\rm vel} \times P_{\rm pm} \times P_{\rm parallax}  \times P_{\rm [Fe/H]}
\end{eqnarray}

\noindent
We next describe in more detail each of these terms.

$P_{\rm CMD}$ is determined by comparing each star's position on the CMD to an old, metal-poor isochrone from the \texttt{PARSEC} stellar evolutionary track models \citep[$\tau = 12$ Gyr, ${\rm [Fe/H]}$ = $-2.1$,][]{bressan2012}. We use the updated distance value of $38.55 \pm 0.45$ kpc based on HST imaging from \citet[][uncertainty via private communication]{Durbin2025}. This value is consistent with that measured in \citet[][]{willman2011}, $38 \pm 7$ kpc. We use this to calculate the minimum distance between the isochrone and the star's location on the CMD. This distance ($d_{\rm min}$) and the overall magnitude uncertainty of the star ($\delta_{\rm mag}$, the quadrature sum of the $r$-band and $g$-band magnitude uncertainty), are then used to evaluate the star's CMD membership probability, $P_{\rm CMD}$ (see Figure \ref{fig:membership}a):

\begin{eqnarray}
    P_{\rm CMD} = \exp{ \Biggl( - \frac{d_{\rm min}^2}{2 \bigl[ \sigma_{\rm CMD}^2 + \delta_{\rm mag}^2 \bigr]} \Biggr) },
\end{eqnarray}

\noindent
where $\sigma_{\rm CMD}$ is set to 0.15 mag. This CMD spread was estimated from the span of other possible isochrones with varying ages and metallicities ($\tau = 8$ to $13$ Gyr, ${\rm [Fe/H]}=-1.9$ to $-2.3$) to account for the uncertainty in both parameters.

$P_{\rm vel}$ is based on each star's heliocentric, line-of-sight velocity and, for consistency's sake, is calculated using a similar probability distribution as $P_{\rm CMD}$. For a star with velocity $v \pm \delta_v$, its membership probability based on velocity, $P_{\rm vel}$, is calculated as (see Figure \ref{fig:membership}b):
\begin{eqnarray}
    P_{\rm vel} = \exp \Biggl( -\frac{(v - v_{\rm  W1})^2}{2 \bigl[ (3\sigma_v)^2 + \delta_v^2 \bigr]} \Biggr),
\end{eqnarray}

\noindent
where $v_{W1} = -12.8$ \kms, the systemic velocity of W1, and $\sigma_v = 4.8$ \kms, the velocity dispersion of W1, as determined by \cite{willman2011}. We multiply the velocity dispersion by 3 to loosen the distribution and avoid artificially biasing the sample toward the priors.

For stars with measured {\it Gaia} proper motions and parallaxes, the corresponding membership probabilities for each of these quantities, $P_{\rm pm}$ and $P_{\rm parallax}$, were determined. \cite{Pace2022} reported proper-motion values of $(\mu_{\alpha, \rm W1}^*, \mu_{\delta, \rm W1}) = (0.255 \pm 0.085, -1.110 \pm 0.096)$ mas yr$^{-1}$ for W1 based on \emph{Gaia} EDR3 astrometry. For a star with proper motion ($\mu_\alpha^*, \mu_\delta$) and overall proper-motion uncertainty $\delta_{\rm pm}$ (calculated by summing the two dimensions in quadrature), the membership probability based on proper motion, $P_{\rm pm}$, is defined as (see Figure \ref{fig:membership}c):

\begin{eqnarray}
    P_{\rm pm} = \exp \Biggl( -\frac{(\mu_\alpha^* - \mu_{\alpha, \rm W1}^*)^2 + (\mu_\delta - \mu_\delta {}_{, \rm  W1})^2}{2 \bigl[ (2\sigma_{\rm pm})^2 + \delta_{\rm pm}^2 \bigr]} \Biggr),
\end{eqnarray}

\noindent
where $\sigma_{\rm pm}$ are the reported \cite{Pace2022} proper-motion uncertainties added in quadrature. Stars with a proper-motion uncertainty more than twice its signal value are treated as non-measurements.

Situated at a distance of $\sim$39 kpc, the parallax of W1 is immeasurable by {\it Gaia}. As we find that the proper-motion membership criterion cleans the sample for the brightest stars, we opt for a simple binary parallax marker: The parallax membership probability is defined as $P_{\rm parallax} = 1$ for stars with parallax-over-error is less than 3 and $P_{\rm parallax} = 0$ for stars with parallax-over-error is greater than 3.

Finally, $P_{\rm [Fe/H]}$ is determined from the CaT-based metallicity, ${\rm [Fe/H]}$, for each star. Similar to the other markers, we determine the metallicity membership probability for a star with iron abundance ${\rm [Fe/H]} \pm \delta_{\rm [Fe/H]}$ as (see Figure \ref{fig:membership}d):
\begin{eqnarray}
    P_{\rm [Fe/H]} = \exp \Biggl( -\frac{({\rm[Fe/H]} - {\rm [Fe/H]}_{\rm W1})^2}{2 \bigl[ (2\sigma_{\rm [Fe/H]})^2 + \delta_{\rm [Fe/H]}^2 \bigr]} \Biggr),
\end{eqnarray}

\noindent
where ${\rm [Fe/H]}_{\rm W1}=-2.1$, the mean spectroscopic metallicity of W1, and $\sigma_{\rm [Fe/H]} = 0.45$ dex, half the distance between the most metal-poor and metal-rich star in the three star spectroscopic sample of \cite{willman2011}. This measurement was chosen as our prior, as it is the largest member sample available with spectroscopic metallicity measurements.

Beyond these five parameters, there are a few spectral lines covered by DEIMOS that can be incorporated into membership selection to further differentiate between member stars and Milky Way dwarf star contaminants. The NaI absorption line at $8234~\AA$ correlates with surface gravity and can be used to identify Milky Way dwarf stars that display systematically stronger lines than giants \citep{Schiavon1997}. All stars with $P_{\rm mem} > 0.5$ have small NaI equivalent widths ($<1~\AA$ within 1$\sigma$). Alternatively, \cite{Battaglia2012} identified a distinction line between RGB stars and dwarf stars in the equivalent width of the MgI line at $8806.8~\AA$ as a function of the equivalent width of the CaT lines that can also be used to cut Milky Way dwarf star interlopers. Again, all stars with $P_{\rm mem} > 0.5$ lie within 1$\sigma$ of the cutoff line, a result that is consistent with the predictions for a metal-poor system. While the use of these NaI and MgI lines is redundant and does not add more layers to our membership selection, they are a useful tool for validating that our selection criterion are efficiently differentiating between W1 members and Milky Way foreground stars.

\subsection{Final Member Sample and Comparison to Literature Samples}\label{ssec:final_members}

The membership probability, $P_{\rm mem}$, of each star is shown in the bottom panels of Figure \ref{fig:membership}. From this, we define stars with $P_{\rm mem} > 0.5$ as members of W1. This results in \ntot\ members, all but one of which lie within 3 $\rell$. We recognize the subjectiveness of this probability cutoff and take this into account when calculating the properties of W1 (see \S\ref{ssec:vel_mass}). Multiple velocity epochs exist for 36 of the \ntot\ stars in our sample, from which we identify six velocity variables including one blue straggler. We infer the physical parameters of W1 from the \nmem\ members within 3 $\rell$, and these properties are presented in Table \ref{table_properties}.

\begin{figure}[ht!]
    \includegraphics[width = 0.45\textwidth, trim=0cm 0cm 0cm 0cm]{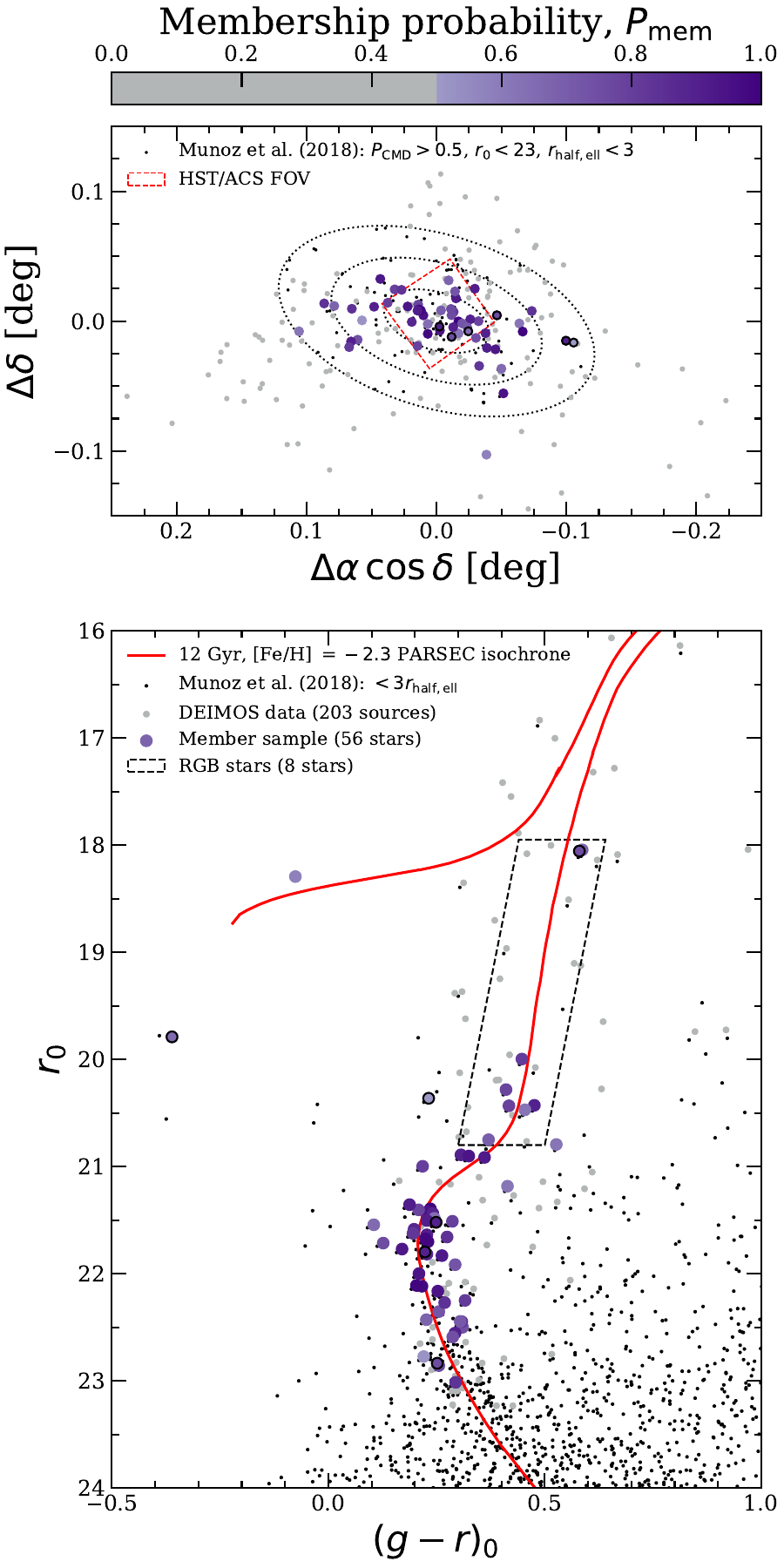}
    \caption{{\it Top panel:} Spatial distribution of the \ntot\ W1 member stars, color-coded according to $P_{\rm mem}$. Members with variable velocities are circled in black, and nonmembers are colored in gray. The three concentric dotted ellipses represent the space within 1, 2, and 3 $\rell$. All stars with photometry from \cite{munoz2018a} within 3 $\rell$, $P_{\rm CMD} > 0.5$, and $r_0 < 23$ are shown in black. The field of view of the HST/ACS imaging is denoted red (see \S\ref{ssec:hst_redux}). {\it Bottom panel:} CMD of W1 member stars overlaid with an old, metal-poor \texttt{PARSEC} isochrone in red. All stars with photometry from \cite{munoz2018a} within 3 $\rell$ are shown in black. The liberally defined red giant branch (RGB) for which we measure spectroscopic metallicities is boxed.}
    \label{fig:sample}
\end{figure}

As shown in Figure \ref{fig:sample}, we report that 28 members are inside the elliptical half-light radius, 19 of which have multiple velocity epochs and nonvariable velocities. We highlight that, despite implementing no spatial cuts, the stellar density distribution of our member sample shows no stars in excess of a Plummer profile. The CMD of W1 members is consistent with that of an old, metal-poor system, and we identify eight RGB stars for which we measure spectroscopic metallicities.

All 45 published members in \cite{willman2011} overlap with the Keck/DEIMOS spectroscopy within 1.5$^{\prime \prime}$.  Due to more stringent quality criteria, seven stars in the 2011 sample lack velocity measurements in this work due to low ${\rm S/N}$ and poor velocity fits; they are thus not considered for membership. A total of 31 stars in the 2011 sample overlap with the \ntot\ member stars in this work. The seven remaining stars in the 2011 sample are labeled as nonmembers: four on the basis of proper-motion/parallax estimates from {\it Gaia} and three due to updated velocities that are inconsistent with the systemic motion of W1. We also find the presence of four stars flagged as binaries by this work in the 2011 sample. Of the 52 foreground Milky Way stars reported in \cite{willman2011}, 51 stars have matches with our measured velocities. One of these stars is included in our member sample based on an updated velocity measurement, and we confirm the remaining stars as nonmembers.

The completeness of our kinematic sample was evaluated by comparing with the CFHT MegaCam photometric catalog from \cite{munoz2018a}. This photometric catalog was restricted to stars within 3 $\rell$ and $P_{\rm CMD} \geq 0.5$, ensuring that we only evaluate the completeness of our sample against likely member stars (see Figure \ref{fig:sample}). All stars with $r_0 \leq 21$ inside 1 half-light radius have spectroscopic measurements and are included in our analysis. For fainter stars ($r_0 \leq 23$), the spectroscopic completeness decreases approximately linearly from 100\% to 70\% from 0 to 3 $\rell$.

\begin{deluxetable}{l c c c}
\tablewidth{\columnwidth}
\tabletypesize{\footnotesize}
\tablecaption{Properties of Willman 1. \label{table_properties}}
\tablehead{\colhead{Parameter} & \colhead{Value} & \colhead{Units} & \colhead{References}}
\startdata
$\alpha_{J2000}$   &  162.3436          &  deg    & \cite{munoz2018a} \\
$\delta_{J2000}$   &  51.0501           &   deg   & \cite{munoz2018a} \\
$D_{\odot}$        & $38.55 \pm 0.45$  &   kpc   & \cite{Durbin2025}\tablenotemark{a} \\
$(m-M)_0$          & $17.93 \pm 0.03$     &   mag   & \cite{Durbin2025}\tablenotemark{a} \\
$r_{\rm half}$     & $2.51 \pm 0.22$    & arcmin  & \cite{munoz2018b} \\
$r_{\rm half}$     & $26.8 \pm 3.2$     &   pc    & \cite{munoz2018b} \\
$\epsilon$         & 0.47               &   --    & \cite{munoz2018b} \\
Position angle     & 73                 &  deg    & \cite{munoz2018b} \\
$M_V$              & $-2.56 \pm 0.74$   &   mag   & \cite{munoz2018b} \\
$L_V$              & $888 \pm 605$      & $L_{V, \odot}$  &  \cite{munoz2018b} \\ 
\hline
$v_{\rm sys}$      & $\vsys$           & \kms    & \S\ref{ssec:vel_mass} \\
$\sigma_v$         & $\vdisp$           & \kms    & \S\ref{ssec:vel_mass}\tablenotemark{b} \\
$M_{1/2}$          & $\mass$            & \Msun   & \S\ref{ssec:vel_mass} \\
$({\rm M/L})_{V, 1/2}$   & $\ml$              & $M_{\odot} / L_{V, \odot}$ & \S\ref{ssec:vel_mass} \\
${\rm [Fe/H]}$     & $\met$             &   --    & \S\ref{ssec:met} \\
$\sigma_{\rm [Fe/H]}$ & $\metdisp$      &   dex   & \S\ref{ssec:met} \\
\hline
$\mu_\alpha^*$     & $\pmra$            & mas yr$^{-1}$ & \S\ref{ssec:orbit} \\
$\mu_\delta$       & $\pmdec$           & mas yr$^{-1}$ & \S\ref{ssec:orbit} \\
$r_{\rm peri}$     & $\rperi$           & kpc     & \S\ref{ssec:orbit}\tablenotemark{c} \\
$r_{\rm apo}$      & $\rapo$            & kpc     & \S\ref{ssec:orbit}\tablenotemark{c} \\
$e$                & $\e$               &    --   & \S\ref{ssec:orbit}\tablenotemark{c} \\
\enddata
\tablecomments{List of global structural, kinematic, chemical, and orbital parameters of W1.}
\tablenotetext{a}{The reported distance/distance modulus uncertainty is likely underestimated, as it does not include the systematic contributions from isochrones.}
\tablenotetext{b}{We opt to cite the velocity dispersion determined from our two-component Gaussian mixture model, as it incorporates the uncertainty in membership.}
\tablenotetext{c}{We report the orbital properties determined from {\texttt MWPotential2014} of \cite{galpy}, with the dark matter halo mass scaled up by 50\% and the addition of the LMC.}
\end{deluxetable}

\section{Results}\label{sec:results}

With a sample of \nmem\ W1 members within 3 $\rell$ with measured velocities, we next probe the kinematic (\S\ref{ssec:vel_mass}) and chemical (\S\ref{ssec:met}) properties of this system.

\subsection{Velocity Dispersion and Dynamical Mass} \label{ssec:vel_mass}

\begin{figure}[t!]
\includegraphics[width = 0.5\textwidth, trim=0.75cm 2cm 0.5cm 4.75cm]{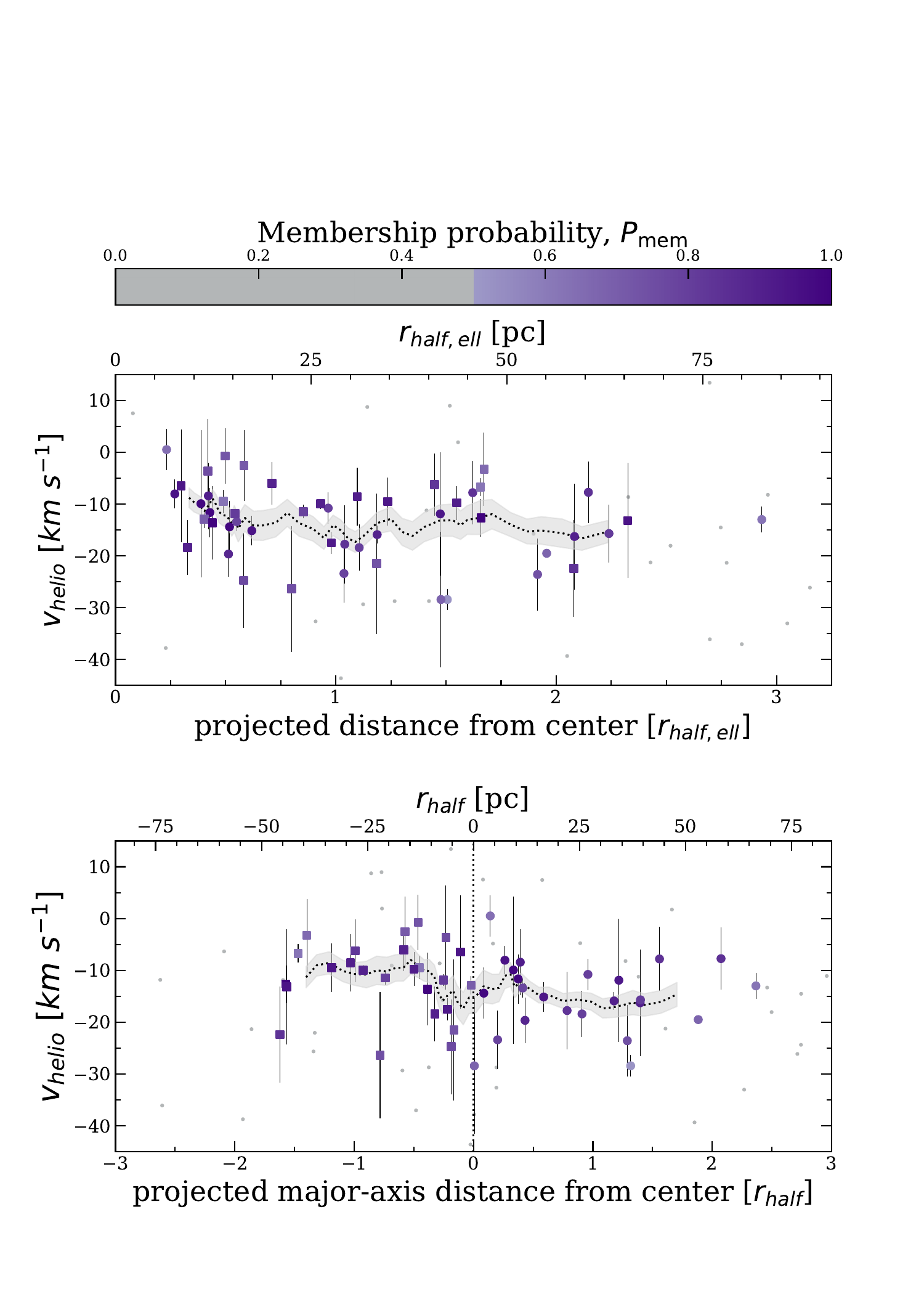}
\caption{Heliocentric radial velocities and uncertainties of the \nmem\ members as a function of elliptical half-light radius (\emph{top panel}) and projected distance with respect to the major axis (\emph{bottom panel}). For both plots, stars to the left and the right of the center axis of W1 are plotted as squares and circles, respectively. All \nmem\ stars are color-coded purple according to their membership probability, and nonmembers are in gray. In both panels, the dotted black line represents the membership probability weighted seven-star rolling velocity average, and the gray shaded region represents the uncertainty in the rolling average calculated via bootstrapping.}
\label{fig:sample_vel_disp}
\end{figure}

Based on our member sample, we present the velocity profile of W1 in Figure \ref{fig:sample_vel_disp}. \citet[][]{willman2011} reported a peculiar velocity profile where stars within $\sim$0.75 $\rell$ are offset from those in the outskirts, which they note as difficult to to explain with an equilibrium model. This phenomenon is still observed by visual inspection in our updated member sample (see Figure \ref{fig:sample_vel_disp}, top panel). Quantitatively, the rolling average decreases by about $8.5 \pm 3.0$ \kms\ from the center to the outskirts of W1. We find that this gradient is not obviously explained by ordered rotation (see Figure \ref{fig:sample_vel_disp}, bottom panel).

We calculate the systemic velocity and velocity dispersion of W1 using a two-component Gaussian mixture model to simultaneously model the Milky Way and W1. We opt for this approach, as it separates the two populations without imposing a strict membership threshold, allowing the inferred velocity dispersion to marginalize over contamination from Milky Way foreground stars \citep[e.g.,][]{Martinez2011}.

We begin by eliminating obvious nonmembers from our full dataset using the membership probabilities determined in \S\ref{ssec:membership}. We define obvious nonmembers as stars whose membership probability without velocity ($P_{\rm mem, no v} \propto P_{\rm CMD} \times P_{\rm pm} \times P_{\rm parallax} \times P_{\rm [Fe/H]}$) is $<0.1$ or stars with large radial velocities ($|v| > 150$ \kms). This results in \nmix\ stars within 3 $\rell$. Our results are not sensitive to our membership or velocity cutoff choices within the reported uncertainties. We then apply a two-component Gaussian maximum likelihood function, whose total log likelihood takes the form \citep{Walker2006}:

\begin{equation*}
\begin{split}
    \ln \mathcal{L} = \sum_i \ln \Bigl[f \cdot \mathcal{N}(v_i \, | \, v_{\rm W1}, \sigma_{\rm W1}^2 + \delta v_i^2) + \\ (1-f) \cdot \mathcal{N}(v_i \, | \, v_{\rm MW}, \sigma_{\rm MW}^2 + \delta v_i^2) \Bigr],
\end{split}
\end{equation*}

\noindent
where $f$ is the fraction of W1 members in our dataset, $v_{\rm W1}$ is the systemic velocity of W1, $\sigma_{\rm W1}$ is the velocity dispersion of W1, $v_{\rm MW}$ is the systemic velocity of the Milky Way, and $\sigma_{\rm MW}$ is the velocity dispersion of the Milky Way. We account for the velocity measurement uncertainty of each star, $\delta v_i$, by adding it in quadrature to the intrinsic dispersion of each Gaussian. 

\begin{figure}[t!]
\centering
\includegraphics[width = 0.45\textwidth, trim=0cm 0cm 0cm 0cm]{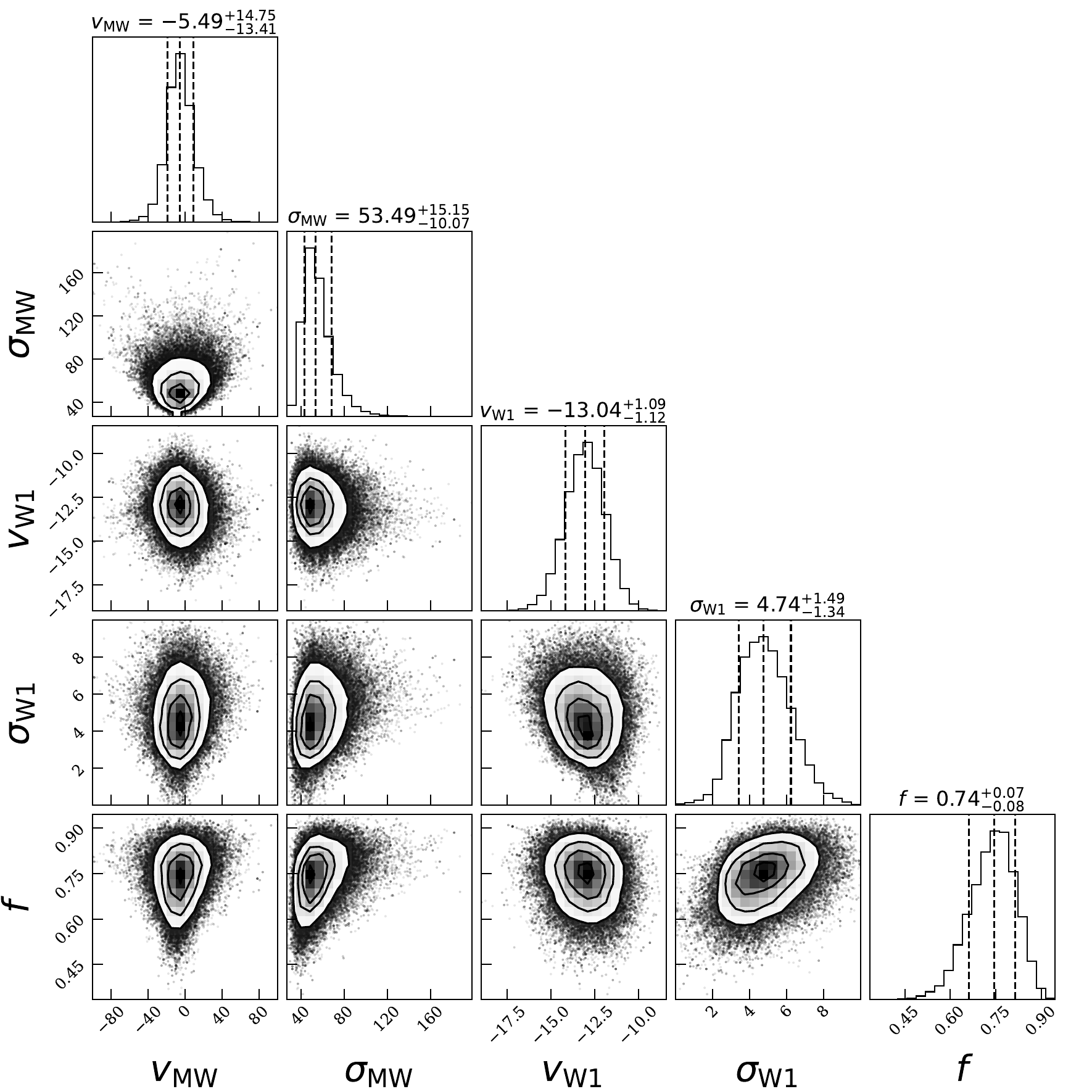}
\includegraphics[width = 0.45\textwidth, trim=0cm 0cm 0cm 0cm]{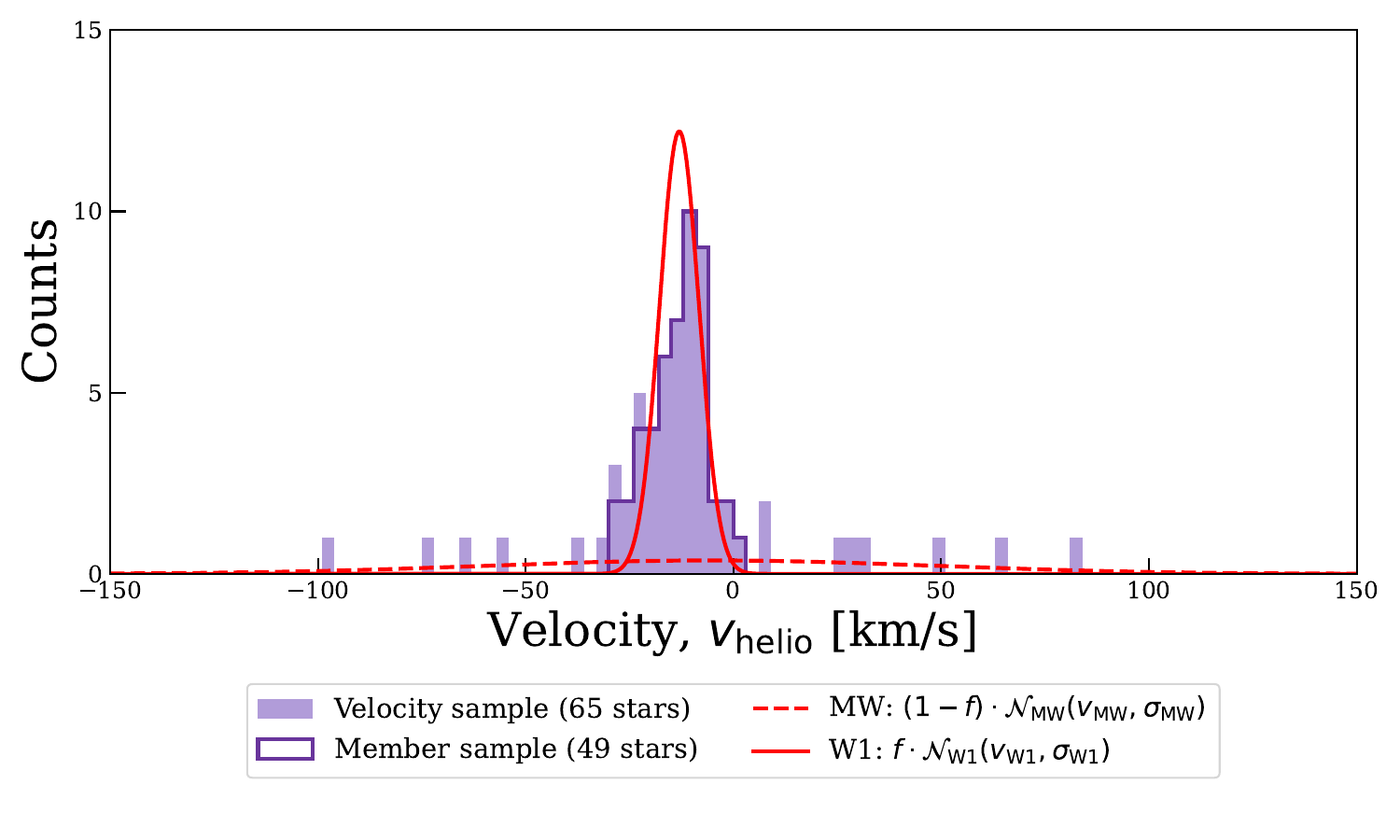}
\caption{{\it Top panel}: Posterior samples for the two-component Gaussian mixture model to measure the systemic velocity and velocity dispersion of W1 based on a velocity sample of \nmix\ potential member stars within 3 $\rell$. The dashed lines represent the median value and 16th and 84th percentiles. {\it Bottom panel}: Best-fit mixture model of two-component Gaussian representing the Milky Way and W1. The \nmix\ potential members used for our fitting are overplotted, and the member sample (\S\ref{ssec:final_members}) is outlined in purple. We note the asymmetric velocity distribution and comment on this in \S\ref{ssec:tidal}.}
\label{fig:mcmc_vel}
\end{figure}

To sample this model, the MCMC sampler \texttt{emcee} \citep{emcee} was run using 20 walkers. We initialize our walkers with the values from \cite{willman2011} for W1 ($v_{\rm W1} = -12.8$ \kms, $\sigma_{\rm W1} = 4.8$ \kms), reasonable values for the Milky Way ($v_{\rm MW} = 0$ \kms, $\sigma_{\rm W1} = 50$ \kms), and a member fraction equivalent to the fraction of our determined member sample ($f = 0.75$), but our results are not dependent on these choices. We adopt the following uniform priors for our five parameters: $-100 < v_{\rm MW} < 100$ \kms, $0 < \sigma_{\rm MW} < 200$ \kms, $-30 < v_{\rm W1} < 10$ \kms, $0 < \sigma_{\rm W1} < 10$ \kms, and $0 < f < 1$. The distribution of posterior samples is shown in the top panel of Figure \ref{fig:mcmc_vel}. We measure the systemic velocity as the sample median and the associated uncertainties from 16th and 84th percentiles.

The systemic velocity of W1 is estimated to be $v_{\rm sys} = \vsys$ \kms\ with a velocity dispersion of $\sigma_v = \vdisp$ \kms, which is consistent with measurements by \citet{willman2011} and \citet{geha_paper2}. We highlight that the velocity dispersion is degenerate with the member fraction, implying that the large uncertainty in W1's velocity dispersion is driven by the uncertainty in identifying a pure member sample. This result is unchanged within the reported uncertainty when restricting the sample to anywhere within 2 to 4 $\rell$, or when restricting the sample to only stars with multiple velocity epochs. When cutting the sample to 1 $\rell$, the uncertainty on the velocity dispersion grows such that the system is not conclusively resolved due to the small sample size of stars. Notably, the kinematic distribution of W1 is asymmetric and non-Gaussian, suggesting that a single global velocity dispersion value is insufficient to fully characterize the complex kinematic profile of W1 (see Figure \ref{fig:mcmc_vel}, bottom panel).

Assuming dynamical equilibrium, the dynamical mass of W1 within its half-light radius can be inferred from this radial velocity dispersion. Following the methods of \cite{wolf2010}, the mass of W1 within the half-light radius is estimated to be $\mass\Msun$, corresponding to a central mass-to-light ratio of $({\rm M/L})_V = \ml$. This is consistent with the dynamical mass and mass-to-light ratio previously determined by \cite{willman2011} ($\sim3.9^{+2.5}_{-1.6} \times 10^5$\Msun; $({\rm M/L})_V = 770^{+930}_{-440}$; note that we use an updated $M_V$ value causing the mass-to-light ratio in this work to be smaller despite a larger mass).

The member fraction estimated from our mixture model is $f = 0.74^{+0.07}_{-0.08}$, which is consistent with the number of members determined in \S\ref{sec_mem}. We find similar results when employing a single Gaussian maximum likelihood model on our member sample of \nmem\ stars within 3 $\rell$. From this method, the systemic velocity of W1 is measured to be $v_{\rm sys} = -12.8^{+1.0}_{-1.0}$ \kms\ with a velocity dispersion of $\sigma_v = 4.8^{+0.9}_{-0.8}$ \kms. Requiring multiple velocity epochs or reducing the half-light radius cutoff down to 2 $\rell$ does not alter the systemic velocity or velocity dispersion beyond the uncertainties determined. These results are consistent with the two-component Gaussian, but with likely underestimated uncertainties. As such, we emphasize our values from the mixture model, which incorporate the uncertainty in membership as well. Table \ref{tab:members} presents the measured properties of all 72 stars considered in our analysis, including the \nmix\ stars used in the mixture model and the \ntot\ stars in the final member sample.

\subsection{Metallicity Distribution}\label{ssec:met}

\begin{figure}[t!]
\includegraphics[width = 0.45\textwidth]{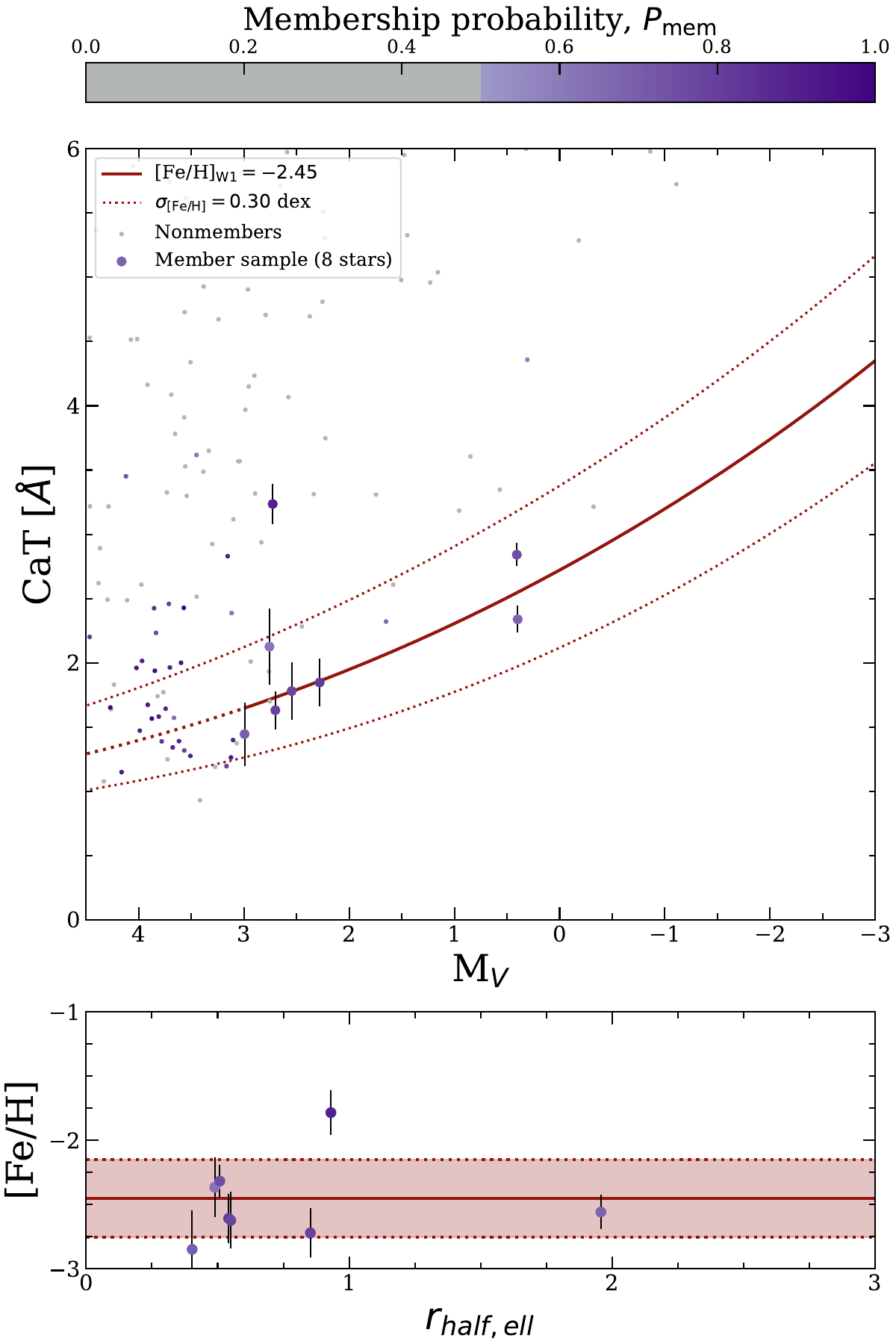}
\caption{CaT equivalent width as a function of $V$-band magnitude ({\it top panel}) and the CaT-based spatial metallicity distribution ({\it bottom panel}) for the eight RGB stars in our member sample. The solid line denotes the calibration range of the empirical relationship of \cite{Carrera2013} plotted at the mean metallicity of W1 while the dotted lines represents the metallicity dispersion determined from MCMC sampling.}
\label{fig:metallicity}
\end{figure}

As detailed in \S\ref{ssec:deimos_redux_feh}, we are able to infer ${\rm [Fe/H]}$ measurements for the stars in our member sample based on CaT equivalent width measurements. This calibration is only valid for RGB stars ($M_V < 3$), significantly reducing our member sample to eight stars with measured metallicities, all of which are within 2 $\rell$ (see Figure \ref{fig:sample}, bottom panel; Figure \ref{fig:metallicity}, top panel). From this sample, we determine the CaT-based mean metallicity and metallicity dispersion using an MCMC sampler with 20 walkers to sample from a Gaussian maximum likelihood function \citep{emcee,Walker2006}. We opt not to use a mixture model as performed in \S\ref{ssec:vel_mass} due to the small sample size (see Figure \ref{fig:membership}h). We initialize our walkers using the values determined by \cite{willman2011} ($[{\rm Fe/H}] = -2.1$, $\sigma_{\rm [Fe/H]} = 0.45$ dex). The median and 16th and 84th percentiles of the resulting posterior distribution function are presented.

The metallicity of W1 is measured to be ${\rm [Fe/H]}=\met$ with a metallicity dispersion of $\sigma_{\rm [Fe/H]} = \metdisp$ dex (see Figure \ref{fig:metallicity}). The removal of the highest-metallicity star slightly lowers the measured average metallicity and substantially reduces the measured metallicity spread (to ${\rm [Fe/H]} = -2.53^{+0.08}_{-0.09}$, $\sigma_{\rm [Fe/H]} = 0.12^{+0.12}_{-0.08}$ dex). We emphasize that this star has a high membership probability ($P_{\rm mem} \sim 0.9$) and that the mean metallicity without this star is still within the uncertainties derived with the full sample.

\cite{willman2011} measured a metallicity spread in W1 based on Keck/DEIMOS spectroscopy of three RGB members (${\rm [Fe/H]} = -1.73 \pm 0.12$, $-2.65 \pm 0.12$, and $-1.92 \pm 0.21$). These three stars are in our member sample, although we report slightly lower metallicity values for two of these stars (${\rm [Fe/H]} = -2.32 \pm 0.13$, $-2.56 \pm 0.13$, and $-2.61 \pm 0.19$, respectively). In comparison with the photometric metallicity values derived from HST CaHK imaging from \cite{fu2023} (${\rm [Fe/H]} = -2.53^{+0.11}_{-0.11}$, $\sigma_{\rm [Fe/H]} = 0.65^{+0.10}_{-0.09}$ dex), we measure W1 to be a slightly more metal-rich system with a substantially smaller metallicity dispersion.

\section{Mass Segregation}\label{sec:mass_segregation}

We now shift our attention to the HST/ACS data (\S\ref{ssec:hst_redux}), evaluating the presence of stellar mass segregation, which can be used to differentiate between a galactic and star cluster origin for W1 (\S\ref{ssec:mass_seg_analysis}).

\subsection{HST/ACS Imaging}\label{ssec:hst_redux}

W1 was observed by HST in 2017 as part of the Treasury Program GO-14734 (PI: Kallivayalil). Deep photometry of W1 was obtained with ACS \citep[][]{Ford1998} using the Wide Field Channel (WFC). The ACS/WFC field approximately covers the inner half-light radius of W1 (see Figure \ref{fig:sample}, top panel). Observations were split evenly between the F606W and F814W filters with a combined exposure time of 9254 s.

For the analysis below, we use the same HST data reduction as described in \citet{Richstein2024}.  Briefly, the individual exposures from each filter were aligned and coadded using DrizzlePac, an HST software package. The combined output images were then masked with the \texttt{segmentation} routine from the \texttt{photutils} package \citep{photutils}. Point-spread function (PSF)-fitting photometry was performed on the combined images using DAOPHOT-II and ALLSTAR \citep{Stetson1987, Stetson1992}. The source lists from F606W and F814W were matched using DAOMATCH and DAOMASTER to create a preliminary PSF source catalog. An empirically derived aperture correction was applied to the magnitudes, which were then converted to the VegaMag system and adjusted for the exposure time. Finally, to address star--galaxy separation, each source was assigned a quality flag from 0 to 1 based on how closely its parameters match those in artificial star tests across different filters.

We also note that \cite{Durbin2025} used HST imaging to measure the star formation histories of 36 ultra-faint dwarf galaxies, including W1. Their results indicate that 50\% of W1's stellar mass formed $\tau_{50} = 13.44^{+0.00}_{-0.81}$ Gyr and 90\% of W1's stellar mass formed before $\tau_{90} = 8.37^{+0.08}_{-0.93}$ Gyr, indicating a uniformly ancient stellar population formed from an early burst of star formation. We obtain consistent results when performing a similar star formation history analysis on this dataset.

\subsection{Mass Segregation Determination}\label{ssec:mass_seg_analysis}

The presence of mass segregation can be used as a diagnostic for determining whether a system is a star cluster or a dwarf galaxy \citep{baumgardt2022}. Due to the equipartition of kinetic energy whereby high-mass stars transfer energy to low-mass stars over time, high-mass stars tend to aggregate near the center of systems while low-mass stars drift toward the outer regions. This process occurs over the course of a few relaxation periods, and is thus a significant feature in old stellar objects with relatively short relaxation periods compared to their age, such as globular clusters \citep[e.g.,][]{Tripathi2023,Kim2015,Weatherford2020}. For systems with substantial amounts of dark matter, such as ultra-faint dwarf galaxies, the system's relaxation period is prolonged beyond a Hubble time such that mass segregation is unobservable \citep[][]{Longeard2018}.

\cite{baumgardt2022} first evaluated the presence of mass segregation in W1 using an independent reduction of the same HST/ACS imaging. They compared the cumulative radial distribution of stars in the upper and lower regions of the main sequence, constructing two samples: one of bright, higher-mass stars and the other of faint, lower-mass stars. In highly segregated systems, such as globular clusters, the ratio of half-light radii for these two samples usually lies around 0.7. For W1, \cite{baumgardt2022} measured a ratio of $R_{\rm half, bright} / R_{\rm half, faint} = 1.05 \pm 0.08$, indicating a lack of mass segregation and evidence toward a dwarf galaxy classification. 

Loosely following the methods of \cite{baumgardt2022}, we re-analyze the same HST/ACS data to rigorously evaluate the possibility of mass segregation in W1. We begin by removing extragalactic sources, defined as those with star quality flags $< 0.7$. To isolate the probable member stars in W1, we compared the CMD of the 1055 remaining stars to an old, metal-poor isochrone from the \texttt{MIST} stellar evolutionary track models ($\tau = 12$ Gyr, ${\rm [Fe/H]}=-2.1$, shifted to a distance modulus of $m-M = 17.93$ which corresponds to a distance of 38.55 kpc), accounting for interstellar reddening \citep{Schlegel1998, Sirianni2005, Choi2016, Dotter2016}.   We define members as stars with the greater of either 0.15 mag or twice their photometric uncertainty from the isochrone. Finally, due to the limited field of view of the ACS/WFC compared to the extended spatial distribution of W1, only stars contained within 1 $\rell$ were included to ensure that the mass segregation profile was not controlled by just a few outlying stars. Our final sample consists of 635 stars.

\begin{figure*}[t!]
\centering
\includegraphics[width = \textwidth,trim=0cm 0cm 0cm 0cm]{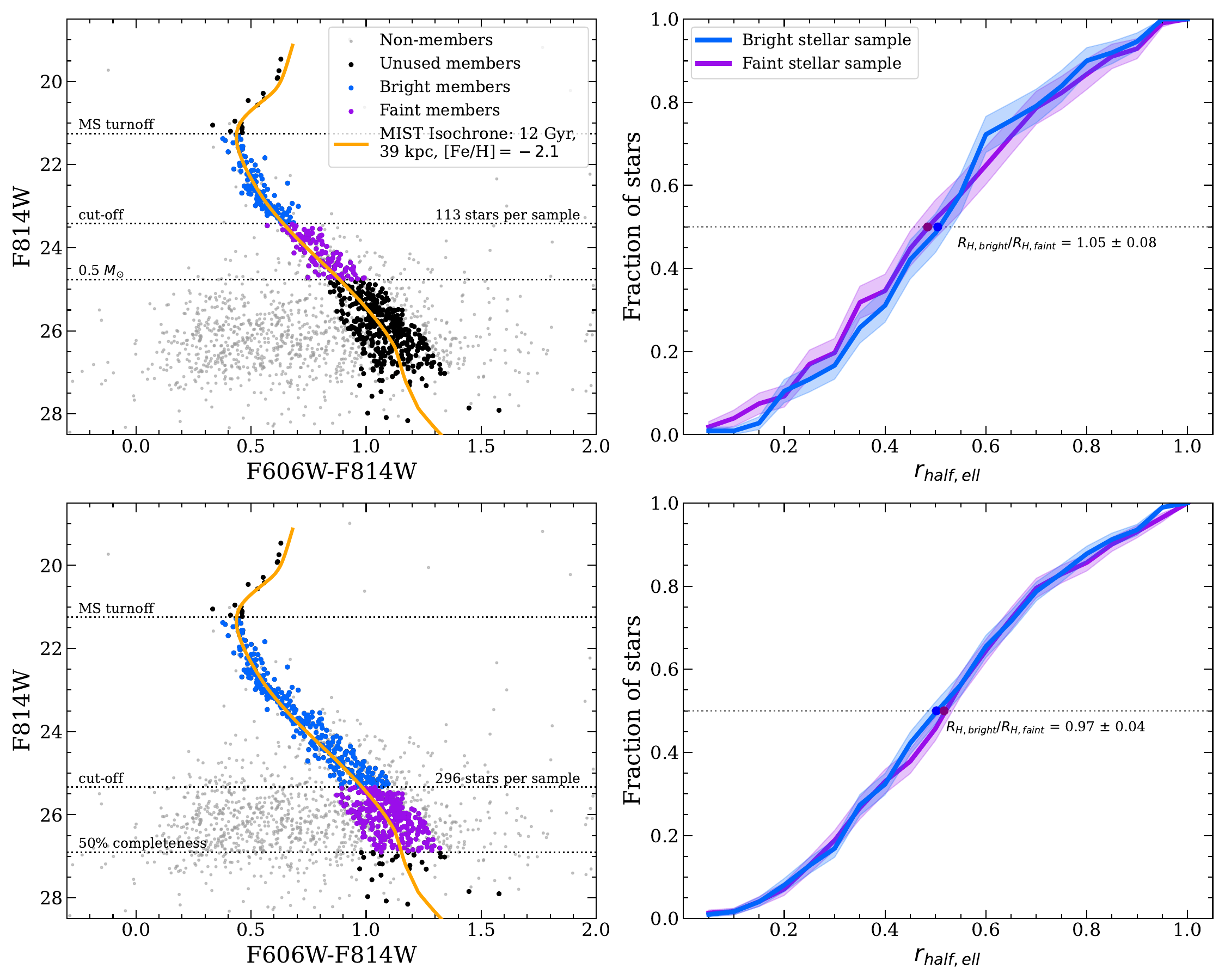}
\caption{Mass segregation analysis of W1 from HST/ACS observations. \emph{Left panels:} CMD of W1 HST data overlaid with an old, metal-poor \texttt{MIST} isochrone in orange. Blue and purple stars show the stars selected based on different magnitude cuts ({\it top panels} = $0.5M_{\odot}$; {\it bottom panels} = 50\% completeness threshold) to calculate the mass segregation profile. \emph{Right panels:} Corresponding average cumulative elliptical radial distribution of stars. The shaded region corresponds to the 1$\sigma$ uncertainty calculated by bootstrapping over each sample. No significant signs of mass segregation are detected.}
\label{fig:mass_seg}
\end{figure*}

We first use the MS turnoff and the estimated location of $0.5~M_{\odot}$ stars included in the synthetic isochrone as upper and lower photometric bounds. The stars that passed the spatial and CMD cuts within these bounds were divided into two groups of equal size based on stellar brightness to create two separate samples of bright and faint stars (see Figure \ref{fig:mass_seg}, top-left panel). The radial cumulative distribution of stars for each sample was determined using an elliptical aperture with half-light radius, position angle, and ellipticity values from \citet[][]{Martin2008} (see Table \ref{table_properties}). The half-light radius for each sample was defined as the elliptical radius within which half of the stars of each sample are contained.

\cite{baumgardt2022} estimated that the expected half-light ratio between bright and faint stars for a star cluster the size of W1 is 0.90, a value that would only be more extreme if W1 is tidally disturbed as both mass segregation and tidal stripping would preferentially remove low-mass stars from the central regions \citep{Paust2009}. We obtained a half-light radius ratio of $R_{\rm half, bright} / R_{\rm half, faint} = 1.05 \pm 0.08$, where the uncertainty is calculated by bootstrapping over each sample (see Figure \ref{fig:mass_seg}, top-right panel). A standard two-sample Kolmogorov--Smirnov (K-S) test shows a <1\% probability of mass segregation, which we highlight as a more comprehensive test statistic that takes into account the full distributions rather than evaluating the discrepancy between distributions at a single location. This process was repeated using varying lower-bound cutoffs all the way down to the 50\% photometric completeness limit (see Figure \ref{fig:mass_seg}, bottom panels). 

The results of these varied samples are consistent with an unsegregated system, with only a few samples showing inconclusive results (that is, with values compatible with both a segregated and unsegregated system). These results are unchanged by the use of a circular aperture, as originally performed by \cite{baumgardt2022}. All K-S tests performed had probabilities of mass segregation <29.7\%, the value determined by \cite{baumgardt2022}.

It should be noted that \cite{willman2006} presented evidence for mass segregation using ground-based photometry to compare the luminosity functions (LFs) of stars in the central and tail regions. Observing an excess of low-luminosity stars in the tail region, they performed a K-S test that showed a 68\% chance that the two stellar LFs were drawn from different populations, which was interpreted as evidence for mass segregation. To revisit this claim using a larger sample size and deeper photometry, the 635 HST member stars were divided into two samples by implementing similar cuts to \citet[][see their Figure 4]{willman2006}: central stars within 0.5 $\rell$ and outer stars beyond 0.5 $\rell$. As carried out by \cite{willman2006}, the LF for each sample was determined, and an offset value was added to the central region sample to correct for a difference in sample size and facilitate direct comparison. No surplus of low-luminosity stars is observed in the outer stars, and performing a two-sample K-S test on the inner and outer sample LFs shows a $< 10\%$ probability that the two LFs were drawn from different populations, which we interpret as additional evidence that W1 is unsegregated. As one final check, this process was repeated for \ntot\ member stars from our analysis of the DEIMOS data (\S\ref{ssec:final_members}), which also show no excess of low-luminosity stars at the outskirts.   Thus, we find no evidence for mass segregation in W1, consistent with the results of \citet{baumgardt2022}. This supports a dwarf galaxy classification for the system.

\section{Dynamical history}\label{kinematics}

We next investigate the possibility of past tidal interactions between the Milky Way and W1 by examining the orbital history of W1 (\S\ref{ssec:orbit}) and simulating mock stellar streams (\S\ref{ssec:streams}).

\subsection{Orbit}\label{ssec:orbit}

We compute the orbit of W1 using the analytic potential \texttt{MWPotential2014} of the \texttt{galpy} package by \cite{galpy}. This model of the Milky Way's gravitational potential consists of a power-law spherical potential representing the bulge \citep{Binney}, a Miyamoto--Nagai disk potential \citep{Miyamoto}, and a Navarro--Frenk--White dark matter halo potential \citep{Navarro}. From this, we integrate the orbit of W1 backwards in time based on its six-dimensional phase-space coordinates (see Table \ref{table_properties} for values). We measure a proper motion of $(\mu_{\alpha}^*, \mu_{\delta}) = (\pmra, \pmdec)$ mas yr$^{-1}$ by using a single Gaussian maximum likelihood model on the seven stars in our member sample with measured {\it Gaia} proper motions (see Figure \ref{fig:membership}g). This is consistent with the results of \cite{Pace2022}, who applied background mixture models to {\it Gaia} EDR3 data, both in measurement value and the number of {\it Gaia} members.

Previous orbital analyses of Milky Way dwarf galaxies have reported moderately small pericenter distance values for W1 \citep[ranging from 16 to 44 kpc;][]{Pace2022,Simon2018,Armstrong2021}, but are highly dependent on membership selection and the assumed Milky Way potential. To probe the effect of different assumed Milky Way potentials, we also model W1's orbit using two other potentials. The estimated mass of the Milky Way's dark matter halo for \texttt{MWPotential2014} is on the lower end with a value of $0.8 \times 10^{12}$\Msun. To account for this uncertainty, the second potential we adopt is the same \texttt{MWPotential2014} with a 50\% larger dark matter halo mass ($1.2 \times 10^{12}$\Msun). The Large Magellanic Cloud (LMC) has also been known to affect the orbits of objects, even those distant from it, due to the reflex motion of the Milky Way's orbit \citep[e.g.,][]{Erkal2021,Ji2021,Magnus2022,Patel2024}. This MW+LMC potential is the third and final potential we use to calculate the orbit of W1, where the LMC is modeled by a moving Hernquist potential of mass $1.38 \times 10^{11} \Msun$ and scale radius of 8.7 kpc, taking into account dynamical friction \citep{Hernquist1990, Marel2014, galpy}. We find the dynamical friction of W1 to have a negligible impact on its orbit. Due to uncertainty in orbit integration models beyond one orbital period \citep[especially for satellites with pericenter distances $<30$ kpc,][]{DSouza2022,Machado2025}, we use these potentials to model the three-dimensional orbit of W1 over only the last 1 Gyr with a time resolution of 1 Myr.

\begin{figure}[t!]
\includegraphics[width = 0.5\textwidth, trim = 0.5cm 1.5cm 0.25cm 2.5cm]{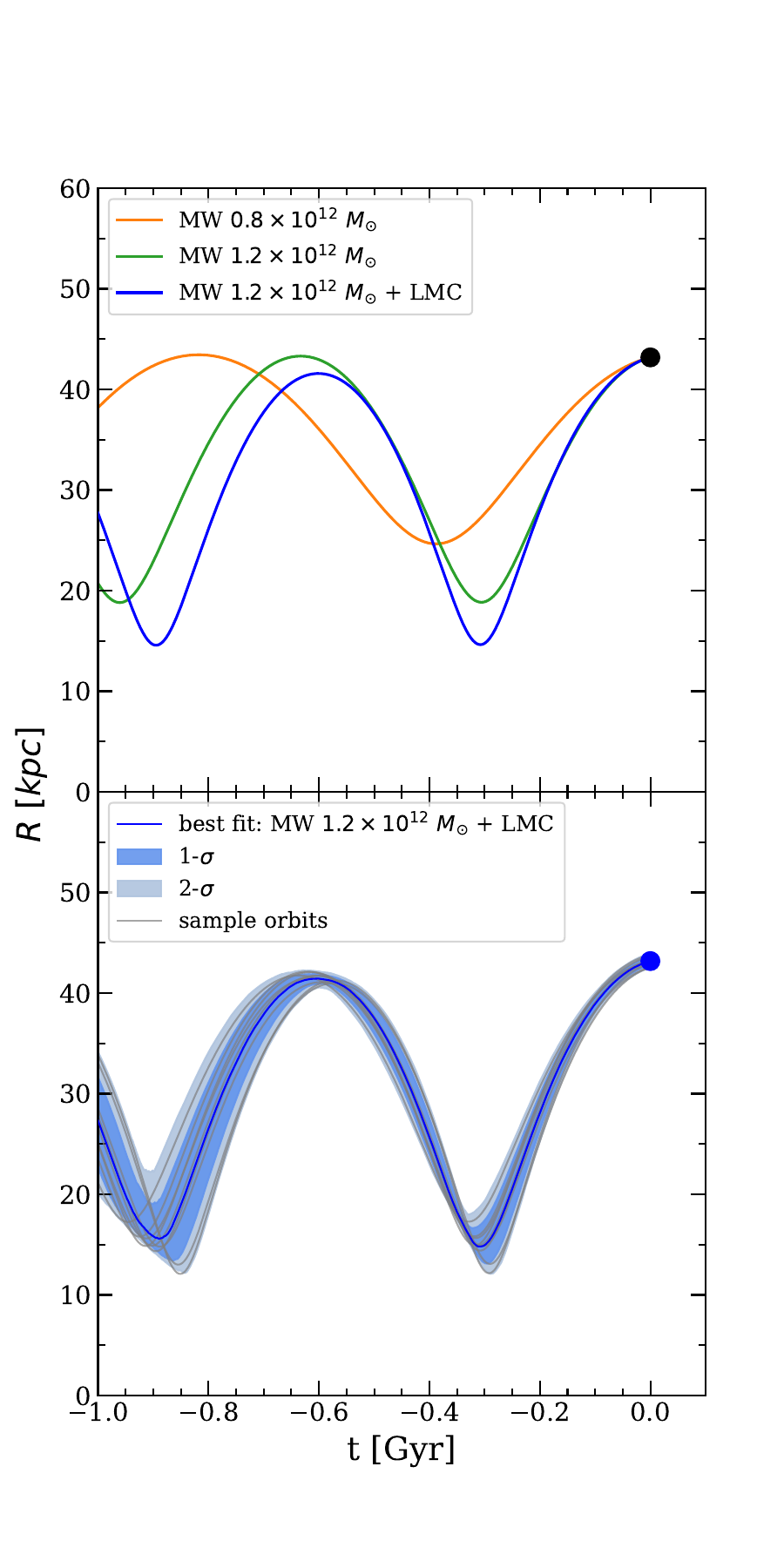}
\caption{Orbit of W1. \emph{Top panel:} Distance of W1 to the galactic center as a function of time using three different Milky Way potentials from \cite{galpy} orbital simulations. \emph{Bottom panel:} Same as above but only showing the orbital simulations using \texttt{MWPotential2014} of \cite{galpy}, with the dark matter halo mass scaled up by 50\% and the addition of the LMC. The blue line represents the 50th percentile orbit from Monte Carlo sampling over the six-dimensional phase-space parameter uncertainties; the dark blue shaded region represents the space within the 16th and 84th percentile orbits; and the light blue region represents the space between the 5th and 95th percentile orbits.}
\label{fig:orbit_comp}
\end{figure}

Figure \ref{fig:orbit_comp} compares the modeled orbits of W1 using these three different Milky Way potentials. The orbital period predicted by the original \texttt{MWPotential2014} is larger than the other two potentials by $\sim 0.2$ Gyr. In all cases, W1 is at apocenter, coming from a position closer to the Galactic center, although the exact pericenter distance of the orbit depends on the potential used.

Due to the uncertainties in distance, velocity, and proper-motion measurements, the uncertainty in the orbit of W1 was calculated through Monte Carlo sampling. We create 1000 six-dimensional phase-space coordinates by sampling distance, velocity, and proper-motion values from Gaussian distributions based on each parameter's average estimated value and 1$\sigma$ standard deviation (see Table \ref{table_properties}). The reported distance uncertainty from \cite{Durbin2025} is likely underestimated, as it does not include systematic contributions from the isochrones. Our orbital solutions remain conclusively consistent with a system moving away from a position closer in to the Milky Way up to distance uncertainties of $\sim$3 kpc.

This procedure was performed for both the regular \texttt{MWPotential2014} and the potential with the addition of the LMC (see Figure \ref{fig:orbit_comp}). The orbit of W1 under the regular \texttt{MWPotential2014} has a pericenter distance of $r_{\rm peri} = 24.9^{+3.5}_{-3.1}$ kpc, an apocenter distance of $r_{\rm apo} = 43.5^{+0.4}_{-0.5}$ kpc, and an eccentricity $e = 0.27^{+0.06}_{-0.06}$. The addition of the LMC modifies these predictions, simulating a pericenter distance of $r_{\rm peri} = \rperi$ kpc, an apocenter distance of $r_{\rm apo} = \rapo$ kpc, and an eccentricity of $e = \e$. These reported parameters represent the median of our sampling procedure, and the uncertainties represent the 16th and 84th percentiles. None of the orbital samples for both potentials have pericenter distances >40 kpc, indicating that the trajectory of W1 is most likely to be coming out of the Milky Way approaching apocenter.

To quantitatively verify that W1 is at apocenter, we calculate the ratio $f_{\rm peri}$ of \cite{Fritz2018}:
\begin{eqnarray}
    f_{\rm peri} = \frac{r_{\rm GC} - r_{\rm peri}}{r_{\rm apo} - r_{\rm peri}},
\end{eqnarray}

\noindent
which is a proxy for the orbital phase in the radial direction. $f_{\rm peri} = 0$ and $1$ indicate an object at pericenter and apocenter, respectively. For both the regular \texttt{MWPotential2014} and with the addition of the LMC, the ratio was determined to be $f_{\rm peri} = 0.99 \pm 0.01$, consistent with our conclusion that W1 is being observed at or very near apocenter after spending time closer to the Galactic center.

From the estimate of the central dynamical mass of W1 ($M_{\rm W1}$), the Jacobi tidal radius of W1 ($R_{J}$) as a function of orbital radius ($R_{\rm W1}$) can be approximated \citep{Innanen1983}:
\begin{eqnarray}
    R_{J}(R_{\rm W1}) &= \Bigl( \frac{M_{\rm W1}}{3 M_{\rm MW}} \Bigr)^{1/3} R_{\rm W1}
\end{eqnarray}

\noindent
The mass of the Milky Way enclosed within the orbital radius of the satellite ($M_{\rm MW}$) was determined from \texttt{MWPotential2014} with a dark matter halo mass enlarged by 50\% ($1.2 \times 10^{12}$\Msun). From this, the current Jacobi tidal radius of W1 is estimated to be $R_{J} = 220^{+120}_{-140}$ pc, which corresponds to $7.7^{+1.6}_{-1.8}$ half-light radii. In comparison, the tidal radius of W1 at pericenter is $R_{J} = 88^{+52}_{-57}$ pc, which is equivalent to $3.1^{+0.8}_{-0.8}$ half-light radii. These orbit integrations suggest that, at pericenter, the Jacobi radius of W1 was within a few half-light radii of its center.

The current dynamical properties of a satellite also correspond to their infall time, as shown by \cite{Rocha2012} using subhalos from the Via Lactea II cosmological simulations. Their work found a tight correlation between orbital energy and infall time where subhalos that were accreted earlier had more time to sink into the gravitational potential of the Milky Way, resulting in energies corresponding to more tightly bound systems \citep[][see their Figure 2]{Rocha2012}. Tightly bound objects cluster at small Galactocentric radii and low speeds \citep[][see their Figure 3]{Rocha2012}, consistent with the properties of W1 (see Figure \ref{fig:comparison2}, center panel). Indeed, our orbital model for W1 suggests that W1 is one of the most tightly bound dwarf galaxies based on its total energy. From this relation, we estimate W1's infall time to be roughly 9 Gyr ago. Combining this with an orbital period of $\sim$0.6 Gyr (see Figure \ref{fig:orbit_comp}, bottom panel), we can estimate that W1 has undergone $\sim$15 orbits since it crossed into the virial radius of the Milky Way.

\subsection{Mock tidal streams}\label{ssec:streams}

To further probe the possibility of tidal interactions between the Milky Way and W1, we use the \texttt{gala} particle stream spray to generate simulated stellar streams for comparison with the distribution of W1 stars \citep{gala}. We adopt the same six-dimensional phase-space coordinates and Milky Way gravitational potential as used in the orbit integration \citep[\texttt{MWPotential2014}, 50\% enlarged dark matter halo of][]{galpy}. The mass of the simulated progenitor was set to $5.9 \times 10^5 \Msun$ (the central dynamical mass of W1 determined in \S\ref{ssec:vel_mass}), and a scale radius of 26.8 pc (the half-light radius of W1) was used to represent W1. The self-gravity of W1 was modeled by a Plummer potential \citep{plummer,willman2011}. The orbit of W1 was integrated backward in time over 1 Gyr in 1 Myr intervals, and then re-integrated forward in time to the present spraying particles at each of the two Lagrange points at each time step following the Fardal Stream distribution function \citep{Fardal2015}.

\begin{figure}[t!]
\hskip 0.3 cm
\includegraphics[width = 0.45\textwidth, trim = 1.25cm 2.25cm 1.75cm 1cm]{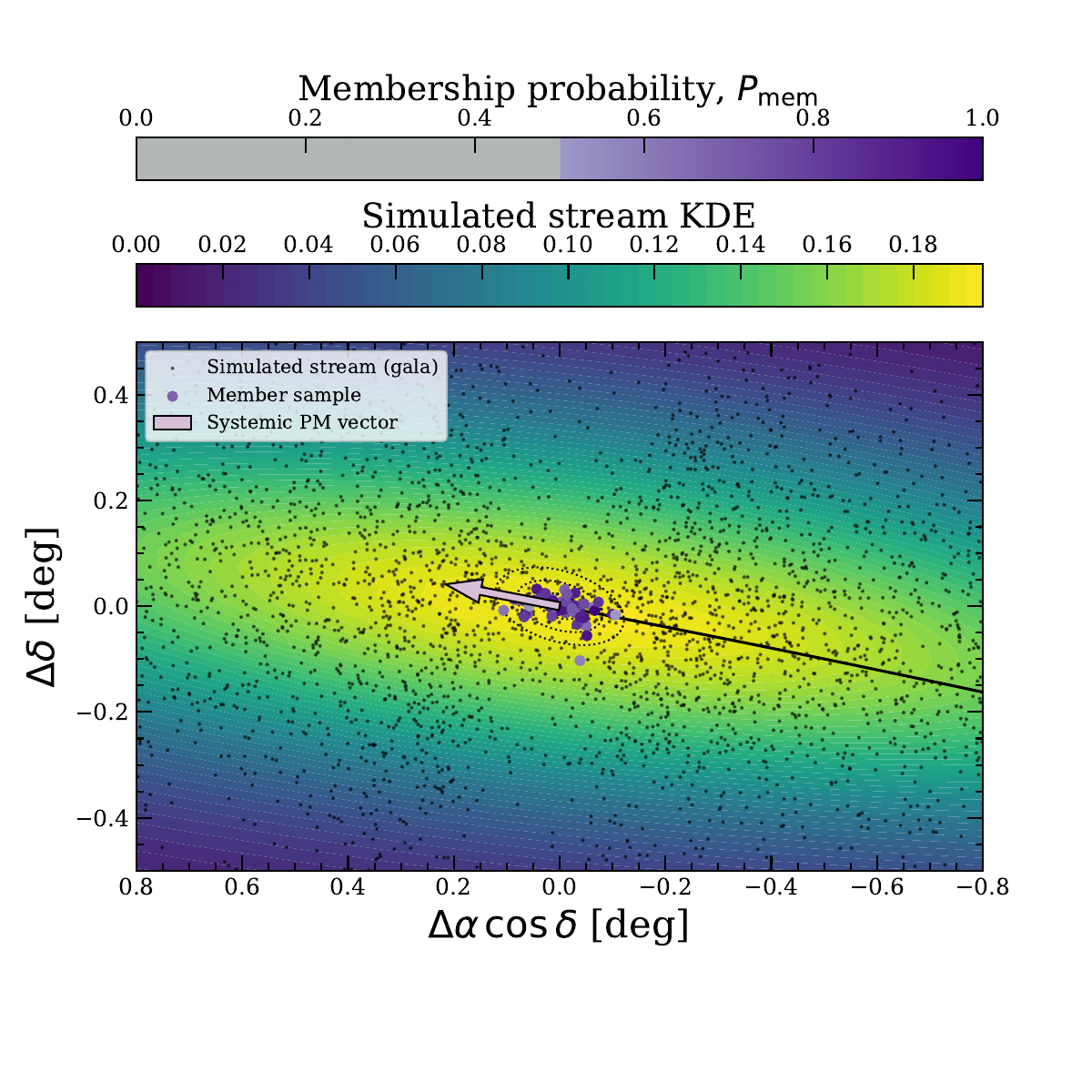}
\caption{Spatial distribution of simulated particles from \texttt{gala} mock stream spray model using \texttt{MWPotential2014} of \cite{galpy} with a 50\% enlarged dark matter halo ($1.2 \times 10^{12} \Msun$). The member sample is overplotted and color-coded according to their membership probability. The systemic proper-motion vector corrected for solar-reflex motion is denoted by the light purple arrow, and the three concentric dotted ellipses represent 1, 2, and 3 $\rell$.}
\label{fig:mock_stream}
\end{figure}

Figure \ref{fig:mock_stream} presents the results of these simulations. The mock streams exhibit a large spatial spread, with the highest-density regions approximately along W1's semi-major axis. This orientation is also aligned with W1's orbital trajectory and average solar-reflex-corrected proper-motion vector, consistent with expectations for tidal stripping along the orbit.

\begin{figure}[t!]
\includegraphics[width = 0.45\textwidth, trim = 1cm 0cm 1cm 0cm]{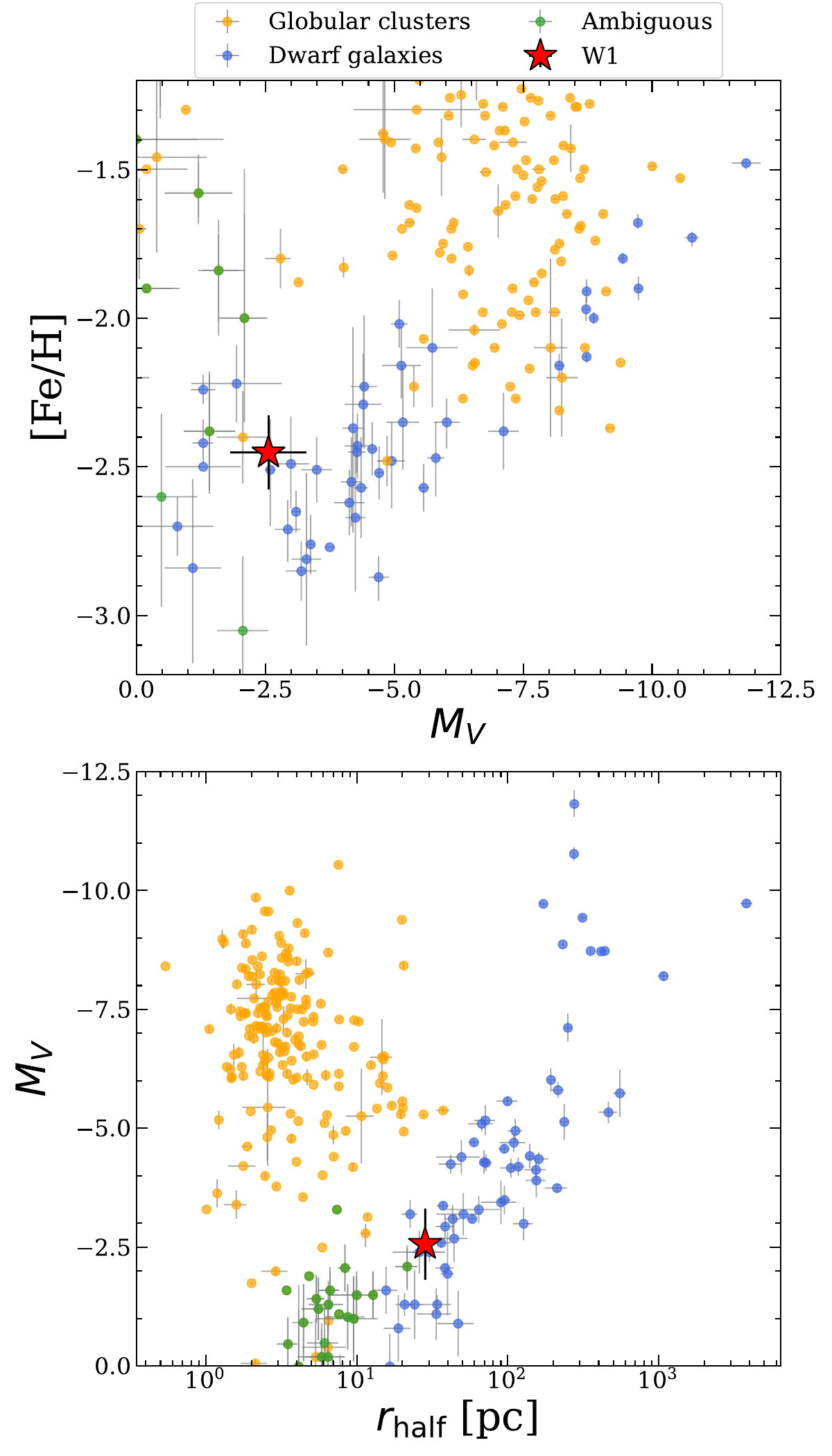}
\caption{Comparison of W1 (red star) with other Milky Way dwarf galaxies (blue circles), globular clusters (orange circles), and ambiguous objects (green circles) taken from \citep{Pace2025}. \emph{Upper:} ${\rm [Fe/H]}$ iron abundance spectroscopic metallicity as a function of absolute V-band magnitude. W1 is significantly fainter compared to other dwarf galaxies in its metallicity range, which could be explained by mass loss due to tidal stripping. \emph{Lower:} Absolute V-band magnitude vs. half-light radius. W1 is one of the faintest dwarf galaxies and has one of the smallest half-light radii.}
\label{fig:comparison1}
\end{figure}

\begin{figure*}[t!]
\includegraphics[width = \textwidth]{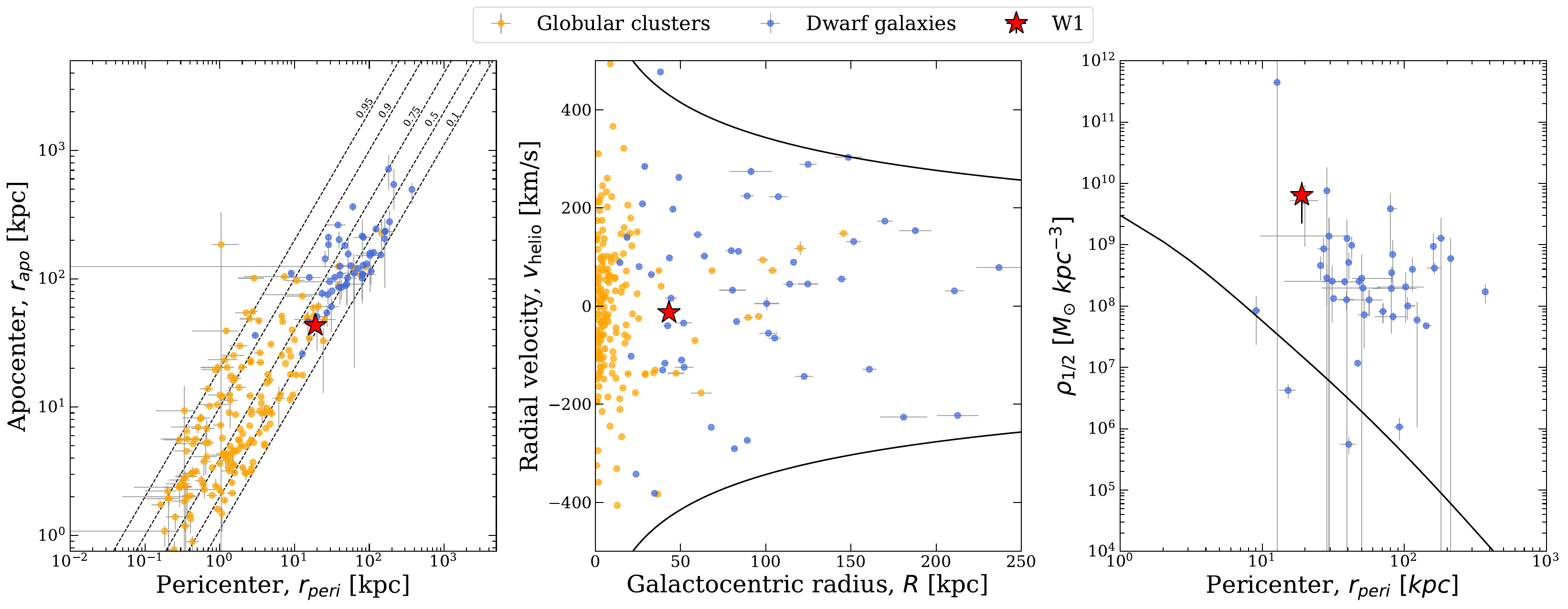}
\caption{Comparison of the orbital properties of W1 with other Milky Way dwarf galaxies and globular clusters. W1 is denoted by a red star (see Table \ref{table_properties}), Milky Way dwarf galaxies as blue circles, and globular clusters as orange circles \citep{Pace2025}. {\it Left panel:} Apocenter distance as a function of pericenter distance. The locations of different orbital eccentricities are labeled as dotted lines. W1 has small pericenter and apocenter distances relative to other Milky Way dwarf galaxies. {\it Center panel:} Radial velocity as a function of Galactocentric radius, which corresponds to its binding energy and infall time \citep{Rocha2012}. The black lines represent the escape velocity curves computed using \texttt{MWPotential2014} of \cite{galpy} with a 50\% enlarged dark matter halo mass. W1 has one of the smallest radial speeds and Galactocentric radii, corresponding to a high binding energy and early infall time (see \S\ref{ssec:orbit}). {\it Right panel:} Average density within one half-light radius as a function of orbital pericenter distance. The black line shows twice the enclosed Milky Way density as a function of radius. According to \cite{Pace2022}, if the satellite sits below this line, its Jacobi radius will be larger than the half-light radius, and it will likely be tidal disrupting.}
\label{fig:comparison2}
\end{figure*}

\section{Discussion}\label{dis}

\subsection{Is W1 a dwarf galaxy?}\label{ssec:dwarf}

We now turn to the question of the nature of W1: is this object a dwarf galaxy or a star cluster? Due to its large mass-to-light ratio, observed metallicity spread, lack of mass segregation, and position on the size-luminosity plane, we  conclude that W1 is a dwarf galaxy.

From our kinematic analysis (\S\ref{ssec:vel_mass}), we measure a mass-to-light ratio of $\ml$ from the \nmem\ nonvariable member stars within 3 $\rell$. This large mass-to-light ratio, far from unity, indicates a dark matter-dominated object \citep[e.g.,][]{Martin2007,simon07a}. We emphasize that cutting to within different half-light radii down to 2 $\rell$ does not change our results. For W1's stellar mass, its mass-to-light ratio is consistent with the larger Milky Way satellite galaxy population \citep[][see their Figure 2]{geha_paper2}. From our metallicity analysis (\S\ref{ssec:met}), we find evidence for a metallicity spread ($\sigma_{\rm [Fe/H]} = \metdisp$ dex) based on eight RGB stars with measured spectroscopic metallicities. This metallicity spread suggests that W1 has self-enriched, a property of satellites with dark matter halos that allow them to retain their gas more efficiently and go through successive epochs of star formation \citep[e.g.,][]{willman2012,Kirby2013}.

We conduct a comprehensive search for the presence of mass segregation in W1 based on HST/ACS imaging (\S\ref{sec:mass_segregation}). We find a lack of mass segregation, which implies a dwarf galaxy classification and is consistent with previous work of \cite{baumgardt2022}. Recent simulations by \cite{Errani2025} suggest that mass segregation can occur in objects with very low stellar masses, even if they host a dark matter halo. However, W1's relatively large size and luminosity compared to the objects studied prolong its relaxation time beyond a Hubble time, making this effect negligible.

Finally, we compare the structural properties of W1 to other Milky Way satellites. We rely on the Local Volume Database of \cite{Pace2025} for a comprehensive repository of the latest literature values of the observed properties of Milky Way dwarf galaxies, globular clusters, and ambiguous objects. The position of W1 on the size--luminosity plane of confirmed Milky Way satellites is more consistent with the the dwarf galaxy population; W1 is significantly less compact than globular clusters with similar luminosities (see Figure \ref{fig:comparison1}; bottom panel).

\subsection{Is W1 tidally disrupted?}\label{ssec:tidal}

The total mass of satellite systems is estimated under the assumption of dynamical equilibrium \citep{wolf2010}. It is thus crucial to understand the dynamical state of W1 in order to determine whether its measured internal velocity dispersion is a true reflection of its dynamical mass. We will now weigh the evidence for and against tidal disruption.

The evidence for tidal disruption is as follows. First, the spatial distribution of W1 is elongated, with an ellipticity of $\epsilon = 0.47$ \citep{munoz2018b}. Crucially, high-probability member stars extend out to 3 $\rell$ and are not confined to only the central regions. Although we do not see the presence of extended spatial features beyond 3 $\rell$ (as suggested by previous work) and we report that the number density of stars is consistent with the expected Plummer distribution, this is not inconsistent with simulations of mildly stripped satellites \citep[e.g.,][]{Penarrubia2009}. 

Second, we continue to observe an irregular velocity profile, as originally noted in \cite{willman2011}, where the velocities of stars within $\sim$0.75 $\rell$ are systematically higher (see Figure \ref{fig:sample_vel_disp}, top panel). This produces a remarkably asymmetric, non-Gaussian velocity distribution (see Figure \ref{fig:mcmc_vel}, bottom panel). Furthermore, we find that the velocity dispersion of the central member stars within 1 $\rell$ is significantly lower than the velocity dispersion measured out to 2--4 $\rell$, even when accounting for Milky Way interlopers via Gaussian mixture modeling. 

The luminosity--metallicity relation has been shown to be tight for dwarf galaxies \citep{Kirby2013,Simon2019,geha_paper2}. For satellites that initially conform to this relation, tidal stripping decreases their luminosity, while the average metallicity is unchanged. The location of satellites to the left of the luminosity--metallicity relation has been used as a diagnostic for tidal stripping \citep[e.g.,][however see \citet{Riley2026} for a more in-depth discussion]{Kirby2013_seg2,Kirby2015_tri,collins2020}. We find evidence for this phenomenon in the case of W1, where the luminosity of W1 is much fainter when compared to dwarf galaxies of similar metallicities (see Figure \ref{fig:comparison1}, top panel).

Our tidal-stream simulations for a W1-like progenitor show that W1's elongated structure is along the same axis as the high-density regions of the mock streams. These features are also aligned with the orbital trajectory of W1, consistent with the hypothesis that tidal stripping from the Milky Way's gravitational field could explain W1’s elongated and elliptical spatial distribution.

Our best-fit orbital solutions conclusively predict W1 to currently be at apocenter, with a pericentric passage $\lesssim 25$ kpc. At this location, the Jacobi radius of W1 would be on the order of a few half-light radii, potentially allowing for the stripping of stars at the outskirts. We also compare the orbital parameters of W1 with those of other Milky Way dwarf galaxies and globular clusters in Figure \ref{fig:comparison2}. These parameters were determined by following the same methods as performed on W1 using the MW+LMC potential (\S\ref{ssec:orbit}). W1 has one of the smallest radial speeds and Galactocentric radii of the Milky Way's dwarf galaxies, consistent with a system with a high binding energy and early infall time (see Figure \ref{fig:comparison2}, center panel). In combination with its small pericenter and apocenter distances (see Figure \ref{fig:comparison2}, left panel), this suggests that slow tidal stripping from W1's sustained passage near the Milky Way over the course of its orbit could explain W1's elongated structural properties and irregular kinematic profile.

The evidence against tidal disruption is W1's large density, which implies that its half-light radius is smaller than its Jacobi radius (see Figure \ref{fig:comparison2}, right panel). However, we note multiple other dwarf galaxies with evidence for tidal disruption also lie above this line \citep[e.g., Bo\"otes I, Hercules;][]{Filion2021,Ou2024}. If W1 is indeed tidally disrupted and exhibits a large density, this could suggest that tidal stripping is more efficient than previously thought.

Overall, we find that the evidence for tidal disruption outweighs the evidence against it. The morphological and kinematic properties of W1 support a scenario in which W1 has experienced sustained tidal stripping due to its proximity to the Milky Way over the course of its orbit. Importantly, the presence of tidal stripping does not invalidate the evidence for W1's classification as a dwarf galaxy. Even if its current velocity dispersion is inflated or not fully representative of its dynamical mass, we emphasize the observed metallicity spread, lack of mass segregation, and position on the size--luminosity plane as continued evidence for W1's dwarf galaxy classification. Therefore, while tidal effects may complicate the interpretation of the measured velocity dispersion, we still conclude that W1 is a dwarf galaxy that is likely tidally evolving with the Milky Way.

\section{Conclusion}

We have presented a comprehensive study of the structural, kinematic, chemical, and orbital properties of W1. Our main results are summarized as follows:
\begin{itemize}
    \setlength\itemsep{0pt}
    \setlength\parskip{0pt}
    \setlength\parsep{0pt}
    
    \item We identify a member sample of \ntot\ stars, including six binaries, from new and improved Keck/DEIMOS spectroscopy and {\it Gaia} astrometry (\S\ref{sec_data}, \S\ref{ssec:membership}). We measure the physical properties of W1 from the \nmem\ of these stars within 3 half-light radii.
    \item We measure a systemic velocity of $v_{\rm sys} = \vsys$ \kms~ and a velocity dispersion of $\sigma_v = \vdisp$ \kms~ based on a two-component Gaussian mixture model (\S\ref{ssec:vel_mass}). If W1 is in equilibrium, this implies a dynamical mass of $\mass$ \Msun\ and a mass-to-light ratio of $({\rm M/L})_V = \ml$.
    \item We measure an average metallicity of ${\rm [Fe/H]} = \met$ and metallicity dispersion $\sigma_{\rm [Fe/H]} = \metdisp$ dex from eight member stars with CaT-based metallicity measurements (\S\ref{ssec:met}).
    \item We conclude that W1 is a dwarf galaxy based on its dynamical mass-to-light ratio (\S\ref{ssec:vel_mass}), observed metallicity spread (\S\ref{ssec:met}), lack of mass segregation within $\sim$1 $\rell$ (\S\ref{ssec:mass_seg_analysis}), and position on the size--luminosity plane (\S\ref{ssec:dwarf}).
    \item Evidence for tidal disruption outweighs evidence against tidal disruption (\S\ref{ssec:tidal}).  Our best-fit orbital solution (\S\ref{ssec:orbit}) and comparison with our simulations of tidal streams (\S\ref{ssec:streams}) indicate a sustained passage near the Milky Way over the course of its orbit. This history could explain W1's elliptical shape, asymmetric velocity distribution (\S\ref{ssec:vel_mass}), and offset on the luminosity--metallicity plane. 
    \item We therefore caution that the internal velocity dispersion of W1 may not accurately reflect the dynamical mass of the system. The true dynamical nature of W1 may be better understood through deeper wide-field imaging and spectroscopy as well as more detailed dynamical modeling to assess the validity and use of equilibrium models. 
\end{itemize}

Looking ahead, the upcoming Vera C. Rubin Observatory's Legacy Survey of Space and Time (LSST) is anticipated to uncover a population of similarly faint, distant, and low-luminosity Local Group satellites. Many of these objects are likely to resemble W1, with sparse red giant branches, uncertain memberships, and properties that blur the line between star clusters and dwarf galaxies. Accurate characterization will require extensive spectroscopic follow-up in order to secure clean member samples, enabling precise kinematic and chemical measurements. By developing the tools needed to study these systems, we are better equipped to understand the satellites LSST will reveal and their place in the broader Local Group satellite population. From this, we can more comprehensively test the predictions of cosmological models and probe the processes that govern galaxy formation, evolution, and structure on the smallest scales.

\begin{acknowledgements}

We thank the anonymous referee for providing suggestions to our manuscript that improved its quality. We also thank Yasmeen Asali and Sebastian Monzon for their insightful comments. We thank Ricardo Mu{\~n}oz for making his raw Megacam photometric catalogs available.

This research has made extensive use of the Keck Observatory Archive (KOA), which is operated by the W. M. Keck Observatory and the NASA Exoplanet Science Institute (NExScI), under contract with the National Aeronautics and Space Administration. 

This work has made use of data from the European Space Agency (ESA) mission Gaia\footnote{\url{https://www.cosmos.esa.int/gaia}}, processed by the Gaia Data Processing and Analysis Consortium (DPAC\footnote{\url{https://www.cosmos.esa.int/web/gaia/dpac/consortium}}). Funding for the DPAC has been provided by national institutions, in particular the institutions participating in the Gaia Multilateral Agreement. 

\end{acknowledgements}

\software{This research made use of many community-developed or community-maintained software packages, including (in alphabetical order): Astropy \citep{astropy}, Emcee \citep{emcee}, gala \citep{gala}, galpy \citep{galpy}, IPython \citep{ipython}, Jupyter (\hyperlink{jupyter.org}{jupyter.org}), Matplotlib \citep{matplotlib}, NumPy \citep{numpy}, pandas \citep{pandas}, photutils \citep{photutils}, and SciPy \citep{scipy}. This research has also made use of NASA's Astrophysics Data System.}

\facility{Gaia, HST, Keck:II (DEIMOS).}

\bibliography{bib_wil1,software}

@ARTICLE{Willman2005,
       author = {{Willman}, Beth and {Blanton}, Michael R. and {West}, Andrew A. and {Dalcanton}, Julianne J. and {Hogg}, David W. and {Schneider}, Donald P. and {Wherry}, Nicholas and {Yanny}, Brian and {Brinkmann}, Jon},
        title = "{A New Milky Way Companion: Unusual Globular Cluster or Extreme Dwarf Satellite?}",
      journal = {\aj},
         year = 2005,
        month = jun,
       volume = {129},
       number = {6},
        pages = {2692-2700},
          doi = {10.1086/430214},
archivePrefix = {arXiv},
       eprint = {astro-ph/0410416},
 primaryClass = {astro-ph},
       adsurl = {https://ui.adsabs.harvard.edu/abs/2005AJ....129.2692W}
}

@ARTICLE{willman2006,
       author = {{Willman}, Beth and {Masjedi}, Morad and {Hogg}, David W. and {Dalcanton}, Julianne J. and {Martinez-Delgado}, David and {Blanton}, Michael and {West}, Andrew A. and {Dotter}, Aaron and {Chaboyer}, Brian},
        title = "{Willman 1 - A Galactic Satellite at 40 kpc With Multiple Stellar Tails}",
      journal = {arXiv e-prints},
         year = 2006,
        month = mar,
          eid = {astro-ph/0603486},
        pages = {astro-ph/0603486},
          doi = {10.48550/arXiv.astro-ph/0603486},
archivePrefix = {arXiv},
       eprint = {astro-ph/0603486},
 primaryClass = {astro-ph},
       adsurl = {https://ui.adsabs.harvard.edu/abs/2006astro.ph..3486W}
}

@ARTICLE{Martin2008,
       author = {{Martin}, Nicolas F. and {de Jong}, Jelte T.~A. and {Rix}, Hans-Walter},
        title = "{A Comprehensive Maximum Likelihood Analysis of the Structural Properties of Faint Milky Way Satellites}",
      journal = {\apj},
         year = 2008,
        month = sep,
       volume = {684},
       number = {2},
        pages = {1075-1092},
          doi = {10.1086/590336},
archivePrefix = {arXiv},
       eprint = {0805.2945},
 primaryClass = {astro-ph},
       adsurl = {https://ui.adsabs.harvard.edu/abs/2008ApJ...684.1075M}
}

@ARTICLE{Siegel2008,
       author = {{Siegel}, Michael H. and {Shetrone}, Matthew D. and {Irwin}, Michael},
        title = "{Trimming Down the Willman 1 dSph}",
      journal = {\aj},
         year = 2008,
        month = jun,
       volume = {135},
       number = {6},
        pages = {2084-2094},
          doi = {10.1088/0004-6256/135/6/2084},
archivePrefix = {arXiv},
       eprint = {0803.2489},
 primaryClass = {astro-ph},
       adsurl = {https://ui.adsabs.harvard.edu/abs/2008AJ....135.2084S}
}

@ARTICLE{willman2011,
       author = {{Willman}, Beth and {Geha}, Marla and {Strader}, Jay and {Strigari}, Louis E. and {Simon}, Joshua D. and {Kirby}, Evan and {Ho}, Nhung and {Warres}, Alex},
        title = "{Willman 1{\textemdash}A Probable Dwarf Galaxy with an Irregular Kinematic Distribution}",
      journal = {\aj},
         year = 2011,
        month = oct,
       volume = {142},
       number = {4},
          eid = {128},
        pages = {128},
          doi = {10.1088/0004-6256/142/4/128},
archivePrefix = {arXiv},
       eprint = {1007.3499},
 primaryClass = {astro-ph.GA},
       adsurl = {https://ui.adsabs.harvard.edu/abs/2011AJ....142..128W}
}

@ARTICLE{willman2012,
       author = {{Willman}, B. and {Strader}, J.},
        title = "{``Galaxy,'' Defined}",
      journal = {\aj},
         year = 2012,
        month = sep,
       volume = {144},
       number = {3},
          eid = {76},
        pages = {76},
          doi = {10.1088/0004-6256/144/3/76},
archivePrefix = {arXiv},
       eprint = {1203.2608},
 primaryClass = {astro-ph.CO},
       adsurl = {https://ui.adsabs.harvard.edu/abs/2012AJ....144...76W}
}

@ARTICLE{saeedi2020,
       author = {{Saeedi}, Sara and {Sasaki}, Manami},
        title = "{XMM-Newton study of X-ray sources in the field of Willman 1 dwarf spheroidal galaxy}",
      journal = {\mnras},
         year = 2020,
        month = dec,
       volume = {499},
       number = {3},
        pages = {3111-3129},
          doi = {10.1093/mnras/staa2846},
archivePrefix = {arXiv},
       eprint = {2009.06994},
 primaryClass = {astro-ph.GA},
       adsurl = {https://ui.adsabs.harvard.edu/abs/2020MNRAS.499.3111S}
}

@ARTICLE{Loewenstein2010,
       author = {{Loewenstein}, Michael and {Kusenko}, Alexander},
        title = "{Dark Matter Search Using Chandra Observations of Willman 1 and a Spectral Feature Consistent with a Decay Line of a 5 keV Sterile Neutrino}",
      journal = {\apj},
         year = 2010,
        month = may,
       volume = {714},
       number = {1},
        pages = {652-662},
          doi = {10.1088/0004-637X/714/1/652},
archivePrefix = {arXiv},
       eprint = {0912.0552},
 primaryClass = {astro-ph.HE},
       adsurl = {https://ui.adsabs.harvard.edu/abs/2010ApJ...714..652L}
}

@ARTICLE{Navabi2025,
       author = {{Navabi}, M. and {Carrera}, R. and {No{\"e}l}, N.~E.~D. and {Gallart}, C. and {Pancino}, E. and {De Leo}, M.},
        title = "{Revisiting the near-infrared calcium triplet as metallicity indicator}",
      journal = {\mnras},
         year = 2026,
        month = feb,
       volume = {546},
       number = {2},
          eid = {stag019},
        pages = {stag019},
          doi = {10.1093/mnras/stag019},
       adsurl = {https://ui.adsabs.harvard.edu/abs/2026MNRAS.546ag019N}
}

@ARTICLE{Nieto2010,
       author = {{Nieto}, D. and {Mirabal}, N.},
        title = "{Willman 1: An X-ray shot in the dark with Chandra}",
      journal = {arXiv e-prints},
         year = 2010,
        month = mar,
          eid = {arXiv:1003.3745},
        pages = {arXiv:1003.3745},
          doi = {10.48550/arXiv.1003.3745},
archivePrefix = {arXiv},
       eprint = {1003.3745},
 primaryClass = {astro-ph.CO},
       adsurl = {https://ui.adsabs.harvard.edu/abs/2010arXiv1003.3745N}
}

@ARTICLE{Torrealba2016,
       author = {{Torrealba}, G. and {Koposov}, S.~E. and {Belokurov}, V. and {Irwin}, M. and {Collins}, M. and {Spencer}, M. and {Ibata}, R. and {Mateo}, M. and {Bonaca}, A. and {Jethwa}, P.},
        title = "{At the survey limits: discovery of the Aquarius 2 dwarf galaxy in the VST ATLAS and the SDSS data}",
      journal = {\mnras},
         year = 2016,
        month = nov,
       volume = {463},
       number = {1},
        pages = {712-722},
          doi = {10.1093/mnras/stw2051},
archivePrefix = {arXiv},
       eprint = {1605.05338},
 primaryClass = {astro-ph.GA},
       adsurl = {https://ui.adsabs.harvard.edu/abs/2016MNRAS.463..712T}
}

@ARTICLE{Rocha2012,
       author = {{Rocha}, Miguel and {Peter}, Annika H.~G. and {Bullock}, James},
        title = "{Infall times for Milky Way satellites from their present-day kinematics}",
      journal = {\mnras},
         year = 2012,
        month = sep,
       volume = {425},
       number = {1},
        pages = {231-244},
          doi = {10.1111/j.1365-2966.2012.21432.x},
archivePrefix = {arXiv},
       eprint = {1110.0464},
 primaryClass = {astro-ph.CO},
       adsurl = {https://ui.adsabs.harvard.edu/abs/2012MNRAS.425..231R}
}

@ARTICLE{Kirby2013,
       author = {{Kirby}, Evan N. and {Cohen}, Judith G. and {Guhathakurta}, Puragra and {Cheng}, Lucy and {Bullock}, James S. and {Gallazzi}, Anna},
        title = "{The Universal Stellar Mass-Stellar Metallicity Relation for Dwarf Galaxies}",
      journal = {\apj},
         year = 2013,
        month = dec,
       volume = {779},
       number = {2},
          eid = {102},
        pages = {102},
          doi = {10.1088/0004-637X/779/2/102},
archivePrefix = {arXiv},
       eprint = {1310.0814},
 primaryClass = {astro-ph.GA},
       adsurl = {https://ui.adsabs.harvard.edu/abs/2013ApJ...779..102K}
}

@ARTICLE{Simon2019,
       author = {{Simon}, Joshua D.},
        title = "{The Faintest Dwarf Galaxies}",
      journal = {\araa},
         year = 2019,
        month = aug,
       volume = {57},
        pages = {375-415},
          doi = {10.1146/annurev-astro-091918-104453},
archivePrefix = {arXiv},
       eprint = {1901.05465},
 primaryClass = {astro-ph.GA},
       adsurl = {https://ui.adsabs.harvard.edu/abs/2019ARA&A..57..375S}
}

@ARTICLE{Richstein2024,
       author = {{Richstein}, Hannah and {Kallivayalil}, Nitya and {Simon}, Joshua D. and {Garling}, Christopher T. and {Wetzel}, Andrew and {Warfield}, Jack T. and {van der Marel}, Roeland P. and {Jeon}, Myoungwon and {Rose}, Jonah C. and {Torrey}, Paul and {Engelhardt}, Anna Claire and {Besla}, Gurtina and {Choi}, Yumi and {Geha}, Marla and {Guhathakurta}, Puragra and {Kirby}, Evan N. and {Patel}, Ekta and {Sacchi}, Elena and {Sohn}, Sangmo Tony},
        title = "{Deep Hubble Space Telescope Photometry of Large Magellanic Cloud and Milky Way Ultrafaint Dwarfs: A Careful Look into the Magnitude{\textendash}Size Relation}",
      journal = {\apj},
         year = 2024,
        month = may,
       volume = {967},
       number = {1},
          eid = {72},
        pages = {72},
          doi = {10.3847/1538-4357/ad393c},
archivePrefix = {arXiv},
       eprint = {2402.08731},
 primaryClass = {astro-ph.GA},
       adsurl = {https://ui.adsabs.harvard.edu/abs/2024ApJ...967...72R}
}

@ARTICLE{Stetson1987,
       author = {{Stetson}, Peter B.},
        title = "{DAOPHOT: A Computer Program for Crowded-Field Stellar Photometry}",
      journal = {\pasp},
         year = 1987,
        month = mar,
       volume = {99},
        pages = {191},
          doi = {10.1086/131977},
       adsurl = {https://ui.adsabs.harvard.edu/abs/1987PASP...99..191S}
}

@INPROCEEDINGS{Stetson1992,
       author = {{Stetson}, Peter B.},
        title = "{More Experiments with DAOPHOT II and WF/PC Images}",
    booktitle = {Astronomical Data Analysis Software and Systems I},
         year = 1992,
       editor = {{Worrall}, Diana M. and {Biemesderfer}, Chris and {Barnes}, Jeannette},
       series = {Astronomical Society of the Pacific Conference Series},
       volume = {25},
        month = jan,
        pages = {297},
       adsurl = {https://ui.adsabs.harvard.edu/abs/1992ASPC...25..297S}
}

@INPROCEEDINGS{Ford1998,
       author = {{Ford}, Holland C. and {Bartko}, Frank and {Bely}, Pierre Y. and {Broadhurst}, Tom and {Burrows}, Christopher J. and {Cheng}, Edward S. and {Clampin}, Mark and {Crocker}, James H. and {Feldman}, Paul D. and {Golimowski}, David A. and {Hartig}, George F. and {Illingworth}, Garth and {Kimble}, Randy A. and {Lesser}, Michael P. and {Miley}, George and {Neff}, Susan G. and {Postman}, Marc and {Sparks}, William B. and {Tsvetanov}, Zlatan and {White}, Richard L. and {Sullivan}, Pamela and {Krebs}, Carolyn A. and {Leviton}, Douglas B. and {La Jeunesse}, Tom and {Burmester}, William and {Fike}, Sherri and {Johnson}, Rich and {Slusher}, Robert B. and {Volmer}, Paul and {Woodruff}, Robert A.},
        title = "{Advanced camera for the Hubble Space Telescope}",
    booktitle = {Space Telescopes and Instruments V},
         year = 1998,
       editor = {{Bely}, Pierre Y. and {Breckinridge}, James B.},
       series = {Society of Photo-Optical Instrumentation Engineers (SPIE) Conference Series},
       volume = {3356},
        month = aug,
        pages = {234-248},
          doi = {10.1117/12.324464},
       adsurl = {https://ui.adsabs.harvard.edu/abs/1998SPIE.3356..234F}
}

@ARTICLE{Cerny2024,
       author = {{Cerny}, W. and {Chiti}, A. and {Geha}, M. and {Mutlu-Pakdil}, B. and {Drlica-Wagner}, A. and {Tan}, C.~Y. and {Adam{\'o}w}, M. and {Pace}, A.~B. and {Simon}, J.~D. and {Sand}, D.~J. and {Ji}, A.~P. and {Li}, T.~S. and {Vivas}, A.~K. and {Bell}, E.~F. and {Carlin}, J.~L. and {Carballo-Bello}, J.~A. and {Chaturvedi}, A. and {Choi}, Y. and {Doliva-Dolinsky}, A. and {Gnedin}, O.~Y. and {Limberg}, G. and {Mart{\'\i}nez-V{\'a}zquez}, C.~E. and {Mau}, S. and {Medina}, G.~E. and {Navabi}, M. and {No{\"e}l}, N.~E.~D. and {Placco}, V.~M. and {Riley}, A.~H. and {Roederer}, I.~U. and {Stringfellow}, G.~S. and {Bom}, C.~R. and {Ferguson}, P.~S. and {James}, D.~J. and {Mart{\'\i}nez-Delgado}, D. and {Massana}, P. and {Nidever}, D.~L. and {Sakowska}, J.~D. and {Santana-Silva}, L. and {Sherman}, N.~F. and {Tollerud}, E.~J.},
        title = "{Discovery and Spectroscopic Confirmation of Aquarius III: A Low-Mass Milky Way Satellite Galaxy}",
      journal = {arXiv e-prints},
         year = 2024,
        month = oct,
          eid = {arXiv:2410.00981},
        pages = {arXiv:2410.00981},
          doi = {10.48550/arXiv.2410.00981},
archivePrefix = {arXiv},
       eprint = {2410.00981},
 primaryClass = {astro-ph.GA},
       adsurl = {https://ui.adsabs.harvard.edu/abs/2024arXiv241000981C}
}

@ARTICLE{Belokurov2007,
       author = {{Belokurov}, V. and {Zucker}, D.~B. and {Evans}, N.~W. and {Kleyna}, J.~T. and {Koposov}, S. and {Hodgkin}, S.~T. and {Irwin}, M.~J. and {Gilmore}, G. and {Wilkinson}, M.~I. and {Fellhauer}, M. and {Bramich}, D.~M. and {Hewett}, P.~C. and {Vidrih}, S. and {De Jong}, J.~T.~A. and {Smith}, J.~A. and {Rix}, H. -W. and {Bell}, E.~F. and {Wyse}, R.~F.~G. and {Newberg}, H.~J. and {Mayeur}, P.~A. and {Yanny}, B. and {Rockosi}, C.~M. and {Gnedin}, O.~Y. and {Schneider}, D.~P. and {Beers}, T.~C. and {Barentine}, J.~C. and {Brewington}, H. and {Brinkmann}, J. and {Harvanek}, M. and {Kleinman}, S.~J. and {Krzesinski}, J. and {Long}, D. and {Nitta}, A. and {Snedden}, S.~A.},
        title = "{Cats and Dogs, Hair and a Hero: A Quintet of New Milky Way Companions}",
      journal = {\apj},
         year = 2007,
        month = jan,
       volume = {654},
       number = {2},
        pages = {897-906},
          doi = {10.1086/509718},
archivePrefix = {arXiv},
       eprint = {astro-ph/0608448},
 primaryClass = {astro-ph},
       adsurl = {https://ui.adsabs.harvard.edu/abs/2007ApJ...654..897B}
}

@ARTICLE{Loewenstein2012,
       author = {{Loewenstein}, Michael and {Kusenko}, Alexander},
        title = "{Dark Matter Search Using XMM-Newton Observations of Willman 1}",
      journal = {\apj},
         year = 2012,
        month = jun,
       volume = {751},
       number = {2},
          eid = {82},
        pages = {82},
          doi = {10.1088/0004-637X/751/2/82},
archivePrefix = {arXiv},
       eprint = {1203.5229},
 primaryClass = {astro-ph.CO},
       adsurl = {https://ui.adsabs.harvard.edu/abs/2012ApJ...751...82L}
}

@ARTICLE{geha_paper1,
       author = {{Geha}, Marla and {Pelliccia}, Debora and {Prochaska}, J. Xavier and {Cerny}, William and {Davies}, Frederick B. and {Hennawi}, Joseph and {Holden Dusty Reichwein}, Brad and {Westfall}, Kyle B.},
        title = "{The Keck/DEIMOS Stellar Archive: I. Uniform Velocities and Metallicities for 78 Milky Way Dwarf Galaxies and Globular Clusters}",
      journal = {arXiv e-prints},
         year = 2026,
        month = feb,
          eid = {arXiv:2602.10200},
        pages = {arXiv:2602.10200},
archivePrefix = {arXiv},
       eprint = {2602.10200},
 primaryClass = {astro-ph.GA},
       adsurl = {https://ui.adsabs.harvard.edu/abs/2026arXiv260210200G}
}

@ARTICLE{geha_paper2,
       author = {{Geha}, Marla},
        title = "{The Keck/DEIMOS Stellar Archive: II. Dynamical Masses and Metallicities for a Uniform Sample of Milky Way Satellites}",
      journal = {arXiv e-prints},
         year = 2026,
        month = feb,
          eid = {arXiv:2602.10202},
        pages = {arXiv:2602.10202},
          doi = {10.48550/arXiv.2602.10202},
archivePrefix = {arXiv},
       eprint = {2602.10202},
 primaryClass = {astro-ph.GA},
       adsurl = {https://ui.adsabs.harvard.edu/abs/2026arXiv260210202G}
}

@ARTICLE{Riley2026,
       author = {{Riley}, Alexander H. and {Bieri}, Rebekka and {Deason}, Alis J. and {Shipp}, Nora and {Simpson}, Christine M. and {Fragkoudi}, Francesca and {G{\'o}mez}, Facundo A. and {Grand}, Robert J.~J. and {Marinacci}, Federico},
        title = "{Auriga Streams III: the mass─metallicity relation does not rule out tidal mass-loss in Local Group satellites}",
      journal = {\mnras},
         year = 2026,
        month = mar,
       volume = {546},
       number = {3},
          eid = {stag029},
        pages = {stag029},
          doi = {10.1093/mnras/stag029},
archivePrefix = {arXiv},
       eprint = {2509.06859},
 primaryClass = {astro-ph.GA},
       adsurl = {https://ui.adsabs.harvard.edu/abs/2026MNRAS.546ag029R}
}

@ARTICLE{Aliu2009,
       author = {{Aliu}, E. and {Anderhub}, H. and {Antonelli}, L.~A. and {Antoranz}, P. and {Backes}, M. and {Baixeras}, C. and {Balestra}, S. and {Barrio}, J.~A. and {Bartko}, H. and {Bastieri}, D. and {Becerra Gonz{\'a}lez}, J. and {Becker}, J.~K. and {Bednarek}, W. and {Berger}, K. and {Bernardini}, E. and {Biland}, A. and {Bock}, R.~K. and {Bonnoli}, G. and {Bordas}, P. and {Tridon}, D. Borla and {Bosch-Ramon}, V. and {Bose}, D. and {Bretz}, T. and {Britvitch}, I. and {Camara}, M. and {Carmona}, E. and {Commichau}, S. and {Contreras}, J.~L. and {Cortina}, J. and {Costado}, M.~T. and {Covino}, S. and {Curtef}, V. and {Dazzi}, F. and {DeAngelis}, A. and {DeCea del Pozo}, E. and {de los Reyes}, R. and {DeLotto}, B. and {DeMaria}, M. and {DeSabata}, F. and {Mendez}, C. Delgado and {Dominguez}, A. and {Dorner}, D. and {Doro}, M. and {Elsaesser}, D. and {Errando}, M. and {Ferenc}, D. and {Fern{\'a}ndez}, E. and {Firpo}, R. and {Fonseca}, M.~V. and {Font}, L. and {Galante}, N. and {Garc{\'\i}a L{\'o}pez}, R.~J. and {Garczarczyk}, M. and {Gaug}, M. and {Goebel}, F. and {Hadasch}, D. and {Hayashida}, M. and {Herrero}, A. and {H{\"o}hne-M{\"o}nch}, D. and {Hose}, J. and {Hsu}, C.~C. and {Huber}, S. and {Jogler}, T. and {Kranich}, D. and {La Barbera}, A. and {Laille}, A. and {Leonardo}, E. and {Lindfors}, E. and {Lombardi}, S. and {Longo}, F. and {L{\'o}pez}, M. and {Lorenz}, E. and {Majumdar}, P. and {Maneva}, G. and {Mankuzhiyil}, N. and {Mannheim}, K. and {Maraschi}, L. and {Mariotti}, M. and {Mart{\'\i}nez}, M. and {Mazin}, D. and {Meucci}, M. and {Meyer}, M. and {Miranda}, J.~M. and {Mirzoyan}, R. and {Mold{\'o}n}, J. and {Moles}, M. and {Moralejo}, A. and {Nieto}, D. and {Nilsson}, K. and {Ninkovic}, J. and {Otte}, N. and {Oya}, I. and {Paoletti}, R. and {Paredes}, J.~M. and {Pasanen}, M. and {Pascoli}, D. and {Pauss}, F. and {Pegna}, R.~G. and {Perez-Torres}, M.~A. and {Persic}, M. and {Peruzzo}, L. and {Prada}, F. and {Prandini}, E. and {Puchades}, N. and {Rhode}, W. and {Rib{\'o}}, M. and {Rico}, J. and {Rissi}, M. and {Robert}, A. and {R{\"u}gamer}, S. and {Saggion}, A. and {Saito}, T.~Y. and {Salvati}, M. and {Sanchez-Conde}, M. and {Satalecka}, K. and {Scalzotto}, V. and {Scapin}, V. and {Schweizer}, T. and {Shayduk}, M. and {Shinozaki}, K. and {Shore}, S.~N. and {Sidro}, N. and {Sierpowska-Bartosik}, A. and {Sillanp{\"a}{\"a}}, A. and {Sitarek}, J. and {Sobczynska}, D. and {Spanier}, F. and {Stamerra}, A. and {Stark}, L.~S. and {Takalo}, L. and {Tavecchio}, F. and {Temnikov}, P. and {Tescaro}, D. and {Teshima}, M. and {Tluczykont}, M. and {Torres}, D.~F. and {Turini}, N. and {Vankov}, H. and {Wagner}, R.~M. and {Wittek}, W. and {Zabalza}, V. and {Zandanel}, F. and {Zanin}, R. and {Zapatero}, J.},
        title = "{Upper Limits on the VHE Gamma-Ray Emission from the Willman 1 Satellite Galaxy with the Magic Telescope}",
      journal = {\apj},
         year = 2009,
        month = jun,
       volume = {697},
       number = {2},
        pages = {1299-1304},
          doi = {10.1088/0004-637X/697/2/1299},
archivePrefix = {arXiv},
       eprint = {0810.3561},
 primaryClass = {astro-ph},
       adsurl = {https://ui.adsabs.harvard.edu/abs/2009ApJ...697.1299A}
}

@ARTICLE{Durbin2025,
       author = {{Durbin}, Meredith J. and {Choi}, Yumi and {Savino}, Alessandro and {Weisz}, Daniel and {Dolphin}, Andrew E. and {Dalcanton}, Julianne J. and {Jeon}, Myoungwon and {Kallivayalil}, Nitya and {Li}, Ting S. and {Pace}, Andrew B. and {Patel}, Ekta and {Sacchi}, Elena and {Skillman}, Evan D. and {Sohn}, Sangmo Tony and {van der Marel}, Roeland P. and {Wetzel}, Andrew and {Williams}, Benjamin F.},
        title = "{The HST Legacy Archival Uniform Reduction of Local Group Imaging (LAURELIN). I. Photometry and Star Formation Histories for 36 Ultra-faint Dwarf Galaxies}",
      journal = {arXiv e-prints},
         year = 2025,
        month = may,
          eid = {arXiv:2505.18252},
        pages = {arXiv:2505.18252},
          doi = {10.48550/arXiv.2505.18252},
archivePrefix = {arXiv},
       eprint = {2505.18252},
 primaryClass = {astro-ph.GA},
       adsurl = {https://ui.adsabs.harvard.edu/abs/2025arXiv250518252D}
}

@ARTICLE{Bringmann2009,
       author = {{Bringmann}, Torsten and {Doro}, Michele and {Fornasa}, Mattia},
        title = "{Dark matter signals from Draco and Willman 1: prospects for MAGIC II and CTA}",
      journal = {\jcap},
         year = 2009,
        month = jan,
       volume = {2009},
       number = {1},
          eid = {016},
        pages = {016},
          doi = {10.1088/1475-7516/2009/01/016},
archivePrefix = {arXiv},
       eprint = {0809.2269},
 primaryClass = {astro-ph},
       adsurl = {https://ui.adsabs.harvard.edu/abs/2009JCAP...01..016B}
}

@ARTICLE{Li2021,
       author = {{Li}, Shang and {Liang}, Yun-Feng and {Fan}, Yi-Zhong},
        title = "{Search for gamma-ray emission from the 12 nearby dwarf spheroidal galaxies with 12 years of Fermi-LAT data}",
      journal = {\prd},
         year = 2021,
        month = oct,
       volume = {104},
       number = {8},
          eid = {083037},
        pages = {083037},
          doi = {10.1103/PhysRevD.104.083037},
archivePrefix = {arXiv},
       eprint = {2110.01157},
 primaryClass = {astro-ph.HE},
       adsurl = {https://ui.adsabs.harvard.edu/abs/2021PhRvD.104h3037L}
}

@ARTICLE{McDaniel2023,
       author = {{McDaniel}, Alex and {Ajello}, Marco and {Karwin}, Christopher M. and {Di Mauro}, Mattia and {Drlica-Wagner}, Alex and {Sanchez-Conde}, Miguel A.},
        title = "{Legacy Analysis of Dark Matter Annihilation from the Milky Way Dwarf Spheroidal Galaxies with 14 Years of Fermi-LAT Data}",
      journal = {arXiv e-prints},
         year = 2023,
        month = nov,
          eid = {arXiv:2311.04982},
        pages = {arXiv:2311.04982},
archivePrefix = {arXiv},
       eprint = {2311.04982},
 primaryClass = {astro-ph.HE},
       adsurl = {https://ui.adsabs.harvard.edu/abs/2023arXiv231104982M}
}

@ARTICLE{Koposov2015,
       author = {{Koposov}, Sergey E. and {Belokurov}, Vasily and {Torrealba}, Gabriel and {Evans}, N. Wyn},
        title = "{Beasts of the Southern Wild: Discovery of Nine Ultra Faint Satellites in the Vicinity of the Magellanic Clouds.}",
      journal = {\apj},
         year = 2015,
        month = jun,
       volume = {805},
       number = {2},
          eid = {130},
        pages = {130},
          doi = {10.1088/0004-637X/805/2/130},
archivePrefix = {arXiv},
       eprint = {1503.02079},
 primaryClass = {astro-ph.GA},
       adsurl = {https://ui.adsabs.harvard.edu/abs/2015ApJ...805..130K}
}

@ARTICLE{Drlica2015,
       author = {{Drlica-Wagner}, A. and {Bechtol}, K. and {Rykoff}, E.~S. and {Luque}, E. and {Queiroz}, A. and {Mao}, Y.-Y. and {Wechsler}, R.~H. and {Simon}, J.~D. and {Santiago}, B. and {Yanny}, B. and {Balbinot}, E. and {Dodelson}, S. and {Fausti Neto}, A. and {James}, D.~J. and {Li}, T.~S. and {Maia}, M.~A.~G. and {Marshall}, J.~L. and {Pieres}, A. and {Stringer}, K. and {Walker}, A.~R. and {Abbott}, T.~M.~C. and {Abdalla}, F.~B. and {Allam}, S. and {Benoit-L{\'e}vy}, A. and {Bernstein}, G.~M. and {Bertin}, E. and {Brooks}, D. and {Buckley-Geer}, E. and {Burke}, D.~L. and {Carnero Rosell}, A. and {Carrasco Kind}, M. and {Carretero}, J. and {Crocce}, M. and {da Costa}, L.~N. and {Desai}, S. and {Diehl}, H.~T. and {Dietrich}, J.~P. and {Doel}, P. and {Eifler}, T.~F. and {Evrard}, A.~E. and {Finley}, D.~A. and {Flaugher}, B. and {Fosalba}, P. and {Frieman}, J. and {Gaztanaga}, E. and {Gerdes}, D.~W. and {Gruen}, D. and {Gruendl}, R.~A. and {Gutierrez}, G. and {Honscheid}, K. and {Kuehn}, K. and {Kuropatkin}, N. and {Lahav}, O. and {Martini}, P. and {Miquel}, R. and {Nord}, B. and {Ogando}, R. and {Plazas}, A.~A. and {Reil}, K. and {Roodman}, A. and {Sako}, M. and {Sanchez}, E. and {Scarpine}, V. and {Schubnell}, M. and {Sevilla-Noarbe}, I. and {Smith}, R.~C. and {Soares-Santos}, M. and {Sobreira}, F. and {Suchyta}, E. and {Swanson}, M.~E.~C. and {Tarle}, G. and {Tucker}, D. and {Vikram}, V. and {Wester}, W. and {Zhang}, Y. and {Zuntz}, J. and {DES Collaboration}},
        title = "{Eight Ultra-faint Galaxy Candidates Discovered in Year Two of the Dark Energy Survey}",
      journal = {\apj},
         year = 2015,
        month = nov,
       volume = {813},
       number = {2},
          eid = {109},
        pages = {109},
          doi = {10.1088/0004-637X/813/2/109},
archivePrefix = {arXiv},
       eprint = {1508.03622},
 primaryClass = {astro-ph.GA},
       adsurl = {https://ui.adsabs.harvard.edu/abs/2015ApJ...813..109D}
}

@misc{photutils,
  author       = {Larry Bradley and
                  Brigitta Sip{\H o}cz and
                  Thomas Robitaille and
                  Erik Tollerud and
                  Z\`e Vin{\'{\i}}cius and
                  Christoph Deil and
                  Kyle Barbary and
                  Tom J Wilson and
                  Ivo Busko and
                  Axel Donath and
                  Hans Moritz G{\"u}nther and
                  Mihai Cara and
                  P. L. Lim and
                  Sebastian Me{\ss}linger and
                  Zach Burnett and
                  Simon Conseil and
                  Michael Droettboom and
                  Azalee Bostroem and
                  E. M. Bray and
                  Lars Andersen Bratholm and
                  William Jamieson and
                  Adam Ginsburg and
                  Geert Barentsen and
                  Matt Craig and
                  Sergio Pascual and
                  Shivangee Rathi and
                  Marshall Perrin and
                  Brett M. Morris},
  title        = {astropy/photutils: 2.2.0},
  month        = feb,
  year         = 2025,
  publisher    = {Zenodo},
  version      = {2.2.0},
  doi          = {10.5281/zenodo.14889440},
  url          = {https://doi.org/10.5281/zenodo.14889440},
  swhid        = {swh:1:dir:11159107f27a28985192ed1118b1f2055709d093
                   ;origin=https://doi.org/10.5281/zenodo.596036;visi
                   t=swh:1:snp:ae8c4a55d349d43e53cfe9ce92e678fcfe840f
                   3b;anchor=swh:1:rel:0117f67e8888adcdfc85308287dd9c
                   854b466389;path=astropy-photutils-ffb96c5
                  },
}

@ARTICLE{Ou2024,
       author = {{Ou}, Xiaowei and {Chiti}, Anirudh and {Shipp}, Nora and {Simon}, Joshua D. and {Geha}, Marla and {Frebel}, Anna and {Mardini}, Mohammad K. and {Erkal}, Denis and {Necib}, Lina},
        title = "{Signatures of Tidal Disruption of the Hercules Ultrafaint Dwarf Galaxy}",
      journal = {\apj},
         year = 2024,
        month = may,
       volume = {966},
       number = {1},
          eid = {33},
        pages = {33},
          doi = {10.3847/1538-4357/ad2f27},
archivePrefix = {arXiv},
       eprint = {2403.00921},
 primaryClass = {astro-ph.GA},
       adsurl = {https://ui.adsabs.harvard.edu/abs/2024ApJ...966...33O}
}

@ARTICLE{Filion2021,
       author = {{Filion}, Carrie and {Wyse}, Rosemary F.~G.},
        title = "{The Far-away Blues: Exploring the Furthest Extents of the Bo{\"o}tes I Ultra-faint Dwarf Galaxy}",
      journal = {\apj},
         year = 2021,
        month = dec,
       volume = {923},
       number = {2},
          eid = {218},
        pages = {218},
          doi = {10.3847/1538-4357/ac2df1},
archivePrefix = {arXiv},
       eprint = {2110.05468},
 primaryClass = {astro-ph.GA},
       adsurl = {https://ui.adsabs.harvard.edu/abs/2021ApJ...923..218F}
}

@ARTICLE{Patel2024,
       author = {{Patel}, Ekta and {Chatur}, Lipika and {Mao}, Yao-Yuan},
        title = "{Temporal Evolution of the Radial Distribution of Milky Way Satellite Galaxies}",
      journal = {\apj},
         year = 2024,
        month = dec,
       volume = {976},
       number = {2},
          eid = {171},
        pages = {171},
          doi = {10.3847/1538-4357/ad87ee},
archivePrefix = {arXiv},
       eprint = {2409.01991},
 primaryClass = {astro-ph.GA},
       adsurl = {https://ui.adsabs.harvard.edu/abs/2024ApJ...976..171P}
}

@article{gala,
  doi = {10.21105/joss.00388},
  url = {https://doi.org/10.21105%2Fjoss.00388},
  year = 2017,
  month = {oct},
  publisher = {The Open Journal},
  volume = {2},
  number = {18},
  author = {Adrian M. Price-Whelan},
  title = {Gala: A Python package for galactic dynamics},
  journal = {The Journal of Open Source Software}}

@ARTICLE{Homma2024,
       author = {{Homma}, Daisuke and {Chiba}, Masashi and {Komiyama}, Yutaka and {Tanaka}, Masayuki and {Okamoto}, Sakurako and {Tanaka}, Mikito and {Ishigaki}, Miho N. and {Hayashi}, Kohei and {Arimoto}, Nobuo and {Lupton}, Robert H. and {Strauss}, Michael A. and {Miyazaki}, Satoshi and {Wang}, Shiang-Yu and {Murayama}, Hitoshi},
        title = "{Final results of the search for new Milky Way satellites in the Hyper Suprime-Cam Subaru Strategic Program survey: Discovery of two more candidates}",
      journal = {\pasj},
         year = 2024,
        month = aug,
       volume = {76},
       number = {4},
        pages = {733-752},
          doi = {10.1093/pasj/psae044},
archivePrefix = {arXiv},
       eprint = {2311.05439},
 primaryClass = {astro-ph.GA},
       adsurl = {https://ui.adsabs.harvard.edu/abs/2024PASJ...76..733H}
}

@ARTICLE{collins2020,
       author = {{Collins}, Michelle L.~M. and {Tollerud}, Erik J. and {Rich}, R. Michael and {Ibata}, Rodrigo A. and {Martin}, Nicolas F. and {Chapman}, Scott C. and {Gilbert}, Karoline M. and {Preston}, Janet},
        title = "{A detailed study of Andromeda XIX, an extreme local analogue of ultradiffuse galaxies}",
      journal = {\mnras},
         year = 2020,
        month = jan,
       volume = {491},
       number = {3},
        pages = {3496-3514},
          doi = {10.1093/mnras/stz3252},
archivePrefix = {arXiv},
       eprint = {1910.12879},
 primaryClass = {astro-ph.GA},
       adsurl = {https://ui.adsabs.harvard.edu/abs/2020MNRAS.491.3496C}
}

@ARTICLE{Battaglia2012,
   author = {{Battaglia}, G. and {Starkenburg}, E.},
    title = "{Cleaning spectroscopic samples of stars in nearby dwarf galaxies. The use of the nIR Mg I line to weed out Milky Way contaminants}",
  journal = {\aap},
archivePrefix = "arXiv",
   eprint = {1201.3634},
     year = 2012,
    month = mar,
   volume = 539,
      eid = {A123},
    pages = {A123},
      doi = {10.1051/0004-6361/201117557},
   adsurl = {http://adsabs.harvard.edu/abs/2012A%26A...539A.123B}
}

@ARTICLE{Schiavon1997,
   author = {{Schiavon}, R.~P. and {Barbuy}, B. and {Rossi}, S.~C.~F. and 
	{Milone} and {A.}},
    title = "{The Near-Infrared Na I Doublet Feature in M Stars}",
  journal = {\apj},
   eprint = {astro-ph/9610243},
     year = 1997,
    month = apr,
   volume = 479,
    pages = {902-906},
      doi = {10.1086/303907},
   adsurl = {http://adsabs.harvard.edu/abs/1997ApJ...479..902S}
}

@ARTICLE{bressan2012,
       author = {{Bressan}, Alessandro and {Marigo}, Paola and {Girardi}, L{\'e}o. and {Salasnich}, Bernardo and {Dal Cero}, Claudia and {Rubele}, Stefano and {Nanni}, Ambra},
        title = "{PARSEC: stellar tracks and isochrones with the PAdova and TRieste Stellar Evolution Code}",
      journal = {\mnras},
         year = 2012,
        month = nov,
       volume = {427},
       number = {1},
        pages = {127-145},
          doi = {10.1111/j.1365-2966.2012.21948.x},
archivePrefix = {arXiv},
       eprint = {1208.4498},
 primaryClass = {astro-ph.SR},
       adsurl = {https://ui.adsabs.harvard.edu/abs/2012MNRAS.427..127B}
}

@ARTICLE{marz, 
       author = {{Hinton}, S.~R. and {Davis}, Tamara M. and {Lidman}, C. and
         {Glazebrook}, K. and {Lewis}, G.~F.},
        title = "{MARZ: Manual and automatic redshifting software}",
      journal = {Astronomy and Computing},
         year = 2016,
        month = apr,
       volume = {15},
        pages = {61-71},
          doi = {10.1016/j.ascom.2016.03.001},
archivePrefix = {arXiv},
       eprint = {1603.09438},
 primaryClass = {astro-ph.IM},
       adsurl = {https://ui.adsabs.harvard.edu/abs/2016A&C....15...61H}
}

@ARTICLE{baumgardt2022,
       author = {{Baumgardt}, H. and {Faller}, J. and {Meinhold}, N. and {McGovern-Greco}, C. and {Hilker}, M.},
        title = "{Stellar mass segregation as separating classifier between globular clusters and ultrafaint dwarf galaxies}",
      journal = {\mnras},
         year = 2022,
        month = mar,
       volume = {510},
       number = {3},
        pages = {3531-3545},
          doi = {10.1093/mnras/stab3629},
       adsurl = {https://ui.adsabs.harvard.edu/abs/2022MNRAS.510.3531B}
}

@ARTICLE{Sirianni2005,
       author = {{Sirianni}, M. and {Jee}, M.~J. and {Ben{\'\i}tez}, N. and {Blakeslee}, J.~P. and {Martel}, A.~R. and {Meurer}, G. and {Clampin}, M. and {De Marchi}, G. and {Ford}, H.~C. and {Gilliland}, R. and {Hartig}, G.~F. and {Illingworth}, G.~D. and {Mack}, J. and {McCann}, W.~J.},
        title = "{The Photometric Performance and Calibration of the Hubble Space Telescope Advanced Camera for Surveys}",
      journal = {\pasp},
         year = 2005,
        month = oct,
       volume = {117},
       number = {836},
        pages = {1049-1112},
          doi = {10.1086/444553},
archivePrefix = {arXiv},
       eprint = {astro-ph/0507614},
 primaryClass = {astro-ph},
       adsurl = {https://ui.adsabs.harvard.edu/abs/2005PASP..117.1049S}
}

@ARTICLE{Schlegel1998,
       author = {{Schlegel}, David J. and {Finkbeiner}, Douglas P. and {Davis}, Marc},
        title = "{Maps of Dust Infrared Emission for Use in Estimation of Reddening and Cosmic Microwave Background Radiation Foregrounds}",
      journal = {\apj},
         year = 1998,
        month = jun,
       volume = {500},
       number = {2},
        pages = {525-553},
          doi = {10.1086/305772},
archivePrefix = {arXiv},
       eprint = {astro-ph/9710327},
 primaryClass = {astro-ph},
       adsurl = {https://ui.adsabs.harvard.edu/abs/1998ApJ...500..525S}
}

@ARTICLE{Tripathi2023,
       author = {{Tripathi}, Apara and {Panwar}, Neelam and {Sharma}, Saurabh and {Kumar}, Brijesh and {Rastogi}, Shantanu},
        title = "{Photometric and kinematic studies of open cluster NGC 1027}",
      journal = {Journal of Astrophysics and Astronomy},
         year = 2023,
        month = dec,
       volume = {44},
       number = {2},
          eid = {61},
        pages = {61},
          doi = {10.1007/s12036-023-09955-7},
archivePrefix = {arXiv},
       eprint = {2304.05762},
 primaryClass = {astro-ph.GA},
       adsurl = {https://ui.adsabs.harvard.edu/abs/2023JApA...44...61T}
}

@ARTICLE{Longeard2018,
       author = {{Longeard}, Nicolas and {Martin}, Nicolas and {Starkenburg}, Else and {Ibata}, Rodrigo A. and {Collins}, Michelle L.~M. and {Geha}, Marla and {Laevens}, Benjamin P.~M. and {Rich}, R. Michael and {Aguado}, David S. and {Arentsen}, Anke and {Carlberg}, Raymond G. and {C{\^o}t{\'e}}, Patrick and {Hill}, Vanessa and {Jablonka}, Pascale and {Gonz{\'a}lez Hern{\'a}ndez}, Jonay I. and {Navarro}, Julio F. and {S{\'a}nchez-Janssen}, Rub{\'e}n and {Tolstoy}, Eline and {Venn}, Kim A. and {Youakim}, Kris},
        title = "{Pristine dwarf galaxy survey - I. A detailed photometric and spectroscopic study of the very metal-poor Draco II satellite}",
      journal = {\mnras},
         year = 2018,
        month = oct,
       volume = {480},
       number = {2},
        pages = {2609-2627},
          doi = {10.1093/mnras/sty1986},
archivePrefix = {arXiv},
       eprint = {1807.10655},
 primaryClass = {astro-ph.GA},
       adsurl = {https://ui.adsabs.harvard.edu/abs/2018MNRAS.480.2609L}
}

@ARTICLE{Kim2015,
       author = {{Kim}, Dongwon and {Jerjen}, Helmut and {Milone}, Antonino P. and {Mackey}, Dougal and {Da Costa}, Gary S.},
        title = "{Discovery of a Faint Outer Halo Milky Way Star Cluster in the Southern Sky}",
      journal = {\apj},
         year = 2015,
        month = apr,
       volume = {803},
       number = {2},
          eid = {63},
        pages = {63},
          doi = {10.1088/0004-637X/803/2/63},
archivePrefix = {arXiv},
       eprint = {1502.03952},
 primaryClass = {astro-ph.GA},
       adsurl = {https://ui.adsabs.harvard.edu/abs/2015ApJ...803...63K}
}

@ARTICLE{Weatherford2020,
       author = {{Weatherford}, Newlin C. and {Chatterjee}, Sourav and {Kremer}, Kyle and {Rasio}, Frederic A.},
        title = "{A Dynamical Survey of Stellar-mass Black Holes in 50 Milky Way Globular Clusters}",
      journal = {\apj},
         year = 2020,
        month = aug,
       volume = {898},
       number = {2},
          eid = {162},
        pages = {162},
          doi = {10.3847/1538-4357/ab9f98},
archivePrefix = {arXiv},
       eprint = {1911.09125},
 primaryClass = {astro-ph.SR},
       adsurl = {https://ui.adsabs.harvard.edu/abs/2020ApJ...898..162W}
}

@ARTICLE{Dotter2016,
       author = {{Dotter}, Aaron},
        title = "{MESA Isochrones and Stellar Tracks (MIST) 0: Methods for the Construction of Stellar Isochrones}",
      journal = {\apjs},
         year = 2016,
        month = jan,
       volume = {222},
       number = {1},
          eid = {8},
        pages = {8},
          doi = {10.3847/0067-0049/222/1/8},
archivePrefix = {arXiv},
       eprint = {1601.05144},
 primaryClass = {astro-ph.SR},
       adsurl = {https://ui.adsabs.harvard.edu/abs/2016ApJS..222....8D}
}

@ARTICLE{Choi2016,
       author = {{Choi}, Jieun and {Dotter}, Aaron and {Conroy}, Charlie and {Cantiello}, Matteo and {Paxton}, Bill and {Johnson}, Benjamin D.},
        title = "{Mesa Isochrones and Stellar Tracks (MIST). I. Solar-scaled Models}",
      journal = {\apj},
         year = 2016,
        month = jun,
       volume = {823},
       number = {2},
          eid = {102},
        pages = {102},
          doi = {10.3847/0004-637X/823/2/102},
archivePrefix = {arXiv},
       eprint = {1604.08592},
 primaryClass = {astro-ph.SR},
       adsurl = {https://ui.adsabs.harvard.edu/abs/2016ApJ...823..102C}
}

@ARTICLE{Paust2009,
       author = {{Paust}, Nathaniel E.~Q. and {Aparicio}, Antonio and {Piotto}, Giampaolo and {Reid}, I. Neill and {Anderson}, Jay and {Sarajedini}, Ata and {Bedin}, Luigi R. and {Chaboyer}, Brian and {Dotter}, Aaron and {Hempel}, Maren and {Majewski}, Steven and {Mar{\'\i}n-Franch}, A. and {Milone}, Antonino and {Rosenberg}, Alfred and {Siegel}, Michael},
        title = "{The ACS Survey of Galactic Globular Clusters. VI. NGC 6366: A Heavily Stripped Galactic Globular Cluster}",
      journal = {\aj},
         year = 2009,
        month = jan,
       volume = {137},
       number = {1},
        pages = {246-256},
          doi = {10.1088/0004-6256/137/1/246},
       adsurl = {https://ui.adsabs.harvard.edu/abs/2009AJ....137..246P}
}

@ARTICLE{wolf2010,
       author = {{Wolf}, Joe and {Martinez}, Gregory D. and {Bullock}, James S. and {Kaplinghat}, Manoj and {Geha}, Marla and {Mu{\~n}oz}, Ricardo R. and {Simon}, Joshua D. and {Avedo}, Frank F.},
        title = "{Accurate masses for dispersion-supported galaxies}",
      journal = {\mnras},
         year = 2010,
        month = aug,
       volume = {406},
       number = {2},
        pages = {1220-1237},
          doi = {10.1111/j.1365-2966.2010.16753.x},
archivePrefix = {arXiv},
       eprint = {0908.2995},
 primaryClass = {astro-ph.CO},
       adsurl = {https://ui.adsabs.harvard.edu/abs/2010MNRAS.406.1220W}
}

@ARTICLE{Errani2025,
       author = {{Errani}, Rapha{\"e}l and {Pe{\~n}arrubia}, Jorge and {Walker}, Matthew G.},
        title = "{Stellar Mass Segregation in Dark Matter Halos}",
      journal = {arXiv e-prints},
         year = 2025,
        month = may,
          eid = {arXiv:2505.22717},
        pages = {arXiv:2505.22717},
          doi = {10.48550/arXiv.2505.22717},
archivePrefix = {arXiv},
       eprint = {2505.22717},
 primaryClass = {astro-ph.GA},
       adsurl = {https://ui.adsabs.harvard.edu/abs/2025arXiv250522717E}
}

@ARTICLE{Pace2022,
       author = {{Pace}, Andrew B. and {Erkal}, Denis and {Li}, Ting S.},
        title = "{Proper Motions, Orbits, and Tidal Influences of Milky Way Dwarf Spheroidal Galaxies}",
      journal = {\apj},
         year = 2022,
        month = dec,
       volume = {940},
       number = {2},
          eid = {136},
        pages = {136},
          doi = {10.3847/1538-4357/ac997b},
archivePrefix = {arXiv},
       eprint = {2205.05699},
 primaryClass = {astro-ph.GA},
       adsurl = {https://ui.adsabs.harvard.edu/abs/2022ApJ...940..136P}
}

@ARTICLE{Simon2018,
       author = {{Simon}, Joshua D.},
        title = "{Gaia Proper Motions and Orbits of the Ultra-faint Milky Way Satellites}",
      journal = {\apj},
         year = 2018,
        month = aug,
       volume = {863},
       number = {1},
          eid = {89},
        pages = {89},
          doi = {10.3847/1538-4357/aacdfb},
archivePrefix = {arXiv},
       eprint = {1804.10230},
 primaryClass = {astro-ph.GA},
       adsurl = {https://ui.adsabs.harvard.edu/abs/2018ApJ...863...89S}
}

@ARTICLE{Armstrong2021,
       author = {{Armstrong}, Benjamin M. and {Bekki}, Kenji and {Ludlow}, Aaron D.},
        title = "{The orbital evolution of UFDs and GCs in an evolving Galactic potential}",
      journal = {\mnras},
         year = 2021,
        month = jan,
       volume = {500},
       number = {3},
        pages = {2937-2957},
          doi = {10.1093/mnras/staa3391},
archivePrefix = {arXiv},
       eprint = {2011.12535},
 primaryClass = {astro-ph.GA},
       adsurl = {https://ui.adsabs.harvard.edu/abs/2021MNRAS.500.2937A}
}

@ARTICLE{galpy,
       author = {{Bovy}, Jo},
        title = "{galpy: A python Library for Galactic Dynamics}",
      journal = {\apjs},
         year = 2015,
        month = feb,
       volume = {216},
       number = {2},
          eid = {29},
        pages = {29},
          doi = {10.1088/0067-0049/216/2/29},
archivePrefix = {arXiv},
       eprint = {1412.3451},
 primaryClass = {astro-ph.GA},
       adsurl = {https://ui.adsabs.harvard.edu/abs/2015ApJS..216...29B}
}

@BOOK{Binney,
       author = {{Binney}, James and {Tremaine}, Scott},
        title = "{Galactic Dynamics: Second Edition}",
         year = 2008,
       adsurl = {https://ui.adsabs.harvard.edu/abs/2008gady.book.....B}
}

@ARTICLE{Miyamoto,
       author = {{Miyamoto}, M. and {Nagai}, R.},
        title = "{Three-dimensional models for the distribution of mass in galaxies.}",
      journal = {\pasj},
         year = 1975,
        month = jan,
       volume = {27},
        pages = {533-543},
       adsurl = {https://ui.adsabs.harvard.edu/abs/1975PASJ...27..533M}
}

@article{Navarro,
	doi = {10.1086/304888},
	url = {https://doi.org/10.1086%2F304888},
	year = 1997,
	month = {dec},
	publisher = {{IOP} Publishing},
	volume = {490},
	number = {2},
	pages = {493--508},
	author = {Julio F. Navarro and Carlos S. Frenk and Simon D. M. White},
	title = {A Universal Density Profile from Hierarchical Clustering},
	journal = {The Astrophysical Journal}
}

@ARTICLE{GaiaDR3,
       author = {{Gaia Collaboration} and {Vallenari}, A. and {Brown}, A.~G.~A. and {Prusti}, T. and {de Bruijne}, J.~H.~J. and {Arenou}, F. and {Babusiaux}, C. and {Biermann}, M. and {Creevey}, O.~L. and {Ducourant}, C. and {Evans}, D.~W. and {Eyer}, L. and {Guerra}, R. and {Hutton}, A. and {Jordi}, C. and {Klioner}, S.~A. and {Lammers}, U.~L. and {Lindegren}, L. and {Luri}, X. and {Mignard}, F. and {Panem}, C. and {Pourbaix}, D. and {Randich}, S. and {Sartoretti}, P. and {Soubiran}, C. and {Tanga}, P. and {Walton}, N.~A. and {Bailer-Jones}, C.~A.~L. and {Bastian}, U. and {Drimmel}, R. and {Jansen}, F. and {Katz}, D. and {Lattanzi}, M.~G. and {van Leeuwen}, F. and {Bakker}, J. and {Cacciari}, C. and {Casta{\~n}eda}, J. and {De Angeli}, F. and {Fabricius}, C. and {Fouesneau}, M. and {Fr{\'e}mat}, Y. and {Galluccio}, L. and {Guerrier}, A. and {Heiter}, U. and {Masana}, E. and {Messineo}, R. and {Mowlavi}, N. and {Nicolas}, C. and {Nienartowicz}, K. and {Pailler}, F. and {Panuzzo}, P. and {Riclet}, F. and {Roux}, W. and {Seabroke}, G.~M. and {Sordo}, R. and {Th{\'e}venin}, F. and {Gracia-Abril}, G. and {Portell}, J. and {Teyssier}, D. and {Altmann}, M. and {Andrae}, R. and {Audard}, M. and {Bellas-Velidis}, I. and {Benson}, K. and {Berthier}, J. and {Blomme}, R. and {Burgess}, P.~W. and {Busonero}, D. and {Busso}, G. and {C{\'a}novas}, H. and {Carry}, B. and {Cellino}, A. and {Cheek}, N. and {Clementini}, G. and {Damerdji}, Y. and {Davidson}, M. and {de Teodoro}, P. and {Nu{\~n}ez Campos}, M. and {Delchambre}, L. and {Dell'Oro}, A. and {Esquej}, P. and {Fern{\'a}ndez-Hern{\'a}ndez}, J. and {Fraile}, E. and {Garabato}, D. and {Garc{\'\i}a-Lario}, P. and {Gosset}, E. and {Haigron}, R. and {Halbwachs}, J. -L. and {Hambly}, N.~C. and {Harrison}, D.~L. and {Hern{\'a}ndez}, J. and {Hestroffer}, D. and {Hodgkin}, S.~T. and {Holl}, B. and {Jan{\ss}en}, K. and {Jevardat de Fombelle}, G. and {Jordan}, S. and {Krone-Martins}, A. and {Lanzafame}, A.~C. and {L{\"o}ffler}, W. and {Marchal}, O. and {Marrese}, P.~M. and {Moitinho}, A. and {Muinonen}, K. and {Osborne}, P. and {Pancino}, E. and {Pauwels}, T. and {Recio-Blanco}, A. and {Reyl{\'e}}, C. and {Riello}, M. and {Rimoldini}, L. and {Roegiers}, T. and {Rybizki}, J. and {Sarro}, L.~M. and {Siopis}, C. and {Smith}, M. and {Sozzetti}, A. and {Utrilla}, E. and {van Leeuwen}, M. and {Abbas}, U. and {{\'A}brah{\'a}m}, P. and {Abreu Aramburu}, A. and {Aerts}, C. and {Aguado}, J.~J. and {Ajaj}, M. and {Aldea-Montero}, F. and {Altavilla}, G. and {{\'A}lvarez}, M.~A. and {Alves}, J. and {Anders}, F. and {Anderson}, R.~I. and {Anglada Varela}, E. and {Antoja}, T. and {Baines}, D. and {Baker}, S.~G. and {Balaguer-N{\'u}{\~n}ez}, L. and {Balbinot}, E. and {Balog}, Z. and {Barache}, C. and {Barbato}, D. and {Barros}, M. and {Barstow}, M.~A. and {Bartolom{\'e}}, S. and {Bassilana}, J. -L. and {Bauchet}, N. and {Becciani}, U. and {Bellazzini}, M. and {Berihuete}, A. and {Bernet}, M. and {Bertone}, S. and {Bianchi}, L. and {Binnenfeld}, A. and {Blanco-Cuaresma}, S. and {Blazere}, A. and {Boch}, T. and {Bombrun}, A. and {Bossini}, D. and {Bouquillon}, S. and {Bragaglia}, A. and {Bramante}, L. and {Breedt}, E. and {Bressan}, A. and {Brouillet}, N. and {Brugaletta}, E. and {Bucciarelli}, B. and {Burlacu}, A. and {Butkevich}, A.~G. and {Buzzi}, R. and {Caffau}, E. and {Cancelliere}, R. and {Cantat-Gaudin}, T. and {Carballo}, R. and {Carlucci}, T. and {Carnerero}, M.~I. and {Carrasco}, J.~M. and {Casamiquela}, L. and {Castellani}, M. and {Castro-Ginard}, A. and {Chaoul}, L. and {Charlot}, P. and {Chemin}, L. and {Chiaramida}, V. and {Chiavassa}, A. and {Chornay}, N. and {Comoretto}, G. and {Contursi}, G. and {Cooper}, W.~J. and {Cornez}, T. and {Cowell}, S. and {Crifo}, F. and {Cropper}, M. and {Crosta}, M. and {Crowley}, C. and {Dafonte}, C. and {Dapergolas}, A. and {David}, M. and {David}, P. and {de Laverny}, P. and {De Luise}, F. and {De March}, R. and {De Ridder}, J. and {de Souza}, R. and {de Torres}, A. and {del Peloso}, E.~F. and {del Pozo}, E. and {Delbo}, M. and {Delgado}, A. and {Delisle}, J. -B. and {Demouchy}, C. and {Dharmawardena}, T.~E. and {Di Matteo}, P. and {Diakite}, S. and {Diener}, C. and {Distefano}, E. and {Dolding}, C. and {Edvardsson}, B. and {Enke}, H. and {Fabre}, C. and {Fabrizio}, M. and {Faigler}, S. and {Fedorets}, G. and {Fernique}, P. and {Fienga}, A. and {Figueras}, F. and {Fournier}, Y. and {Fouron}, C. and {Fragkoudi}, F. and {Gai}, M. and {Garcia-Gutierrez}, A. and {Garcia-Reinaldos}, M. and {Garc{\'\i}a-Torres}, M. and {Garofalo}, A. and {Gavel}, A. and {Gavras}, P. and {Gerlach}, E. and {Geyer}, R. and {Giacobbe}, P. and {Gilmore}, G. and {Girona}, S. and {Giuffrida}, G. and {Gomel}, R. and {Gomez}, A. and {Gonz{\'a}lez-N{\'u}{\~n}ez}, J. and {Gonz{\'a}lez-Santamar{\'\i}a}, I. and {Gonz{\'a}lez-Vidal}, J.~J. and {Granvik}, M. and {Guillout}, P. and {Guiraud}, J. and {Guti{\'e}rrez-S{\'a}nchez}, R. and {Guy}, L.~P. and {Hatzidimitriou}, D. and {Hauser}, M. and {Haywood}, M. and {Helmer}, A. and {Helmi}, A. and {Sarmiento}, M.~H. and {Hidalgo}, S.~L. and {Hilger}, T. and {H{\l}adczuk}, N. and {Hobbs}, D. and {Holland}, G. and {Huckle}, H.~E. and {Jardine}, K. and {Jasniewicz}, G. and {Jean-Antoine Piccolo}, A. and {Jim{\'e}nez-Arranz}, {\'O}. and {Jorissen}, A. and {Juaristi Campillo}, J. and {Julbe}, F. and {Karbevska}, L. and {Kervella}, P. and {Khanna}, S. and {Kontizas}, M. and {Kordopatis}, G. and {Korn}, A.~J. and {K{\'o}sp{\'a}l}, {\'A}. and {Kostrzewa-Rutkowska}, Z. and {Kruszy{\'n}ska}, K. and {Kun}, M. and {Laizeau}, P. and {Lambert}, S. and {Lanza}, A.~F. and {Lasne}, Y. and {Le Campion}, J. -F. and {Lebreton}, Y. and {Lebzelter}, T. and {Leccia}, S. and {Leclerc}, N. and {Lecoeur-Taibi}, I. and {Liao}, S. and {Licata}, E.~L. and {Lindstr{\o}m}, H.~E.~P. and {Lister}, T.~A. and {Livanou}, E. and {Lobel}, A. and {Lorca}, A. and {Loup}, C. and {Madrero Pardo}, P. and {Magdaleno Romeo}, A. and {Managau}, S. and {Mann}, R.~G. and {Manteiga}, M. and {Marchant}, J.~M. and {Marconi}, M. and {Marcos}, J. and {Marcos Santos}, M.~M.~S. and {Mar{\'\i}n Pina}, D. and {Marinoni}, S. and {Marocco}, F. and {Marshall}, D.~J. and {Martin Polo}, L. and {Mart{\'\i}n-Fleitas}, J.~M. and {Marton}, G. and {Mary}, N. and {Masip}, A. and {Massari}, D. and {Mastrobuono-Battisti}, A. and {Mazeh}, T. and {McMillan}, P.~J. and {Messina}, S. and {Michalik}, D. and {Millar}, N.~R. and {Mints}, A. and {Molina}, D. and {Molinaro}, R. and {Moln{\'a}r}, L. and {Monari}, G. and {Mongui{\'o}}, M. and {Montegriffo}, P. and {Montero}, A. and {Mor}, R. and {Mora}, A. and {Morbidelli}, R. and {Morel}, T. and {Morris}, D. and {Muraveva}, T. and {Murphy}, C.~P. and {Musella}, I. and {Nagy}, Z. and {Noval}, L. and {Oca{\~n}a}, F. and {Ogden}, A. and {Ordenovic}, C. and {Osinde}, J.~O. and {Pagani}, C. and {Pagano}, I. and {Palaversa}, L. and {Palicio}, P.~A. and {Pallas-Quintela}, L. and {Panahi}, A. and {Payne-Wardenaar}, S. and {Pe{\~n}alosa Esteller}, X. and {Penttil{\"a}}, A. and {Pichon}, B. and {Piersimoni}, A.~M. and {Pineau}, F. -X. and {Plachy}, E. and {Plum}, G. and {Poggio}, E. and {Pr{\v{s}}a}, A. and {Pulone}, L. and {Racero}, E. and {Ragaini}, S. and {Rainer}, M. and {Raiteri}, C.~M. and {Rambaux}, N. and {Ramos}, P. and {Ramos-Lerate}, M. and {Re Fiorentin}, P. and {Regibo}, S. and {Richards}, P.~J. and {Rios Diaz}, C. and {Ripepi}, V. and {Riva}, A. and {Rix}, H. -W. and {Rixon}, G. and {Robichon}, N. and {Robin}, A.~C. and {Robin}, C. and {Roelens}, M. and {Rogues}, H.~R.~O. and {Rohrbasser}, L. and {Romero-G{\'o}mez}, M. and {Rowell}, N. and {Royer}, F. and {Ruz Mieres}, D. and {Rybicki}, K.~A. and {Sadowski}, G. and {S{\'a}ez N{\'u}{\~n}ez}, A. and {Sagrist{\`a} Sell{\'e}s}, A. and {Sahlmann}, J. and {Salguero}, E. and {Samaras}, N. and {Sanchez Gimenez}, V. and {Sanna}, N. and {Santove{\~n}a}, R. and {Sarasso}, M. and {Schultheis}, M. and {Sciacca}, E. and {Segol}, M. and {Segovia}, J.~C. and {S{\'e}gransan}, D. and {Semeux}, D. and {Shahaf}, S. and {Siddiqui}, H.~I. and {Siebert}, A. and {Siltala}, L. and {Silvelo}, A. and {Slezak}, E. and {Slezak}, I. and {Smart}, R.~L. and {Snaith}, O.~N. and {Solano}, E. and {Solitro}, F. and {Souami}, D. and {Souchay}, J. and {Spagna}, A. and {Spina}, L. and {Spoto}, F. and {Steele}, I.~A. and {Steidelm{\"u}ller}, H. and {Stephenson}, C.~A. and {S{\"u}veges}, M. and {Surdej}, J. and {Szabados}, L. and {Szegedi-Elek}, E. and {Taris}, F. and {Taylor}, M.~B. and {Teixeira}, R. and {Tolomei}, L. and {Tonello}, N. and {Torra}, F. and {Torra}, J. and {Torralba Elipe}, G. and {Trabucchi}, M. and {Tsounis}, A.~T. and {Turon}, C. and {Ulla}, A. and {Unger}, N. and {Vaillant}, M.~V. and {van Dillen}, E. and {van Reeven}, W. and {Vanel}, O. and {Vecchiato}, A. and {Viala}, Y. and {Vicente}, D. and {Voutsinas}, S. and {Weiler}, M. and {Wevers}, T. and {Wyrzykowski}, {\L}. and {Yoldas}, A. and {Yvard}, P. and {Zhao}, H. and {Zorec}, J. and {Zucker}, S. and {Zwitter}, T.},
        title = "{Gaia Data Release 3. Summary of the content and survey properties}",
      journal = {\aap},
         year = 2023,
        month = jun,
       volume = {674},
          eid = {A1},
        pages = {A1},
          doi = {10.1051/0004-6361/202243940},
archivePrefix = {arXiv},
       eprint = {2208.00211},
 primaryClass = {astro-ph.GA},
       adsurl = {https://ui.adsabs.harvard.edu/abs/2023A&A...674A...1G}
}

@ARTICLE{simon07a,
   author = {{Simon}, J.~D. and {Geha}, M.},
    title = "{The Kinematics of the Ultra-faint Milky Way Satellites: Solving the Missing Satellite Problem}",
  journal = {\apj},
   eprint = {arXiv:0706.0516},
     year = 2007,
    month = nov,
   volume = 670,
    pages = {313-331},
      doi = {10.1086/521816},
   adsurl = {http://adsabs.harvard.edu/abs/2007ApJ...670..313S}
}

@ARTICLE{Penarrubia2009,
       author = {{Pe{\~n}arrubia}, Jorge and {Navarro}, Julio F. and {McConnachie}, Alan W. and {Martin}, Nicolas F.},
        title = "{The Signature of Galactic Tides in Local Group Dwarf Spheroidals}",
      journal = {\apj},
         year = 2009,
        month = jun,
       volume = {698},
       number = {1},
        pages = {222-232},
          doi = {10.1088/0004-637X/698/1/222},
archivePrefix = {arXiv},
       eprint = {0811.1579},
 primaryClass = {astro-ph},
       adsurl = {https://ui.adsabs.harvard.edu/abs/2009ApJ...698..222P}
}

@ARTICLE{Kirby2015_tri,
       author = {{Kirby}, Evan N. and {Cohen}, Judith G. and {Simon}, Joshua D. and {Guhathakurta}, Puragra},
        title = "{Triangulum II: Possibly a Very Dense Ultra-faint Dwarf Galaxy}",
      journal = {\apjl},
         year = 2015,
        month = nov,
       volume = {814},
       number = {1},
          eid = {L7},
        pages = {L7},
          doi = {10.1088/2041-8205/814/1/L7},
archivePrefix = {arXiv},
       eprint = {1510.03856},
 primaryClass = {astro-ph.GA},
       adsurl = {https://ui.adsabs.harvard.edu/abs/2015ApJ...814L...7K}
}

@ARTICLE{Kirby2013_seg2,
       author = {{Kirby}, Evan N. and {Boylan-Kolchin}, Michael and {Cohen}, Judith G. and {Geha}, Marla and {Bullock}, James S. and {Kaplinghat}, Manoj},
        title = "{Segue 2: The Least Massive Galaxy}",
      journal = {\apj},
         year = 2013,
        month = jun,
       volume = {770},
       number = {1},
          eid = {16},
        pages = {16},
          doi = {10.1088/0004-637X/770/1/16},
archivePrefix = {arXiv},
       eprint = {1304.6080},
 primaryClass = {astro-ph.CO},
       adsurl = {https://ui.adsabs.harvard.edu/abs/2013ApJ...770...16K}
}

@ARTICLE{Fritz2018,
       author = {{Fritz}, T.~K. and {Battaglia}, G. and {Pawlowski}, M.~S. and {Kallivayalil}, N. and {van der Marel}, R. and {Sohn}, S.~T. and {Brook}, C. and {Besla}, G.},
        title = "{Gaia DR2 proper motions of dwarf galaxies within 420 kpc. Orbits, Milky Way mass, tidal influences, planar alignments, and group infall}",
      journal = {\aap},
         year = 2018,
        month = nov,
       volume = {619},
          eid = {A103},
        pages = {A103},
          doi = {10.1051/0004-6361/201833343},
archivePrefix = {arXiv},
       eprint = {1805.00908},
 primaryClass = {astro-ph.GA},
       adsurl = {https://ui.adsabs.harvard.edu/abs/2018A&A...619A.103F}
}

@ARTICLE{Pace2025,
       author = {{Pace}, Andrew B},
        title = "{The Local Volume Database: a library of the observed properties of nearby dwarf galaxies and star clusters}",
      journal = {The Open Journal of Astrophysics},
         year = 2025,
        month = sep,
       volume = {8},
          eid = {142},
        pages = {142},
          doi = {10.33232/001c.144859},
archivePrefix = {arXiv},
       eprint = {2411.07424},
 primaryClass = {astro-ph.GA},
       adsurl = {https://ui.adsabs.harvard.edu/abs/2025OJAp....8E.142P}
}

@ARTICLE{fu2023,
       author = {{Fu}, Sal Wanying and {Weisz}, Daniel R. and {Starkenburg}, Else and {Martin}, Nicolas and {Savino}, Alessandro and {Boylan-Kolchin}, Michael and {Cote}, Patrick and {Dolphin}, Andrew E. and {Ji}, Alexander P. and {Longeard}, Nicolas and {Mateo}, Mario L. and {Patel}, Ekta and {Sandford}, Nathan R.},
        title = "{Metallicity Distribution Functions of 13 Ultra-Faint Dwarf Galaxy Candidates from Hubble Space Telescope Narrowband Imaging}",
      journal = {arXiv e-prints},
         year = 2023,
        month = jun,
          eid = {arXiv:2306.06260},
        pages = {arXiv:2306.06260},
          doi = {10.48550/arXiv.2306.06260},
archivePrefix = {arXiv},
       eprint = {2306.06260},
 primaryClass = {astro-ph.GA},
       adsurl = {https://ui.adsabs.harvard.edu/abs/2023arXiv230606260F}
}

@ARTICLE{plummer,
       author = {{Plummer}, H.~C.},
        title = "{On the problem of distribution in globular star clusters}",
      journal = {\mnras},
         year = 1911,
        month = mar,
       volume = {71},
        pages = {460-470},
          doi = {10.1093/mnras/71.5.460},
       adsurl = {https://ui.adsabs.harvard.edu/abs/1911MNRAS..71..460P}
}

@ARTICLE{DSouza2022,
       author = {{D'Souza}, Richard and {Bell}, Eric F.},
        title = "{Uncertainties associated with the backward integration of dwarf satellites using simple parametric potentials}",
      journal = {\mnras},
         year = 2022,
        month = may,
       volume = {512},
       number = {1},
        pages = {739-760},
          doi = {10.1093/mnras/stac404},
archivePrefix = {arXiv},
       eprint = {2202.05707},
 primaryClass = {astro-ph.GA},
       adsurl = {https://ui.adsabs.harvard.edu/abs/2022MNRAS.512..739D}
}

@ARTICLE{Fardal2015,
       author = {{Fardal}, Mark A. and {Huang}, Shuiyao and {Weinberg}, Martin D.},
        title = "{Generation of mock tidal streams}",
      journal = {\mnras},
         year = 2015,
        month = sep,
       volume = {452},
       number = {1},
        pages = {301-319},
          doi = {10.1093/mnras/stv1198},
archivePrefix = {arXiv},
       eprint = {1410.1861},
 primaryClass = {astro-ph.GA},
       adsurl = {https://ui.adsabs.harvard.edu/abs/2015MNRAS.452..301F}
}

@ARTICLE{Ji2021,
       author = {{Ji}, Alexander P. and {Koposov}, Sergey E. and {Li}, Ting S. and {Erkal}, Denis and {Pace}, Andrew B. and {Simon}, Joshua D. and {Belokurov}, Vasily and {Cullinane}, Lara R. and {Da Costa}, Gary S. and {Kuehn}, Kyler and {Lewis}, Geraint F. and {Mackey}, Dougal and {Shipp}, Nora and {Simpson}, Jeffrey D. and {Zucker}, Daniel B. and {Hansen}, Terese T. and {Bland-Hawthorn}, Joss and {S5 Collaboration}},
        title = "{Kinematics of Antlia 2 and Crater 2 from the Southern Stellar Stream Spectroscopic Survey (S$^{5}$)}",
      journal = {\apj},
         year = 2021,
        month = nov,
       volume = {921},
       number = {1},
          eid = {32},
        pages = {32},
          doi = {10.3847/1538-4357/ac1869},
archivePrefix = {arXiv},
       eprint = {2106.12656},
 primaryClass = {astro-ph.GA},
       adsurl = {https://ui.adsabs.harvard.edu/abs/2021ApJ...921...32J}
}

@ARTICLE{Magnus2022,
       author = {{Correa Magnus}, Lilia and {Vasiliev}, Eugene},
        title = "{Measuring the Milky Way mass distribution in the presence of the LMC}",
      journal = {\mnras},
         year = 2022,
        month = apr,
       volume = {511},
       number = {2},
        pages = {2610-2630},
          doi = {10.1093/mnras/stab3726},
archivePrefix = {arXiv},
       eprint = {2110.00018},
 primaryClass = {astro-ph.GA},
       adsurl = {https://ui.adsabs.harvard.edu/abs/2022MNRAS.511.2610C}
}

@ARTICLE{Erkal2021,
       author = {{Erkal}, Denis and {Deason}, Alis J. and {Belokurov}, Vasily and {Xue}, Xiang-Xiang and {Koposov}, Sergey E. and {Bird}, Sarah A. and {Liu}, Chao and {Simion}, Iulia T. and {Yang}, Chengqun and {Zhang}, Lan and {Zhao}, Gang},
        title = "{Detection of the LMC-induced sloshing of the Galactic halo}",
      journal = {\mnras},
         year = 2021,
        month = sep,
       volume = {506},
       number = {2},
        pages = {2677-2684},
          doi = {10.1093/mnras/stab1828},
archivePrefix = {arXiv},
       eprint = {2010.13789},
 primaryClass = {astro-ph.GA},
       adsurl = {https://ui.adsabs.harvard.edu/abs/2021MNRAS.506.2677E}
}

@ARTICLE{Hernquist1990,
       author = {{Hernquist}, Lars},
        title = "{An Analytical Model for Spherical Galaxies and Bulges}",
      journal = {\apj},
         year = 1990,
        month = jun,
       volume = {356},
        pages = {359},
          doi = {10.1086/168845},
       adsurl = {https://ui.adsabs.harvard.edu/abs/1990ApJ...356..359H}
}

@ARTICLE{Marel2014,
       author = {{van der Marel}, Roeland P. and {Kallivayalil}, Nitya},
        title = "{Third-epoch Magellanic Cloud Proper Motions. II. The Large Magellanic Cloud Rotation Field in Three Dimensions}",
      journal = {\apj},
         year = 2014,
        month = feb,
       volume = {781},
       number = {2},
          eid = {121},
        pages = {121},
          doi = {10.1088/0004-637X/781/2/121},
archivePrefix = {arXiv},
       eprint = {1305.4641},
 primaryClass = {astro-ph.CO},
       adsurl = {https://ui.adsabs.harvard.edu/abs/2014ApJ...781..121V}
}

@ARTICLE{Innanen1983,
       author = {{Innanen}, K.~A. and {Harris}, W.~E. and {Webbink}, R.~F.},
        title = "{Globular cluster orbits and the galactic mass distribution.}",
      journal = {\aj},
         year = 1983,
        month = mar,
       volume = {88},
        pages = {338-360},
          doi = {10.1086/113320},
       adsurl = {https://ui.adsabs.harvard.edu/abs/1983AJ.....88..338I}
}

@ARTICLE{Walker2006,
       author = {{Walker}, Matthew G. and {Mateo}, Mario and {Olszewski}, Edward W. and {Bernstein}, Rebecca and {Wang}, Xiao and {Woodroofe}, Michael},
        title = "{Internal Kinematics of the Fornax Dwarf Spheroidal Galaxy}",
      journal = {\aj},
         year = 2006,
        month = apr,
       volume = {131},
       number = {4},
        pages = {2114-2139},
          doi = {10.1086/500193},
archivePrefix = {arXiv},
       eprint = {astro-ph/0511465},
 primaryClass = {astro-ph},
       adsurl = {https://ui.adsabs.harvard.edu/abs/2006AJ....131.2114W}
}

@ARTICLE{Dey2019,
       author = {{Dey}, Arjun and others},
        title = "{Overview of the DESI Legacy Imaging Surveys}",
      journal = {\aj},
         year = 2019,
        month = may,
       volume = {157},
       number = {5},
          eid = {168},
        pages = {168},
          doi = {10.3847/1538-3881/ab089d},
archivePrefix = {arXiv},
       eprint = {1804.08657},
 primaryClass = {astro-ph.IM},
       adsurl = {https://ui.adsabs.harvard.edu/abs/2019AJ....157..168D}
}

@misc{Machado2025,
      title={Accuracy of analytic potentials for orbits of satellites around a Milky Way-like galaxy: comparison with $N$-body simulations}, 
      author={Rubens E. G. Machado and Giovanni C. Tauil and Nicholas Schweder-Souza},
      year={2025},
      eprint={2506.13813},
      archivePrefix={arXiv},
      primaryClass={astro-ph.GA},
      url={https://arxiv.org/abs/2506.13813}, 
}

@ARTICLE{maxted2001,
       author = {{Maxted}, P.~F.~L. and {Heber}, U. and {Marsh}, T.~R. and {North}, R.~C.},
        title = "{The binary fraction of extreme horizontal branch stars}",
      journal = {\mnras},
         year = 2001,
        month = oct,
       volume = {326},
       number = {4},
        pages = {1391-1402},
          doi = {10.1111/j.1365-2966.2001.04714.x},
archivePrefix = {arXiv},
       eprint = {astro-ph/0103342},
 primaryClass = {astro-ph},
       adsurl = {https://ui.adsabs.harvard.edu/abs/2001MNRAS.326.1391M}
}

@ARTICLE{Smith2024,
       author = {{Smith}, Simon E.~T. and {Cerny}, William and {Hayes}, Christian R. and {Sestito}, Federico and {Jensen}, Jaclyn and {McConnachie}, Alan W. and {Geha}, Marla and {Navarro}, Julio F. and {Li}, Ting S. and {Cuillandre}, Jean-Charles and {Errani}, Rapha{\"e}l and {Chambers}, Ken and {Gwyn}, Stephen and {Hammer}, Francois and {Hudson}, Michael J. and {Magnier}, Eugene and {Martin}, Nicolas},
        title = "{The Discovery of the Faintest Known Milky Way Satellite Using UNIONS}",
      journal = {\apj},
         year = 2024,
        month = jan,
       volume = {961},
       number = {1},
          eid = {92},
        pages = {92},
          doi = {10.3847/1538-4357/ad0d9f},
archivePrefix = {arXiv},
       eprint = {2311.10147},
 primaryClass = {astro-ph.GA},
       adsurl = {https://ui.adsabs.harvard.edu/abs/2024ApJ...961...92S}
}

@ARTICLE{Husser2013,
       author = {{Husser}, T. -O. and {Wende-von Berg}, S. and {Dreizler}, S. and {Homeier}, D. and {Reiners}, A. and {Barman}, T. and {Hauschildt}, P.~H.},
        title = "{A new extensive library of PHOENIX stellar atmospheres and synthetic spectra}",
      journal = {\aap},
         year = 2013,
        month = may,
       volume = {553},
          eid = {A6},
        pages = {A6},
          doi = {10.1051/0004-6361/201219058},
archivePrefix = {arXiv},
       eprint = {1303.5632},
 primaryClass = {astro-ph.SR},
       adsurl = {https://ui.adsabs.harvard.edu/abs/2013A&A...553A...6H}
}

@ARTICLE{pypeit2020,
       author = {{Prochaska}, J. Xavier and {Hennawi}, Joseph F. and {Westfall}, Kyle B. and
         {Cooke}, Ryan J. and {Wang}, Feige and {Hsyu}, Tiffany and
         {Farina}, Emanuele Paolo},
        title = "{PypeIt: The Python Spectroscopic Data Reduction Pipeline}",
      journal = {arXiv e-prints},
         year = 2020,
        month = may,
          eid = {arXiv:2005.06505},
        pages = {arXiv:2005.06505},
archivePrefix = {arXiv},
       eprint = {2005.06505},
 primaryClass = {astro-ph.IM},
       adsurl = {https://ui.adsabs.harvard.edu/abs/2020arXiv200506505P}
}

@ARTICLE{emcee,
       author = {{Foreman-Mackey}, Daniel and {Hogg}, David W. and {Lang}, Dustin and
         {Goodman}, Jonathan},
        title = "{emcee: The MCMC Hammer}",
      journal = {\pasp},
         year = 2013,
        month = mar,
       volume = {125},
       number = {925},
        pages = {306},
          doi = {10.1086/670067},
archivePrefix = {arXiv},
       eprint = {1202.3665},
 primaryClass = {astro-ph.IM},
       adsurl = {https://ui.adsabs.harvard.edu/abs/2013PASP..125..306F}
}

@ARTICLE{telfit2014,
       author = {{Gullikson}, Kevin and {Dodson-Robinson}, Sarah and {Kraus}, Adam},
        title = "{Correcting for Telluric Absorption: Methods, Case Studies, and Release of the TelFit Code}",
      journal = {\aj},
         year = 2014,
        month = sep,
       volume = {148},
       number = {3},
          eid = {53},
        pages = {53},
          doi = {10.1088/0004-6256/148/3/53},
archivePrefix = {arXiv},
       eprint = {1406.6059},
 primaryClass = {astro-ph.IM},
       adsurl = {https://ui.adsabs.harvard.edu/abs/2014AJ....148...53G}
}

@ARTICLE{Carrera2013,
   author = {{Carrera}, R. and {Pancino}, E. and {Gallart}, C. and {del Pino}, A.
	},
    title = "{The near-infrared Ca II triplet as a metallicity indicator - II. Extension to extremely metal-poor metallicity regimes}",
  journal = {\mnras},
archivePrefix = "arXiv",
   eprint = {1306.3883},
     year = 2013,
    month = sep,
   volume = 434,
    pages = {1681-1691},
      doi = {10.1093/mnras/stt1126},
   adsurl = {http://adsabs.harvard.edu/abs/2013MNRAS.434.1681C}
}

@ARTICLE{sohn2007,
       author = {{Sohn}, Sangmo Tony and {Majewski}, Steven R. and
         {Mu{\~n}oz}, Ricardo R. and {Kunkel}, William E. and
         {Johnston}, Kathryn V. and {Ostheimer}, James C. and
         {Guhathakurta}, Puragra and {Patterson}, Richard J. and
         {Siegel}, Michael H. and {Cooper}, Michael C.},
        title = "{Exploring Halo Substructure with Giant Stars. X. Extended Dark Matter or Tidal Disruption?: The Case for the Leo I Dwarf Spheroidal Galaxy}",
      journal = {\apj},
         year = 2007,
        month = jul,
       volume = {663},
       number = {2},
        pages = {960-989},
          doi = {10.1086/518302},
archivePrefix = {arXiv},
       eprint = {astro-ph/0608151},
 primaryClass = {astro-ph},
       adsurl = {https://ui.adsabs.harvard.edu/abs/2007ApJ...663..960S}
}

@ARTICLE{munoz2018a,
   author = {{Mu{\~n}oz}, R.~R. and {C{\^o}t{\'e}}, P. and {Santana}, F.~A. and 
	{Geha}, M. and {Simon}, J.~D. and {Oyarz{\'u}n}, G.~A. and {Stetson}, P.~B. and 
	{Djorgovski}, S.~G.},
    title = "{A MegaCam Survey of Outer Halo Satellites. I. Description of the Survey}",
  journal = {\apj},
     year = 2018,
    month = jun,
   volume = 860,
      eid = {65},
    pages = {65},
      doi = {10.3847/1538-4357/aac168},
   adsurl = {http://adsabs.harvard.edu/abs/2018ApJ...860...65M}
}

@ARTICLE{munoz2018b,
       author = {{Mu{\~n}oz}, Ricardo R. and {C{\^o}t{\'e}}, Patrick and {Santana}, Felipe A. and {Geha}, Marla and {Simon}, Joshua D. and {Oyarz{\'u}n}, Grecco A. and {Stetson}, Peter B. and {Djorgovski}, S.~G.},
        title = "{A MegaCam Survey of Outer Halo Satellites. III. Photometric and Structural Parameters}",
      journal = {\apj},
         year = 2018,
        month = jun,
       volume = {860},
       number = {1},
          eid = {66},
        pages = {66},
          doi = {10.3847/1538-4357/aac16b},
archivePrefix = {arXiv},
       eprint = {1806.06891},
 primaryClass = {astro-ph.GA},
       adsurl = {https://ui.adsabs.harvard.edu/abs/2018ApJ...860...66M}
}

@ARTICLE{Martinez2011,
   author = {{Martinez}, G.~D. and {Minor}, Q.~E. and {Bullock}, J. and {Kaplinghat}, M. and 
	{Simon}, J.~D. and {Geha}, M.},
    title = "{A Complete Spectroscopic Survey of the Milky Way Satellite Segue 1: Dark Matter Content, Stellar Membership, and Binary Properties from a Bayesian Analysis}",
  journal = {\apj},
archivePrefix = "arXiv",
   eprint = {1008.4585},
 primaryClass = "astro-ph.GA",
     year = 2011,
    month = sep,
   volume = 738,
      eid = {55},
    pages = {55},
      doi = {10.1088/0004-637X/738/1/55},
   adsurl = {http://adsabs.harvard.edu/abs/2011ApJ...738...55M}
}

@INPROCEEDINGS{faber03a,
   author = {{Faber}, S.~M. and others},
    title = "{The DEIMOS spectrograph for the Keck II Telescope: integration and testing}",
booktitle = {Instrument Design and Performance for Optical/Infrared Ground-based Telescopes.  Edited by Iye \& Moorwood, Proceedings of the SPIE, Volume 4841, pp. 1657},
     year = 2003,
    month = mar,
    pages = {},
   adsurl = {http://adsabs.harvard.edu/cgi-bin/nph-bib_query?bibcode=2003SPIE.4841.1657F&db_key=AST}
}

@ARTICLE{Martin2007,
   author = {{Martin}, N.~F. and {Ibata}, R.~A. and {Chapman}, S.~C. and 
	{Irwin}, M. and {Lewis}, G.~F.},
    title = "{A Keck/DEIMOS spectroscopic survey of faint Galactic satellites: searching for the least massive dwarf galaxies}",
  journal = {\mnras},
archivePrefix = "arXiv",
   eprint = {0705.4622},
     year = 2007,
    month = sep,
   volume = 380,
    pages = {281-300},
      doi = {10.1111/j.1365-2966.2007.12055.x},
   adsurl = {http://adsabs.harvard.edu/abs/2007MNRAS.380..281M}
}

@ARTICLE{Cerny2026,
       author = {{Cerny}, William and {Li}, Ting S. and {Pace}, Andrew B. and {Simon}, Joshua D. and {Geha}, Marla and {Ji}, Alexander P. and {Drlica-Wagner}, Alex and {Bruce}, Jordan and {Gnedin}, Oleg Y. and {Bell}, Eric F. and {Mau}, Sidney and {Escala}, Ivanna and {Bissonette}, Daisy and {Savino}, Alessandro and {Chiti}, Anirudh and {Kirby}, Evan N.},
        title = "{A Chemodynamical Census of the Milky Way's Ultra-Faint Compact Satellites. I. A First Population-Level Look at the Internal Kinematics and Metallicities of 19 Extremely-Low-Mass Halo Stellar Systems}",
      journal = {arXiv e-prints},
         year = 2026,
        month = feb,
          eid = {arXiv:2602.17652},
        pages = {arXiv:2602.17652},
          doi = {10.48550/arXiv.2602.17652},
archivePrefix = {arXiv},
       eprint = {2602.17652},
 primaryClass = {astro-ph.GA},
       adsurl = {https://ui.adsabs.harvard.edu/abs/2026arXiv260217652C}
}

@ARTICLE{numpy,
author={S. van der Walt and S. C. Colbert and G. Varoquaux},
journal={Computing in Science Engineering},
title={The NumPy Array: A Structure for Efficient Numerical Computation},
year={2011},
volume={13},
number={2},
pages={22-30},
doi={10.1109/MCSE.2011.37},
ISSN={1521-9615},
month={March},}

@ARTICLE{ipython,
author={F. Perez and B. E. Granger},
journal={Computing in Science Engineering},
title={IPython: A System for Interactive Scientific Computing},
year={2007},
volume={9},
number={3},
pages={21-29},
doi={10.1109/MCSE.2007.53},
ISSN={1521-9615},
month={May},}

@ARTICLE{matplotlib,
author={J. D. Hunter},
journal={Computing in Science Engineering},
title={Matplotlib: A 2D Graphics Environment},
year={2007},
volume={9},
number={3},
pages={90-95},
doi={10.1109/MCSE.2007.55},
ISSN={1521-9615},
month={May},}

@ARTICLE{astropy,
   author = {{Astropy Collaboration} and {Robitaille}, T.~P. and {Tollerud}, E.~J. and
	{Greenfield}, P. and {Droettboom}, M. and {Bray}, E. and {Aldcroft}, T. and
	{Davis}, M. and {Ginsburg}, A. and {Price-Whelan}, A.~M. and
	{Kerzendorf}, W.~E. and {Conley}, A. and {Crighton}, N. and
	{Barbary}, K. and {Muna}, D. and {Ferguson}, H. and {Grollier}, F. and
	{Parikh}, M.~M. and {Nair}, P.~H. and {Unther}, H.~M. and {Deil}, C. and
	{Woillez}, J. and {Conseil}, S. and {Kramer}, R. and {Turner}, J.~E.~H. and
	{Singer}, L. and {Fox}, R. and {Weaver}, B.~A. and {Zabalza}, V. and
	{Edwards}, Z.~I. and {Azalee Bostroem}, K. and {Burke}, D.~J. and
	{Casey}, A.~R. and {Crawford}, S.~M. and {Dencheva}, N. and
	{Ely}, J. and {Jenness}, T. and {Labrie}, K. and {Lim}, P.~L. and
	{Pierfederici}, F. and {Pontzen}, A. and {Ptak}, A. and {Refsdal}, B. and
	{Servillat}, M. and {Streicher}, O.},
    title = "{Astropy: A community Python package for astronomy}",
  journal = {\aap},
archivePrefix = "arXiv",
   eprint = {1307.6212},
 primaryClass = "astro-ph.IM",
     year = 2013,
    month = oct,
   volume = 558,
      eid = {A33},
    pages = {A33},
      doi = {10.1051/0004-6361/201322068},
   adsurl = {http://adsabs.harvard.edu/abs/2013A%26A...558A..33A}
}

@InProceedings{ pandas,
  author    = { Wes McKinney },
  title     = { Data Structures for Statistical Computing in Python },
  booktitle = { Proceedings of the 9th Python in Science Conference },
  pages     = { 51 - 56 },
  year      = {2010 },
  editor    = { St\'efan van der Walt and Jarrod Millman }
}

@Misc{scipy,
  author =    {Eric Jones and Travis Oliphant and Pearu Peterson and others},
  title =     {{SciPy}: Open source scientific tools for {Python}},
  year =      {2001},
  url = "http://www.scipy.org/",
  note = {[Online; \href{http://www.scipy.org/}{scipy.org}]}
}
\bibliographystyle{mnras}

\appendix

\renewcommand{\thetable}{A\arabic{table}}

\section{Properties of Keck/DEIMOS stars}

Table \ref{tab:members} presents the properties of the 72 stars used in \S\ref{ssec:final_members} and \S\ref{ssec:vel_mass}.  This includes the \ntot\ stars in our member sample ($P_{\rm mem} > 0.5$) and the \nmix\ stars used in our mixture model ($P_{\rm mem, nov} > 0.1$, $\rell < 3$, Var $\neq 1$). We refer the reader to \citet{geha_paper1} for a detailed discussion of the data reduction. The full table of stars, including nonmembers, is available in Table A3 of \citet{geha_paper1}.

\startlongtable
\begin{deluxetable*}{cccccccccccccc}
\tablewidth{\columnwidth}
\tabletypesize{\footnotesize}
\tablecaption{Measured properties of Keck/DEIMOS stars used in our analysis. \label{tab:members}}
\tablehead{\colhead{R.A.} & \colhead{Decl.} & \colhead{$v$} & \colhead{$v_{\rm err}$} & \colhead{$r_o$} & \colhead{$r_{\rm err}$} & \colhead{$g_o$} & \colhead{$g_{\rm err}$} & \colhead{{\rm [Fe/H]}} & \colhead{${\rm [Fe/H]}_{\rm err}$} & \colhead{gaia\_source\_id} & \colhead{Var} & \colhead{$P_{\rm mem}$} & \colhead{Sample} \\
\colhead{(deg)} & \colhead{(deg)} & \colhead{(\kms)} & \colhead{(\kms)} & \colhead{(mag)} & \colhead{(mag)} & \colhead{(mag)} & \colhead{(mag)} & \colhead{···} & \colhead{dex} & \colhead{···} & \colhead{···} & \colhead{···} & \colhead{···}}
\startdata
\hline
162.468917 & 51.061806 & $-19.51$ & 0.69 & 18.045 & 0.020 & 18.632 & 0.020 & $-2.56$ & 0.13 & 835977180433181056 & 0 & 0.68 & 1 \\
162.325208 & 51.038000 & $-6.49$ & 0.80 & 18.057 & 0.020 & 18.637 & 0.020 & $-2.31$ & 0.13 & 835971820314006528 & 1 & 0.75 & 1 \\
162.282500 & 50.947306 & $-11.87$ & 1.37 & 18.294 & 0.020 & 18.218 & 0.020 & ··· & ··· & 835970136686823040 & ··· & 0.58 & 1 \\
162.304708 & 51.042306 & $-24.46$ & 3.03 & 19.790 & 0.020 & 19.429 & 0.020 & ··· & ··· & 836722438863176448 & 1 & 0.66 & 1 \\
162.322625 & 51.057194 & $-11.87$ & 1.16 & 19.997 & 0.020 & 20.445 & 0.020 & $-2.60$ & 0.19 & 836722473222915968 & 0 & 0.81 & 1 \\
162.366083 & 51.062861 & $-13.42$ & 1.82 & 20.282 & 0.020 & 20.693 & 0.020 & $-2.62$ & 0.22 & 835977725893619712 & 0 & 0.78 & 1 \\
162.175708 & 51.033611 & $-6.48$ & 1.33 & 20.364 & 0.020 & 20.595 & 0.020 & ··· & ··· & 836722889836066048 & 1 & 0.50 & 1 \\
162.283708 & 51.040833 & $-9.97$ & 0.97 & 20.429 & 0.020 & 20.905 & 0.020 & $-1.79$ & 0.18 & 836722335783961216 & 0 & 0.90 & 1 \\
162.292208 & 51.050083 & $-11.46$ & 1.29 & 20.434 & 0.020 & 20.851 & 0.020 & $-2.72$ & 0.19 & 836722404504886912 & 0 & 0.79 & 1 \\
162.316583 & 51.040694 & $-9.50$ & 2.24 & 20.471 & 0.021 & 20.925 & 0.020 & $-2.36$ & 0.24 & 835971816018616320 & 0 & 0.62 & 1 \\
162.338167 & 51.058389 & $-12.89$ & 1.80 & 20.750 & 0.021 & 21.120 & 0.021 & $-2.85$ & 0.30 & 835971854673199616 & 0 & 0.70 & 1 \\
162.394000 & 51.044389 & $-64.30$ & 1.96 & 20.766 & 0.021 & 21.224 & 0.021 & ··· & ··· & 835977416655970816 & 0 & 0.00 & 0 \\
162.242208 & 51.048306 & $-6.74$ & 1.77 & 20.795 & 0.021 & 21.321 & 0.022 & ··· & ··· & 836722198345008512 & 0 & 0.62 & 1 \\
162.420125 & 51.061194 & $-15.91$ & 1.75 & 20.891 & 0.025 & 21.198 & 0.023 & ··· & ··· & 835977588454665856 & 0 & 0.90 & 1 \\
162.379000 & 51.061389 & $-15.15$ & 2.88 & 20.903 & 0.021 & 21.227 & 0.021 & ··· & ··· & 835977691533881088 & 0 & 0.94 & 1 \\
162.319833 & 51.067667 & $-17.51$ & 2.14 & 20.915 & 0.021 & 21.276 & 0.021 & ··· & ··· & 836722546237412608 & 0 & 0.92 & 1 \\
162.403417 & 51.064250 & $-10.79$ & 3.05 & 20.998 & 0.021 & 21.216 & 0.021 & ··· & ··· & 835977588454666368 & 0 & 0.80 & 1 \\
162.512042 & 51.042278 & $-13.00$ & 2.47 & 21.184 & 0.021 & 21.598 & 0.021 & ··· & ··· & ··· & 0 & 0.62 & 1 \\
162.379208 & 51.026111 & $-73.13$ & 2.79 & 21.270 & 0.021 & 21.699 & 0.021 & ··· & ··· & ··· & ··· & 0.00 & 0 \\
162.242000 & 51.089500 & $-98.13$ & 3.04 & 21.330 & 0.024 & 21.859 & 0.029 & ··· & ··· & ··· & 0 & 0.00 & 0 \\
162.366917 & 51.058194 & $-8.43$ & 6.41 & 21.356 & 0.021 & 21.544 & 0.021 & ··· & ··· & ··· & 0 & 0.91 & 1 \\
162.238083 & 51.042056 & $-12.68$ & 3.64 & 21.395 & 0.022 & 21.631 & 0.021 & ··· & ··· & ··· & 0 & 1.00 & 1 \\
162.366917 & 51.031000 & $-23.41$ & 5.63 & 21.405 & 0.022 & 21.613 & 0.021 & ··· & ··· & ··· & 0 & 0.78 & 1 \\
162.359708 & 51.054556 & $-8.05$ & 2.75 & 21.425 & 0.022 & 21.654 & 0.021 & ··· & ··· & ··· & 0 & 0.95 & 1 \\
162.302292 & 51.051500 & $-6.03$ & 4.13 & 21.438 & 0.022 & 21.679 & 0.021 & ··· & ··· & ··· & 0 & 0.89 & 1 \\
162.464708 & 51.031194 & $-21.26$ & 2.85 & 21.482 & 0.022 & 21.866 & 0.021 & ··· & ··· & ··· & 0 & 0.46 & 0 \\
162.354417 & 51.047889 & $0.50$ & 3.96 & 21.484 & 0.021 & 21.729 & 0.021 & ··· & ··· & ··· & 0 & 0.64 & 1 \\
162.281917 & 51.029500 & $-8.58$ & 5.55 & 21.505 & 0.022 & 21.732 & 0.021 & ··· & ··· & ··· & 0 & 0.91 & 1 \\
162.394917 & 51.074583 & $-18.42$ & 4.47 & 21.512 & 0.022 & 21.798 & 0.021 & ··· & ··· & ··· & 0 & 0.79 & 1 \\
162.185083 & 51.035028 & $-8.09$ & 3.40 & 21.521 & 0.022 & 21.770 & 0.021 & ··· & ··· & ··· & 1 & 0.90 & 1 \\
162.264208 & 51.013500 & $-3.26$ & 7.06 & 21.544 & 0.022 & 21.649 & 0.021 & ··· & ··· & ··· & ··· & 0.66 & 1 \\
162.373917 & 51.049500 & $-19.65$ & 4.41 & 21.586 & 0.022 & 21.784 & 0.021 & ··· & ··· & ··· & 0 & 0.88 & 1 \\
162.291417 & 51.015611 & $-6.23$ & 6.02 & 21.625 & 0.022 & 21.820 & 0.021 & ··· & ··· & ··· & 0 & 0.83 & 1 \\
162.296750 & 51.075083 & $-9.78$ & 3.27 & 21.638 & 0.022 & 21.865 & 0.021 & ··· & ··· & ··· & 0 & 0.91 & 1 \\
162.447375 & 51.059639 & $-7.81$ & 6.17 & 21.658 & 0.022 & 21.933 & 0.021 & ··· & ··· & ··· & 0 & 0.84 & 1 \\
162.365208 & 51.059111 & $-11.66$ & 4.76 & 21.676 & 0.022 & 21.901 & 0.021 & ··· & ··· & ··· & 0 & 0.98 & 1 \\
162.354333 & 51.040333 & $-14.41$ & 4.94 & 21.702 & 0.022 & 21.932 & 0.021 & ··· & ··· & ··· & 0 & 0.98 & 1 \\
162.325250 & 51.054694 & $-3.68$ & 10.09 & 21.715 & 0.023 & 21.842 & 0.021 & ··· & ··· & ··· & ··· & 0.75 & 1 \\
162.203083 & 51.054694 & $63.33$ & 5.23 & 21.751 & 0.022 & 22.223 & 0.022 & ··· & ··· & ··· & ··· & 0.00 & 0 \\
162.271500 & 51.028389 & $-9.53$ & 4.68 & 21.771 & 0.023 & 21.941 & 0.021 & ··· & ··· & ··· & ··· & 0.93 & 1 \\
162.340083 & 51.045917 & $-8.79$ & 6.32 & 21.798 & 0.023 & 22.021 & 0.022 & ··· & ··· & ··· & 1 & 0.90 & 1 \\
162.323292 & 51.045389 & $-18.42$ & 5.28 & 21.818 & 0.023 & 22.046 & 0.022 & ··· & ··· & ··· & 0 & 0.90 & 1 \\
162.412500 & 51.082611 & $-11.91$ & 11.86 & 21.832 & 0.023 & 22.095 & 0.023 & ··· & ··· & ··· & ··· & 0.93 & 1 \\
162.321028 & 51.072971 & $-21.50$ & 13.57 & 21.916 & 0.023 & 22.209 & 0.022 & ··· & ··· & ··· & ··· & 0.73 & 1 \\
162.339500 & 51.042889 & $-6.48$ & 10.91 & 22.000 & 0.023 & 22.209 & 0.022 & ··· & ··· & ··· & ··· & 0.94 & 1 \\
162.307500 & 51.006306 & $24.23$ & 7.34 & 22.090 & 0.023 & 22.368 & 0.022 & ··· & ··· & ··· & ··· & 0.07 & 0 \\
162.316875 & 51.049889 & $-13.64$ & 7.03 & 22.110 & 0.024 & 22.314 & 0.022 & ··· & ··· & ··· & ··· & 0.99 & 1 \\
162.362833 & 51.058361 & $-9.94$ & 14.21 & 22.118 & 0.024 & 22.334 & 0.023 & ··· & ··· & ··· & ··· & 0.98 & 1 \\
162.348625 & 51.051083 & $7.51$ & 4.74 & 22.141 & 0.024 & 22.407 & 0.023 & ··· & ··· & ··· & ··· & 0.38 & 0 \\
162.315500 & 51.022611 & $-29.34$ & 5.03 & 22.142 & 0.024 & 22.478 & 0.023 & ··· & ··· & ··· & ··· & 0.41 & 0 \\
162.261708 & 50.994500 & $-13.21$ & 11.12 & 22.164 & 0.024 & 22.416 & 0.023 & ··· & ··· & ··· & ··· & 0.95 & 1 \\
162.450625 & 51.030111 & $-15.70$ & 5.60 & 22.250 & 0.025 & 22.566 & 0.023 & ··· & ··· & ··· & ··· & 0.81 & 1 \\
162.481266 & 51.063726 & $-7.74$ & 5.97 & 22.271 & 0.025 & 22.540 & 0.024 & ··· & ··· & ··· & ··· & 0.85 & 1 \\
162.312208 & 51.048306 & $-0.75$ & 5.33 & 22.355 & 0.026 & 22.610 & 0.024 & ··· & ··· & ··· & 0 & 0.73 & 1 \\
162.191875 & 51.000000 & $-56.49$ & 14.64 & 22.364 & 0.025 & 22.625 & 0.023 & ··· & ··· & ··· & ··· & 0.10 & 0 \\
162.328792 & 51.081500 & $-28.43$ & 13.10 & 22.430 & 0.026 & 22.656 & 0.024 & ··· & ··· & ··· & ··· & 0.68 & 1 \\
162.306000 & 51.045389 & $-2.56$ & 6.82 & 22.446 & 0.026 & 22.751 & 0.024 & ··· & ··· & ··· & 0 & 0.71 & 1 \\
162.386417 & 51.074194 & $-17.77$ & 7.54 & 22.448 & 0.026 & 22.757 & 0.024 & ··· & ··· & ··· & ··· & 0.83 & 1 \\
162.440208 & 51.035806 & $-23.59$ & 6.94 & 22.501 & 0.027 & 22.811 & 0.024 & ··· & ··· & ··· & ··· & 0.74 & 1 \\
162.448375 & 51.034389 & $-16.27$ & 10.24 & 22.555 & 0.028 & 22.848 & 0.025 & ··· & ··· & ··· & ··· & 0.90 & 1 \\
162.295375 & 51.037194 & $-26.38$ & 12.14 & 22.591 & 0.028 & 22.878 & 0.025 & ··· & ··· & ··· & ··· & 0.74 & 1 \\
162.364917 & 51.033611 & $-32.65$ & 6.73 & 22.615 & 0.028 & 22.855 & 0.025 & ··· & ··· & ··· & 0 & 0.45 & 0 \\
162.341583 & 51.055000 & $-37.80$ & 13.69 & 22.640 & 0.028 & 22.957 & 0.025 & ··· & ··· & ··· & 0 & 0.42 & 0 \\
162.292417 & 51.082694 & $50.93$ & 10.37 & 22.752 & 0.028 & 23.268 & 0.029 & ··· & ··· & ··· & ··· & 0.00 & 0 \\
162.434917 & 51.050694 & $-28.43$ & 2.07 & 22.774 & 0.030 & 22.994 & 0.026 & ··· & ··· & ··· & ··· & 0.52 & 1 \\
162.281203 & 51.055597 & $8.72$ & 8.25 & 22.793 & 0.031 & 23.199 & 0.027 & ··· & ··· & ··· & ··· & 0.29 & 0 \\
162.373000 & 51.062611 & $82.11$ & 8.87 & 22.829 & 0.032 & 23.124 & 0.028 & ··· & ··· & ··· & ··· & 0.00 & 0 \\
162.269708 & 51.054611 & $-23.40$ & 6.03 & 22.836 & 0.031 & 23.088 & 0.026 & ··· & ··· & ··· & 1 & 0.76 & 1 \\
162.326000 & 51.059500 & $-24.76$ & 9.18 & 22.858 & 0.031 & 23.114 & 0.027 & ··· & ··· & ··· & ··· & 0.76 & 1 \\
162.227000 & 51.057889 & $-22.41$ & 9.30 & 23.014 & 0.034 & 23.308 & 0.030 & ··· & ··· & ··· & ··· & 0.84 & 1 \\
162.496000 & 51.048806 & $32.14$ & 6.15 & 23.028 & 0.034 & 23.339 & 0.030 & ··· & ··· & ··· & ··· & 0.02 & 0 \\
162.216547 & 51.040881 & $29.80$ & 8.04 & 23.233 & 0.042 & 23.659 & 0.035 & ··· & ··· & ··· & ··· & 0.03 & 0 \\

\enddata
\end{deluxetable*}

\end{document}